\documentclass[11pt,a4paper]{article}
\usepackage{jheppub}
\usepackage{caption}
\usepackage{amsmath}
\usepackage[dvipsnames]{xcolor}
\usepackage{amsfonts}
\usepackage{tikz}
\usetikzlibrary{positioning}
\usetikzlibrary{calc}
\usepackage{physics}
\usepackage{float}
\usepackage{bbold}
\usepackage{multirow}
\usepackage{longtable}
\usepackage{makecell}  % For line breaks inside table cells
\usepackage{bm} 
\usepackage{graphicx} % Required for inserting images
\usepackage{float}

\floatstyle{plain}
\newfloat{eqfloat}{!htbp}{loe}
\floatname{eqfloat}{{\bf Equation}}

\makeatletter
  \let\c@eqfloat=\c@equation
  
\makeatother

\newcommand{\MarkerCross}{%
  \tikz[baseline=-0.9ex]{%
    \draw (0,0) circle (0.8ex);
    \draw (-0.6ex,-0.6ex)--(0.6ex,0.6ex);
    \draw (-0.6ex,0.6ex)--(0.6ex,-0.6ex);
  }%
}

\newcommand{\MarkerLightGray}{%
  \tikz[baseline=-0.9ex]{%
    \fill[gray!30] (0,0) circle (0.8ex);
    \draw (0,0) circle (0.8ex);
  }%
}

\newcommand{\MarkerDarkGray}{%
  \tikz[baseline=-0.9ex]{%
    \fill[black!70] (0,0) circle (0.8ex);
    \draw (0,0) circle (0.8ex);
  }%
}

\newcommand{\MarkerCrossSmall}{%
  \tikz[baseline=-0.9ex, scale=0.7]{%
    \draw (0,0) circle (0.8ex);
    \draw (-0.6ex,-0.6ex)--(0.6ex,0.6ex);
    \draw (-0.6ex,0.6ex)--(0.6ex,-0.6ex);
  }%
}

\newcommand{\MarkerLightGraySmall}{%
  \tikz[baseline=-0.9ex, scale=0.7]{%
    \fill[gray!30] (0,0) circle (0.8ex);
    \draw (0,0) circle (0.8ex);
  }%
}

\newcommand{\MarkerDarkGraySmall}{%
  \tikz[baseline=-0.9ex, scale=0.7]{%
    \fill[black!70] (0,0) circle (0.8ex);
    \draw (0,0) circle (0.8ex);
  }%
}

\newcommand{\sig}{{\bm\sigma}}
\def\[{\left[}
\def\]{\right]}
\def\({\left(}
\def\){\right)}

\newcommand\beqa{\begin{eqnarray}}

\newcommand{\mycomment}[1]{}
\newcommand\eeqa{\end{eqnarray}}
\newcommand{\beq}{\begin{eqnarray}}
\newcommand{\eeq}{\end{eqnarray}}
\newcommand{\la}[1]{\label{#1}}
\newcommand{\eq}[1]{(\ref{#1})}

\newcommand{\bR}{{\bf R}}
\newcommand{\bB}{{\bf B}}
\newcommand{\mQ}{{\mathbb Q}}
\newcommand{\es}{\emptyset}
\newcommand{\ii}{i}
\newcommand{\betheQ}{\mathbb{Q}}
\newcommand{\bfB}{\mathbf{B}}
\newcommand{\bfR}{\mathbf{R}}

\newcommand{\Wr}{\text{Wr}}
\newcommand{\fullset}{\bar{\emptyset}}
\newcommand{\solId}{\mathtt{solId}}
\newcommand{\algsu}{\mathfrak{su}}
\newcommand{\Leff}{L_{\text{eff}}}
\newcommand{\pf}{\text{pf}}

\newcommand{\bP}{\mathbf{P}}
\newcommand{\bp}{\mathbf{p}}
\newcommand{\bQ}{\mathbf{Q}}

\newcommand{\algsl}{\mathfrak{sl}}

\title{Regge Trajectories of ${\cal N}=4$ SYM Part I:\\ General Asymptotic Baxter-Bethe Ansatz}
\author{Simon Ekhammar${}^{\MarkerLightGraySmall\,\MarkerCrossSmall}$, Nikolay Gromov${}^{\MarkerLightGraySmall}$ and Michelangelo Preti${}^{\MarkerDarkGraySmall}$}
\affiliation{
    $\MarkerLightGray$ Mathematics Department, King's College London,
    The Strand, London WC2R 2LS, UK;\\
    $\MarkerCross$ Department of Physics and Astronomy, Uppsala University,
Box 516, SE-751 20 Uppsala, Sweden \\
    $\MarkerDarkGray$
    Simons centre for Geometry and Physics, Stony Brook University, Stony Brook, New York 11794, USA; 
    C. N. Yang Institute for Theoretical Physics, Stony Brook University, Stony Brook, New York 11794, USA;
    Dipartimento di Fisica, Universit\`a di Torino and INFN - Sezione di Torino, Via P. Giuria 1, Torino 10125, Italy
    }
    
    \emailAdd{simon.ekhammar@kcl.ac.uk}
 \emailAdd{nikolay.gromov@kcl.ac.uk}
 \emailAdd{michelangelo.preti@stonybrook.edu}

\date{February 2024}

\begin{document}

\abstract{
In this work, we derive a novel set of equations---the \emph{Asymptotic Baxter--Bethe Ansatz}---that determine the asymptotic spectrum of Regge trajectories in the BFKL regime of $\mathcal{N}=4$ SYM. In this challenging limit, our method yields multi-loop results in the ’t~Hooft coupling, with the perturbative accuracy increasing as the quantum numbers grow. Our formalism not only provides a straightforward path to obtain multi-loop perturbative data, as we demonstrate, but also enables the classification of trajectories, paving the way for systematic non-perturbative studies up to the strong-coupling regime.
}

\dedicated{
  {\it To the memory of Rodney~J.~Baxter,\\
  whose pioneering work on integrable models\\
  has been a constant inspiration for this research.}
}

\maketitle

\section{Introduction}

Regge trajectories offer a powerful organising  principle in high-energy physics, relating the spin of exchanged states to their squared mass. They play a central role in both hadron spectroscopy and the high-energy behaviour of scattering amplitudes~\cite{Kuraev:1976ge,Kuraev:1977fs,Balitsky:1978ic}. In perturbative quantum chromodynamics (QCD), the Balitsky–Fadin–Kuraev–Lipatov (BFKL) equation resums leading logarithms in the Regge limit, providing predictions for processes such as deep inelastic scattering and forward-jet production up to next-to-leading order~\cite{Fadin:1998py,Ciafaloni:1998gs,Kotikov:2002ab}. However, the intrinsically non-perturbative nature of Regge physics renders any perturbation-theory-based approach both highly intricate and, to some extent, uncertain.

Maximally supersymmetric Yang–Mills theory ($\mathcal{N}=4$ SYM) in four dimensions offers an ideal laboratory for non-perturbative Regge studies thanks to its integrability in the planar limit. While $\mathcal{N}=4$ SYM is a conformal field theory and thus do not feature massive particles, one can still develop conformal Regge trajectory \cite{Cornalba:2007fs,Costa:2012cb}. The quantum spectral curve (QSC) provides a universal description of the spectrum at finite ’t Hooft coupling \cite{Gromov:2013pga,Gromov:2014caa}. While the QSC grants access to non-perturbative data, in many cases one has to rely entirely on numerical methods. This makes the analysis of interesting regimes time-consuming and often requiring a case-by-case study. At the same time, in the asymptotic regime (i.e. the limit of large length or small coupling), the spectral problem for local operators simplifies dramatically to the algebraic all-loop Beisert–Eden–Staudacher (BES) equations \cite{Beisert:2005fw,Beisert:2006ez}, also know as Asymptotic Bethe Ansatz (ABA), which are significantly more analytically tractable and easier to work with. In this paper, we show that a similar simplification can be achieved in the BFKL regime as well, revealing new intricate structures.

The BFKL regime of ${\cal N}=4$ SYM has already been studied in the literature. E.g. the simplest trajectory, known as Pomeron, was investigated at weak coupling in \cite{Kotikov:2007cy,Kotikov:2000pm,Janik:2013nqa} from the side of the local operators. The integrability methods were then extended to the full BFKL regime for the first time in \cite{Alfimov:2014bwa}
using the QSC framework. It was subsequently studied non-perturbatively at NNLO order and further generalised to include conformal spin in \cite{Gromov:2015wca,Gromov:2015vua,Alfimov:2018cms}. The NNLO was thereafter confirmed independently in \cite{Caron-Huot:2016tzz,Velizhanin:2015xsa}. More recently, both numerical and analytical QSC techniques were employed to investigate the Regge Riemann surface that contains twist $3$ operators in \cite{Klabbers:2023zdz}. Novel structures, such as bridge trajectories, were recently investigated in \cite{Brizio:2024nso}. However, a systematic study of $\mathcal{N}=4$ SYM Regge trajectories still requires a more robust and analytically tractable approach.
In this work, we derive and extend the ``long-range" Asymptotic Baxter–Bethe Ansatz (ABBA), originally reported in a letter by the authors of this paper in \cite{Ekhammar:2024neh}, to the most general case of arbitrary quantum numbers. 

While our results share some similarities with the BES equations, 
they have several novel features, indicating the existence of new and exciting physics that is yet to be fully uncovered. In particular, it reveals novel massless modes responsible for the appearance of odd powers of the coupling in the weak-coupling expansion. This provides a clear interpretation of the odd powers found in \cite{Klabbers:2023zdz} as arising from a massless dispersion relation. Another distinctive feature of our framework is that some of the resulting equations cannot be cast in a Bethe-like form, unlike the BES equations.  Instead, they must be expressed in a Baxter form reminiscent of the Baxter equation for the Pomeron studied in \cite{Lipatov:1993yb,Faddeev:1994zg,DeVega:2001pu}. This is the reason we refer to our approach as the Asymptotic Baxter–Bethe Ansatz (ABBA).

The main motivation for studying Regge trajectories in ${\cal N}=4$ SYM is the striking similarity that this regime shares with its QCD counterpart.
In particular, ${\cal N}=4$ SYM provides direct predictions for the most intricate parts of QCD BFKL data, as captured by the Kotikov–Lipatov maximal transcendentality principle \cite{Kotikov:2002ab}. This correspondence is well established in the case of the Pomeron \cite{Gromov:2015vua,Kotikov:2010nd} and applies to many other scenarios \cite{Brandhuber:2012vm,Loebbert_2015,Brandhuber_2016,Loebbert_2016,Brandhuber:2017bkg,Li:2014afw,Li_2017}.
Furthermore, at the leading order in perturbation theory, depending on the particular observable, the BFKL regimes of both theories are frequently indistinguishable due to gluon dominance in this regime.

Another key motivation arises from the growing interest in non-local operators in recent years. At weak coupling, for instance, non-local light-ray operators~\cite{Balitsky:1995ub,Balitsky:1997mj,Balitsky:2013npa,Homrich:2024nwc} realised as null Wilson lines allow for an analytic continuation in Lorentz spin. These operators have recently attracted attention in the context of conformal field theory (CFT) and the conformal bootstrap program~\cite{kravchuk2018light,Kologlu:2019mfz,Caron-Huot:2016tzz}. The BFKL regime involves even more intricate non-local operators, whose structure remains only partially understood~\cite{Chang:2025zib,Moult:2025nhu}. Gaining access to the full spectrum of such operators could unveil the physics beyond the spectral problem as well, e.g., potentially extending the “Bootstrability” program \cite{Cavaglia:2021bnz,Cavaglia:2022qpg,Cavaglia:2022yvv,Caron-Huot:2022sdy,Cavaglia:2023mmu,Caron-Huot:2024tzr,Cavaglia:2024dkk} to this new regime.

\paragraph{Regge trajectories in $\mathcal{N}=4$ SYM.}\label{sec:ReggeInN4}
In a CFT, Regge trajectories describe how the spin $S_1$ depends on the scaling dimension $\Delta$, or equivalently, the more commonly used inverse relation  $\Delta(S_1)$ \footnote{We have included a subscript for the spin since in $\mathcal{N}=4$, there exist $2$ independent spin numbers; $S_1$ and $S_2$. We will discuss the details in Section~\ref{sec:QuantumNumbers}.}. At specific integer values of $S_1$ one can find $\Delta$ by considering local operators. Generic points can then be understood as an analytic continuation away from these isolated points, resulting in a full Riemann surface to explore. This surface will feature a variety of branch points, typically located at complex values of $\Delta$, as well as a $\Delta \leftrightarrow -\Delta$ symmetry originating from the possibility of constructing a shadow operator for each physical operator. By slicing this Riemann surface along the real $S_1$-$\Delta$ plane, one obtains the Chew-Frautschi plot. One can attempt to associate non-local operators, such as light-ray operators, with each point on the Chew-Frautschi plot. However, in this work, we operate at the level of the spectrum using the QSC, allowing us to evade the question of the exact form of such operators.

To orient our discussion and more precisely state the objective of this series of papers, it is useful to take a closer look at the Chew–Frautschi plot. At zero coupling, Regge trajectories organize themselves into two categories, which we will call ``\textit{Diagonal Trajectories}" (DTs) and ``\textit{Horizontal Trajectories}" (HTs). 
The DTs are linear trajectories in the Chew-Frautschi plot with a slope $\pm 1$, the two different species originating from the $\Delta\leftrightarrow -\Delta$ symmetry. Trajectories with a positive slope contain local operators at non-negative values of the spin; those with a negative slope contain shadow operators.
On the other hand, HTs are trajectories with fixed $S_1$ for any value of $\Delta$. They do not include any local operators; rather, each point should be understood as an extended object. These trajectories correspond to the BFKL regime. 

\begin{figure}
    \centering   
  {\includegraphics[width=\columnwidth]{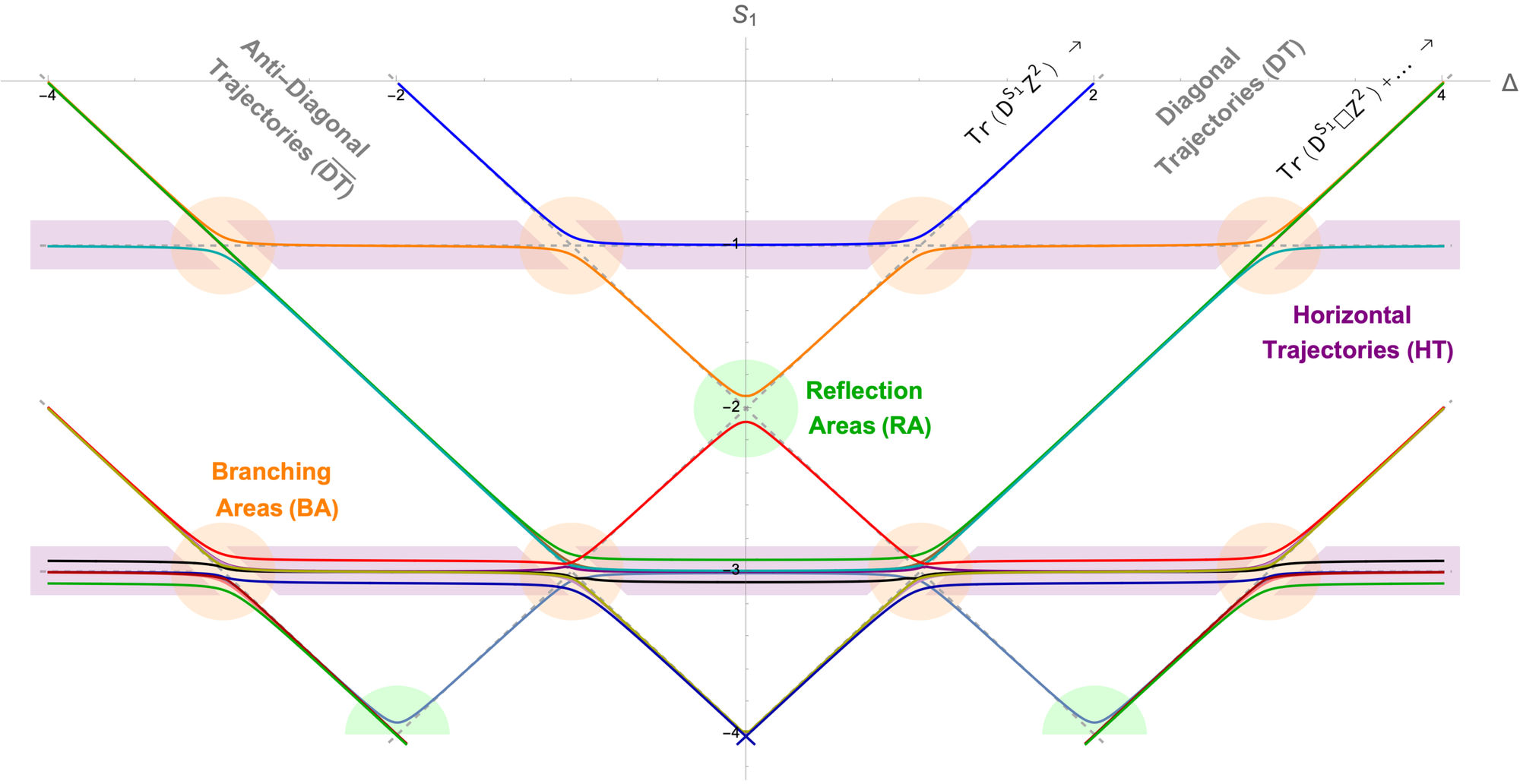}}
    \caption{The real plane slice (Chew–Frautschi plot) of the Regge Riemann surface containing twist-2 operators at the value of the coupling $g=1/50$. The plot represents the analytic continuation from (even) integer values of spin $S_1$ where the dimensions $\Delta$ of local operators of the type $\tr Z\nabla_+^{S_1}Z$ are situated (on the diagonal segment of the top blue curve that we refer to as leading trajectory). By analytically continuing in the complex plane in $\Delta$, one can also reach subleading trajectories where operators of higher-twists sit. For instance twist-4 operators of the type $\tr Z\Box \nabla_+^{S_1}Z+...$ sitting along the orange and green curves (which are no longer in $\frak{sl}_2$ sector). They usually are very complicated linear combinations of fields as explicitly shown in \cite{Klabbers:2023zdz} for subleading twist-3 case. All the operators appear at positive values of the spin $S_1$. Here we present the negative $S_1$ half-plane where the structure of how the Riemann sheets connect is evident. The picture is generated with around 11000 data points computed with the numerical QSC.
    \label{fig:L2full}}
\end{figure}

For a positive spin, there are no HTs, and the DTs do not intersect. However, at negative spin, HTs and DTs coexist, and once a small coupling is turned on, they continuously connect to each other. At negative spin, we identify three distinct types of regions, which we analyse in detail throughout this series of papers.
The first region consists of the HTs themselves. We  study the number of these trajectories, their behaviour at weak coupling, and, using numerical methods, their behaviour at strong coupling.
The other two regions of interest are those where trajectories would naively intersect but instead reconnect at finite coupling. We refer to the region where HTs and DTs cross as a Branching Area (BA), and the region where DTs would collide as a Reflection Area (RA). In both the BA and RA, our goal is to understand how the trajectories connect and reorganise.
To help the reader become familiar with these concepts, we have collected data on the Riemann surface that contains the $\mathcal{N}=4$ Pomeron and highlighted the relevant regions in Figure~\ref{fig:L2full}.

The main tool that will allow for the analysis of these different regimes is the Asymptotic Baxter-Bethe Ansatz (ABBA), introduced earlier. It provides a framework for studying each regime systematically. In this paper (Part I of the series), we focus on the general structure of ABBA and apply it to some of the simplest horizontal trajectories. We will return to the Branching Area and Reflection Area in Part II~\cite{EkhammarGromovPreti:2025b}, and explore more general horizontal trajectories in~\cite{UpcommingArpad}.

\paragraph{Content and structure of the paper.}

The paper is organised as follows.  In Section~\ref{sec:QuantumNumbers} we discuss how to label trajectories of $\mathcal{N}=4$ using quantum numbers and extra charges. In Section~\ref{sec:ABBA} we present the full asymptotic Baxter–Bethe ansatz (ABBA), and in Section~\ref{sec:ABBADerivation} we sketch the derivation from the Quantum Spectral Curve.  Section~\ref{sec:minimal} analyses the leading (minimal) horizontal trajectories, HT$_0$, while Section~\ref{sec:SubleadingHT} discusses in detail how to solve ABBA for HT$_2$ for parity symmetric $\algsl_2$ trajectories, providing the simplest example of subleading horizontal trajectories.  Section~\ref{sec:Intercepts} discusses the computation of intercepts and offers some strong coupling analysis. Finally, Section~\ref{sec:Conclusions} concludes and provides an outlook.

We collect several technical details in the appendices. Appendix~\ref{app:QSCGeneralities} reviews key aspects of the QSC; Appendix~\ref{app:ABBADerivation} gives the full derivation of the ABBA; Appendix~\ref{app:DressingPhases} details the dressing phases; Appendix~\ref{app:StartingUpNumerics} describes the how to find starting points for a numerical analysis from ABBA and Appendix~\ref{App:Qrelations} compiles the Baxter equation and Q‐system relations.

Finally, we have included two notebooks with this submission. The first one solves the ABBA equations for the minimal sector, which will be discussed in Section~\ref{sec:minimal}. The second notebook provides a numerical solution of ABBA for the simplest case of subleading operators, the focus of Section~\ref{sec:SubleadingHT}.

\section{Labelling Trajectories and Areas of Interest} 
\la{sec:QuantumNumbers}

The states of ${\cal N}=4$ SYM can be organised using its global $\mathfrak{psu}(2,2|4)$ symmetry. All unprotected local operators belong to a long multiplet of this algebra, most often specified by giving the weights of the superconformal primary of the multiplet. In this paper, we will instead label the multiplet using a different highest weight state, corresponding to a different choice of positive roots, usually called the ABA-grading. We refer to Appendix C  in \cite{Gromov:2014caa} for all details. The prototypical example is the Konishi multiplet for which $\text{tr} \left(Z\bar{Z}+X\bar{X}+Y\bar{Y}\right)$ is the superconformal primary, while we will label the multiplet by using $\tr \nabla_+^2 Z^{2}$ representative. Here and in the following $X,Y,Z$ are the complex scalars of $\mathcal{N}=4$ SYM.

Generally, all multiplets are partially characterised by the quantum numbers of the bosonic subalgebras of $\mathfrak{psu}(2,2|4)$, that is
\beqa
\Delta,\;S_1,\;S_2 \quad&,&\quad\text{Conformal algebra } \mathfrak{so}(2,4)\\
J_1,\;J_2,\;J_3 \quad&,&\quad\text{R-symmetry } \mathfrak{so}(6) \simeq \mathfrak{su}(4)
\eeqa
where, for local operators, the charges should satisfy
\begin{equation}\la{inequalities}
    J_1 - J_2 \geq 2\,,
    \quad
    J_2 \pm J_{3} \geq 0\,,
    \quad
    S_1 \pm S_{2} \geq 2\,,
\end{equation}
and $\Delta$ is constrained by the unitarity condition
\begin{equation}\label{eq:Unitarity}
    \Delta \geq J_1+J_2 +S_1+\abs{J_3-S_2}\,.
\end{equation}

The trajectories we study in this paper can be obtained by starting from a local operator and then performing analytic continuation in $S_1$, while preserving all other integer quantum numbers\footnote{The analytic continuation in both $S_1$ and $S_2$ can be also constructed naturally within the QSC formalism as is explained and studied numerically in \cite{Alfimov:2018cms}. It would be an interesting extension of our results to consider these more general non-local operators, which should correspond to the light-ray operators further smeared over the orthogonal direction at weak coupling. }. 
The trajectory, which in general is a multi-valued Riemann surface $\Delta(S_1)$ or more conventionally $S_1(\Delta)$, is then parametrised
specified by the four integer quantum numbers
\begin{equation}\label{QNumbers}
S_2,\;J_1,\;J_2,\;J_3\;.
\end{equation}
In general, there are infinitely many distinct trajectories sharing these quantum numbers\footnote{By a \emph{trajectory} we mean the real‐$\Delta$ curve $S_1(\Delta)$.  More formally, one may consider the associated Regge Riemann surface, whose different real sections correspond to these trajectories.  We expect that all such trajectories with the same charges in \eqref{QNumbers} are related by analytic continuation, once two additional discrete $\mathbb{Z}_2$ charges—parity and the left–right symmetry that flips the Dynkin labels $S_2\to -S_2$, $J_3\to -J_3$—are also specified. We thank A.~Zhiboedov for discussing this point. }.

Note that this is similar to local operators; there are infinitely many local operators with the same integer quantum numbers, as one can always add a neutral combination, say $X\bar{X}$. This would, however, shift the bare dimension $\Delta_0$. For the local operators, all quantum numbers are fixed to be integers, whereas $\Delta$ usually changes continuously with $g$; nevertheless, $\Delta_0=\Delta(g=0)$ is a useful additional label \footnote{Keep in mind that $\Delta(g)$ is a multi-valued function of $g$, so $\Delta_0$ is strictly speaking is not uniquely defined.}.
Fixing $\Delta_0$ would lead to a finite number of states.

Similarly, for the case of the trajectories, adding additional labels, which characterise the weak coupling behaviour, reduces the infinite set of
trajectories to a tractable finite set.

Another convenient set of parameters are the spin chain parameters arising in the ABA description of the local operators: the lengths of the spin chain $L$ and the magnon numbers. In our case these are $K_1,\;K_3,\;K_4,\;K_5,\;K_7$, all non-negative integers. In addition, we have a label for the number of massless modes $N$, which we will argue below is $4$ for HTs and $2$ for the RAs. Note that, as we mentioned above, for the Regge trajectories the role of the energy is played by $-S_1$, which becomes a function of the coupling, whereas $\Delta$ will be fixed and plays the role of an external parameter.

The additional labels are related to the global symmetry quantum numbers and the bare ``dimension'' 
\beq
\Sigma\equiv -S_1(g=0)\,,
\eeq
as follows:
\beq\la{magnonnumbers}
\begin{array}{l}
 K_1=\frac{1}{2} \left(L-J_1-J_2+J_3\right) \\
 K_3=\frac{1}{2} \left(-L+N-J_1-J_2+J_3+
 2\Sigma-2\right) \\
K_4=\Sigma-J_1+1 \\
 K_5=\frac{1}{2} \left(-L+N-J_1-J_2-J_3+2\Sigma-2\right) \\
 K_7=\frac{1}{2} \left(L-J_1-J_2-J_3\right) \\
\end{array}\;.
\eeq
As the spin $S_1$ is no longer an integer, it gets an anomalous ``dimension" defined by\footnote{ In the literature, the anomalous ``dimension" is often denoted as $\omega$ with $\omega=-\gamma_S$ as in \cite{Ekhammar:2024neh}.}
\beq\label{gammaSdef}
 -S_1\equiv \Sigma+\gamma_S\,,
\eeq
where $\gamma_S$ vanishes at $g=0$. Its computation is the main purpose of Section \ref{sec:ABBA}. Before that, we label the various areas of interest in the Chew-Frautschi plot.

\begin{figure}[!t]
    \centering
    \includegraphics[width=\columnwidth]{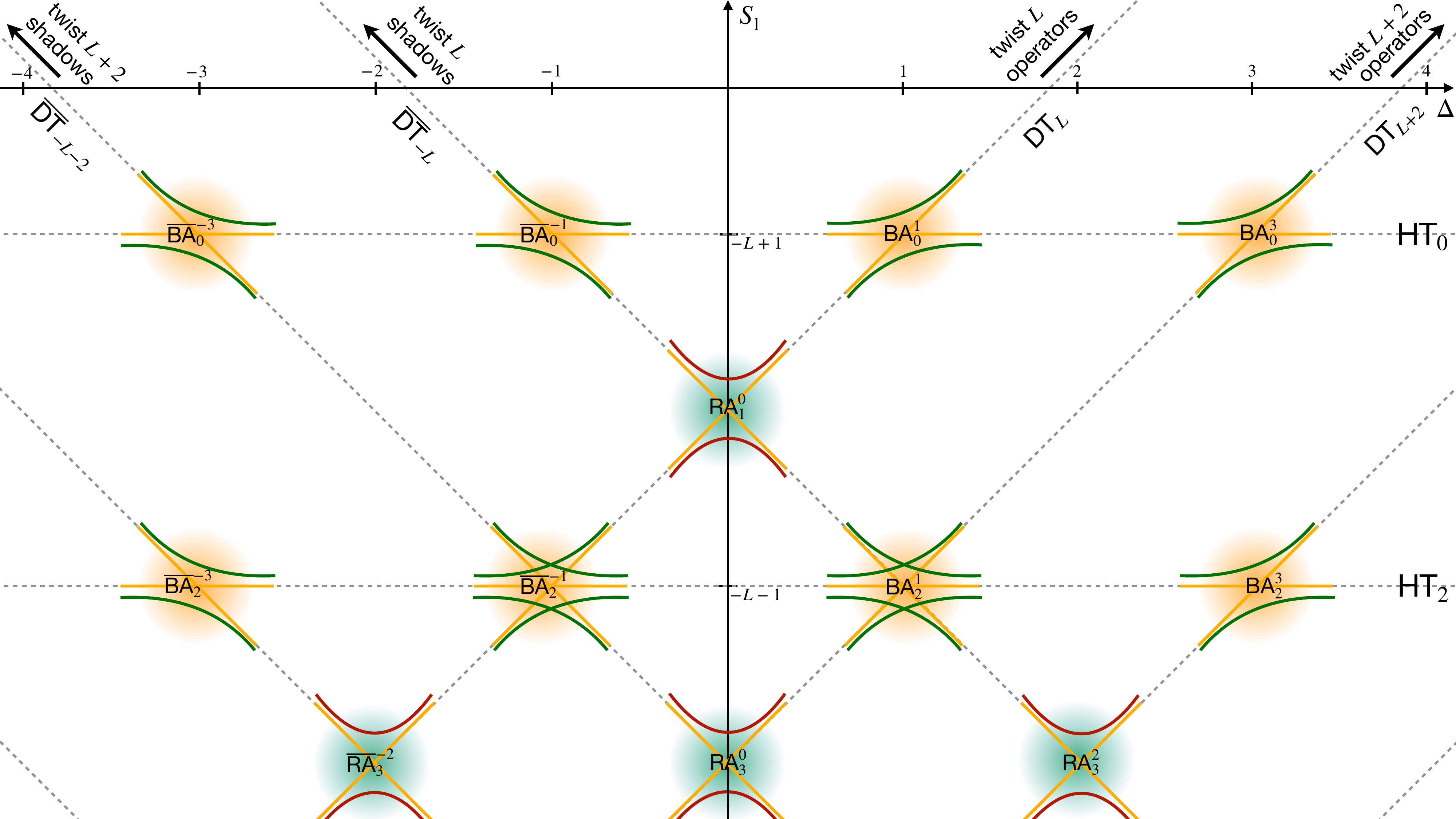}
    \caption{The schematic Chew–Frautschi plot of the Regge Riemann surface containing twist-$L$ parity even operators in $\mathfrak{sl}(2)$ sector. The diagram indicates all main distinct features, this includes Horizontal trajectories (HT), Diagonal trajectories (DT), anti-Diagonal trajectories ($\overline{\rm DT}$), Reflection areas (RA) and Branching areas (BA). They are all labelled by $\Delta$ and the number of massive magnons $K_4$, while the twist $L$ just shift them deeper on the negative $S_1$ plane.
    Dashed gray lines represent the asymptotic shape of the trajectories for $g\rightarrow 0$. Intersection points correspond to BAs (orange) and RAs (green). For any of them, we schematically present how trajectories behave when $g$ is finite: bending from DT to HT (green lines), going through branch-points keeping the same slope (orange lines) and bending from DT to $\overline{\rm DT}$ (red lines).}
    \label{fig:generalTrajSL2}
\end{figure}

\subsection{Horizontal Trajectories, Branching Areas and Reflection Areas}
\la{sec:structure}

In Section~\ref{sec:ReggeInN4}, we identified the three regions: Horizontal Trajectories, Branching Areas, and Reflection Areas. It turns out that the quantum number $K_4$ provides a convenient and natural labelling scheme for these areas. Although we derive the ABBA for the general case in this paper, the typical trajectories we use as examples are parity-symmetric ones containing only $\mathfrak{sl}(2)$ operators, i.e.  $J_1 = L$ and $J_2 = J_3 = S_2 = 0$. To help orient the reader, we have included the schematic Figure~\ref{fig:generalTrajSL2}, which illustrates the typical structure of trajectories of this type.

For clarity, we will use the following notation throughout the paper to refer to the regions of interest in the Chew–Frautschi plot:
\begin{itemize}
    \item ${\rm HT}_{K_4}$ - \textit{Horizontal trajectories} (a.k.a. BFKL regime), the parts of the curve which flatten in $g\to 0 $ limit. The HTs are fully described by the ABBA equations, which we present in this paper to multiple orders in perturbation theory ($\sim g^{L+1}$). In the parity symmetric $\frak{sl}_2$ sector, they are evenly spaced by steps of 2 since $\Sigma=L+2n-1$ with $n\in \mathbb{N}$. We refer to HT$_0$ as the leading horizontal trajectory (studied also in \cite{Ekhammar:2024neh} for a particular class of states) and HT$_{2}$ is the subleading one. In the absence of parity symmetry, we expect that $K_4$ is no longer restricted to be even, allowing for half-integer $n$'s. 
    \item ${\rm DT}_{\Delta+\Sigma}$ - \textit{Diagonal trajectories} (a.k.a. DGLAP regime), the parts of the curve which in the limit $g\to 0$ form $45^\circ$ sloped curve, for sufficiently large $\Delta$ they contain local operators. These are not described by ABBA. Only the vicinity of the intersection with the HTs can be analysed at the moment with this technique. They are labelled by the twist of the operators they contain.    
    \item $\overline{\rm DT}_{-\Delta-\Sigma}$ - \textit{anti-Diagonal trajectories}, they appear reflected to the DT by $\Delta\to -\Delta$ and contain shadow operators, as opposed to the local operators.
    \item ${\rm RA}_{K_4}^{\Delta}$ - \textit{Reflection areas}, they refer to the points where diagonal trajectories transition into anti-diagonal trajectories. In the parity symmetric $\algsl_2$ sector, they exist for odd $K_4$, appearing in-between HTs on the Chew-Frautschi plot. In this sector, the number of reflection areas varies with $K_4$ and is simply equal to $K_4$ itself. Indeed, for a given $K_4$ they sit at $\Delta=-K_4+1,-K_4+3,...,0,...,K_4-1$. Due to the $\Delta\rightarrow -\Delta$ symmetry, RAs at negative values of $\Delta$ (indicated by $\overline{\rm RA}$) are simply reflections of the ones on the positive half-plane.    
    Those regions can be analysed with ABBA, as discussed in part~II~\cite{EkhammarGromovPreti:2025b} of this series of papers.
    \item ${\rm BA}_{K_4}^{\Delta}$ - \textit{Branching areas}, they are the regions in the vicinity of the point where diagonal trajectories transition into horizontal trajectories. 
    For any values of $K_4$, there exist infinitely many BAs that appear in two different fashions depending on whether both DTs and $\overline{\rm DT}$s intersect at the same point. In the parity-symmetric $\frak{sl}_2$ sector, they are located at odd integer values of $\Delta$. 
    Due to the $\Delta\rightarrow -\Delta$ symmetry, BAs at negative values of $\Delta$ (indicated by $\overline{\rm BA}$) are simply reflections of the ones on the positive half-plane. We also reserve the ABBA analysis of branching areas to the part~II~\cite{EkhammarGromovPreti:2025b} of this series of papers.
\end{itemize}

\section{The Asymptotic Baxter-Bethe Ansatz}\label{sec:ABBA}
\begin{eqfloat}[!htbp]
\centering
\beq\hspace{-0.5cm}
\small{
\begin{aligned}
&\begin{tikzpicture}[baseline=(mid23.base),
node distance=1.65cm,
dynkinnode/.style={circle,draw,minimum size=6.8mm,inner sep=0pt}]
  \node[dynkinnode,label=left:{1}] (a1) {};
  \node[dynkinnode,label=left:{2}] [below=of a1] (a2) {};
  \node[dynkinnode,label=left:{3}] [below=of a2] (a3) {};
  \node[dynkinnode,fill=black!70,label=left:{4}] [below=of a3] (a4) {};
  \node[dynkinnode,label=left:{5}] [below=of a4] (a5) {};
  \node[dynkinnode,label=left:{6}] [below=of a5] (a6) {};
  \node[dynkinnode,label=left:{7}] [below=of a6] (a7) {};
  
  \node[inner sep=0pt, minimum size=0pt] (mid23) at ($(a4)!0.99!(a5)$) {};
  \node[dynkinnode,fill=gray!30,label=above:{massless}] [below=of a7] (extra) {};

  \draw[thick] (a1)--(a2)--(a3)--(a4)--(a5)--(a6)--(a7);
  % Additional disconnected node with light blue background
  % Cross marks on nodes 1,3,5,7
  \draw (a1.north west)--(a1.south east);
  \draw (a1.north east)--(a1.south west);
  \draw (a3.north west)--(a3.south east);
  \draw (a3.north east)--(a3.south west);
  \draw (a5.north west)--(a5.south east);
  \draw (a5.north east)--(a5.south west);
  \draw (a7.north west)--(a7.south east);
  \draw (a7.north east)--(a7.south west);
\end{tikzpicture}
\hspace{-0.5cm}
&\quad
\begin{aligned}
% node 1
\nonumber 1&=\frac{{\mathbb Q}_2(u_{1,k}+\tfrac{i}{2})}{{\mathbb Q}_2(u_{1,k}-\tfrac{i}{2})}
\prod_{j=1}^{K_1}\frac{1-\tfrac{1}{x_{1,k} x_{4,j}^+}}{1-\tfrac{1}{x_{1,k} x_{4,j}^-}}
{\color{RoyalBlue}
\prod_{j=1}^{N/2}\frac{x_{1,k}-\tfrac{1}{z_{N/2+j}}}{x_{1,k}-z_{j}}}\\[19pt]
% node 2
\nonumber-{\mathbb T}(u){\mathbb Q}_2(u)&={\mathbb Q}_2(u-i)
\prod_{j=1}^{K_1}(u-u_{1,j}+\tfrac{i}{2})
\prod_{j=1}^{K_3}(u-u_{3,j}+\tfrac{i}{2})
{\color{RoyalBlue}\prod_{j=1}^{N/2}(u-\theta_{j}-\tfrac{i}{2})}+{c.c.}\\[19pt]
% node 3
\nonumber 1&=\frac{{\mathbb Q}_2(u_{3,k}+\tfrac{i}{2})}{{\mathbb Q}_2(u_{3,k}-\tfrac{i}{2})}
\prod_{j=1}^{K_4}\frac{x_{3,k}-x_{4,j}^+}{x_{3,k}-x_{4,j}^-}
{\color{RoyalBlue}
\prod_{j=1}^{N/2}\frac{x_{3,k}-{z_{N/2+j}}}{x_{3,k}-\tfrac{1}{z_{j}}}}
\\[19pt]
% node 4
\(\frac{x^{+}_{4,k}}{x^{-}_{4,k}}\)^{L+2}&=
{\color{ForestGreen}{\sig_{\bullet p}^2(x_{4,k},y_0)}}
\prod_{\substack{j=1\\j\neq i}}^{K_4}
\[\frac{u_{4,k}-u_{4,j}+i}{u_{4,k}-u_{4,j}-i}
\sig_{\bullet\bullet}^2(u_{4,k},u_{4,j})
\]{\color{RoyalBlue}
\prod_{j=1}^N
\[\sig_{\bullet\circ}^2(u_k,z_j)
\frac{u_{4,k}-\theta_{j}+\tfrac{i}{2}}{u_{4,k}-\theta_{j}-\tfrac{i}{2}}
\]}
\\
&\times 
\prod_{j=1}^{K_1}\frac{1-\tfrac{1}{x_{4,k}^-x_{1,j}}}{
1-\tfrac{1}{x_{4,k}^+x_{1,j}}
}
\prod_{j=1}^{K_3}\frac{x_{4,k}^--x_{3,j}}{
x_{4,k}^+-x_{3,j}
}
\prod_{j=1}^{K_5}\frac{x_{4,k}^--x_{5,j}}{
x_{4,k}^+-x_{5,j}
}
\prod_{j=1}^{K_7}\frac{1-\tfrac{1}{x_{4,k}^-x_{7,j}}}{
1-\tfrac{1}{x_{4,k}^+x_{7,j}}
}\\[17pt]
\nonumber 1&=\frac{{\mathbb Q}_6(u_{5,k}+\tfrac{i}{2})}{{\mathbb Q}_6(u_{5,k}-\tfrac{i}{2})}
\prod_{j=1}^{K_4}\frac{x_{5,k}-x_{4,j}^+}{x_{5,k}-x_{4,j}^-}
{\color{RoyalBlue}
\prod_{j=1}^{N/2}\frac{x_{5,k}-{z_{N/2+j}}}{x_{5,k}-\tfrac{1}{z_{j}}}}
\\[21pt]
\nonumber-\dot{\mathbb T}(u){\mathbb Q}_6(u)&={\mathbb Q}_6(u-i)
\prod_{j=1}^{K_1}(u-u_{7,j}+\tfrac{i}{2})
\prod_{j=1}^{K_3}(u-u_{5,j}+\tfrac{i}{2})
{\color{RoyalBlue}\prod_{j=1}^{N/2}(u-\theta_{j}-\tfrac{i}{2})}+{c.c.}
\\[19pt]
\nonumber 1&=\frac{{\mathbb Q}_6(u_{7,k}+\tfrac{i}{2})}{{\mathbb Q}_6(u_{7,k}-\tfrac{i}{2})}
\prod_{j=1}^{K_1}\frac{1-\tfrac{1}{x_{7,k} x_{4,j}^+}}{1-\tfrac{1}{x_{7,k} x_{4,j}^-}}
{\color{RoyalBlue}
\prod_{j=1}^{N/2}\frac{x_{7,k}-\tfrac{1}{z_{N/2+j}}}{x_{7,k}-z_{j}}}
\\[10pt]
\color{RoyalBlue}
\nonumber z_{k}^{L+2}&=
{\color{ForestGreen}{\sig^2_{\circ p}}(z_k,y_0)}
\color{RoyalBlue}
\zeta \frac{
{\mathbb Q}_2(\theta_k-\frac{i}{2})
}{\bar{\mathbb Q}_{\tilde 6}(\theta_k+\frac{i}{2})}\left[\prod_{l=1}^{N/2}\frac{\Gamma(i \theta_k-i\theta_l+1)}{\Gamma(-i \theta_k+i\theta_l+1)}\prod_{l=1+N/2}^{N} \frac{\Gamma(-i \theta_k+i\theta_l+1)}{\Gamma(i \theta_k-i\theta_l+1)}\right]^{1/2} \\
&\times \color{RoyalBlue}\frac{\prod_{j=1}^{K_7}(z_k-{x_{7,j}})
\prod_{j=1}^{K_5}(\tfrac{1}{z_k}-x_{5,j})
}{
\prod_{j=1}^{K_1}(\tfrac{1}{z_k}-x_{1,j})
\prod_{j=1}^{K_3}(z_k-{x_{3,j}})
}
\\
&\times
\color{RoyalBlue}\prod_{j=1}^{K_4}\[
\frac{\theta_k-u_{4,j}+\tfrac{i}{2}}{\theta_k-u_{4,j}-\tfrac{i}{2}}
\sig^2_{\circ\bullet}(z_k,u_{4,j})
\]
\nonumber\prod_{j=1}^N
\sig^2_{\circ\circ}(z_k,z_j)
\end{aligned}
\end{aligned}
%\la{mostgeneralABBA}
}
\eeq
    \caption{ABBA equations. In the equations for nodes $2$ and $6$, the ``c.c.'' terms denote the expression with $i\to-i$ without conjugating the functions and replacing $\theta_j\to\theta_{N/2+j}$; ${\mathbb T}(u)$ and $\dot {\mathbb T}(u)$ are polynomials, that have to be found from self-consistency in the Baxter logic. The terms highlighted in blue are related to the massless particles. And the terms in green are only relevant for the RA cases.}
    \label{fig:ABBAgeneral}
\end{eqfloat}

In this section, we present the main result of this paper: the asymptotic Baxter-Bethe ansatz (ABBA), which determines the spectrum of $\gamma_S$. The goal of this section is to introduce all equations and required notation; the remainder of the paper is dedicated to examples.

Without further ado, we present the full ABBA in \eq{fig:ABBAgeneral}. Readers familiar with integrability in AdS/CFT will recognize that the ABBA shares many features with the ABA of Beisert-Staudacher for local operators, but with some crucial differences. The first is that it features functional equations, not only algebraic equations on roots. The second new feature is the existence of massless roots, which are constrained by their own Bethe-like equation. This is reminiscent of the AdS$_3$ Bethe ansatz case, which also has a similar technical origin \cite{Ekhammar:2024kzp}.

In general, the ABBA contain $K_4$ {\it massive} momentum-carrying Bethe roots $u_{4,k}$ and $N$ {\it massless} momentum-carrying Bethe roots $z_j$, although in practice we currently only know of cases where $N=4$ or $N=2$. It is not clear to us if $N>4$ corresponds to meaningful Regge trajectories. Apart from momentum-carrying roots, the ABBA also features four sets of auxiliary roots
$\{u_{a,k}\}_{k=1}^{K_a}\,, a=1,3,5,7$,
each corresponding to one of the nodes of the $\frak{psu}(2,2|4)$ Dynkin diagram in the \verb|xoxoxox| grading. By contrast, nodes 2 and 6 in \eqref{fig:ABBAgeneral} are not represented by some finite set of Bethe roots—one of the key distinctions compared to the BES equations for local operators. Rather, here we have two meromorphic functions, $\mathbb Q_2$ and $\mathbb Q_6$ constrained by functional Baxter equations and their analytic properties to be detailed in the following paragraphs. We remind the reader that the relation between $K_a$ and quantum numbers was explained in Section~\ref{sec:QuantumNumbers}.

Finally, we introduce some useful shorthand notation
$$
\theta_j = g\(z_j + \frac{1}{z_j}\)
\quad\text{and}\quad
x_{a,k}^{[+b]} \equiv x\bigl(u_{a,k} + \tfrac{i b}{2}\bigr)\,,
$$
where the Zhukovsky map is
$$
x(u)\;=\;\frac{\sqrt{u-2g}\,\sqrt{u+2g}\;+\;u}{2g}\,.
$$
The coming paragraphs will be dedicated to properly defining and discussing the other elements that enter into the ABBA as presented in \eq{fig:ABBAgeneral}.

\paragraph{Dressing phases.} 
We will refer to the functions $\sig_{\bullet \bullet},\sig_{\bullet \circ},\sig_{\circ \bullet}$ and $\sig_{\circ \circ}$ as \textit{dressing phases}.
For the case of RA, where $N=2$, there are additional phases, $\sig_{\bullet p}$ and $\sig_{\circ p}$.
The massive phase, $\sig_{\bullet \bullet}$, is the standard BES-phase \cite{Beisert:2006ez}, an important ingredient in the ABA for local operators. The remaining phases all involve “scattering with massless particles”; they are in a form similar to, but not equal to, the dressing phases that have been proposed to play a role in AdS$_{3}$/CFT$_2$ integrability \cite{Frolov:2021fmj,Ekhammar:2024kzp}. 

To provide an explicit form of the dressing phases, we first introduce
\begin{align}
    \chi(x,y) &= -\oint \frac{dz}{2\pi \ii} \frac{1}{x-z} \oint \frac{dw}{2\pi \ii} \frac{1}{y-w}\log \left( \frac{\Gamma(1+\ii u_z-\ii u_w)}{\Gamma(1-\ii u_z +\ii u_w)}\right)\,,\\
    \chi_B(x,y) &= \oint \frac{dz}{2\pi \ii}\frac{1}{x-z} \log \frac{\sinh{\pi (u_z-u_{y})}}{\pi (u_{z}-u_{y})}\,, \\
    \chi_p(x,y) &= \oint\frac{dz}{2\pi \ii}\frac{1}{x-z}\log \cosh{\pi(u_z-u_y)} = \log\(1-\frac{1}{x\, y^\pm}\)+\chi_{B}(x,y^{\pm})\,,
\end{align}
where $u_{x} = g(x+\frac{1}{x})$ and thereafter also
\begin{equation}
\begin{split}
    \varsigma_{\bullet}(x,y) &= e^{\chi(x,y^{+})-\chi(x,y^{-})}\,,
    \quad
    \varsigma_{\circ}(x,y) = e^{\frac{1}{2}\chi(x,y)-\frac{1}{2}\chi(x,1/y)}\,,
    \\
    \varsigma_B(x,y) &= e^{\frac{1}{2}\chi_B(x,y)}\,,
    \quad\quad\quad\;\;\,
    \varsigma_p(x,y) = e^{\chi_p(x,y)}\;.
\end{split}
\end{equation}
From these building blocks, we find the full dressing phases as
\begin{equation}\label{eq:DressingPhasesIntegralDef}
\begin{split}
   \sig_{\bullet\bullet}(x,y) &= \frac{\varsigma_{\bullet}(x^+,y)}{\varsigma_{\bullet}(x^-,y)}\,,
   \quad\quad\quad\quad\quad\quad\quad\;\,
   \sig_{\bullet\circ}(x,y) = \frac{1}
   {\sig_{\circ\bullet}(y, x)} = \frac{\varsigma_{\circ}(x^+,y)}{\varsigma_{\circ}(x^-,y)}\frac{\varsigma_{B}(x^+,y)}{\varsigma_{B}(x^-,y)},
   \\
   \sig^2_{\circ\circ}(x,y) &= \frac{\varsigma_{\circ}(x,y)}{\varsigma_{\circ}(\frac{1}{x},y)}\frac{\varsigma_{B}(x,y)}{\varsigma_{B}(\frac{1}{x},y)}\frac{\varsigma_B(\frac{1}{y},x)}{\varsigma_B(y,x)}\,,
   \quad\!
   \sig_{\bullet p}(x,y) \!=\! \frac{\varsigma_{p}(x^+,y)}{\varsigma_p(x^-,y)}\,,   
   \quad
   \sig^2_{\circ p}(z,y) \!=\! \frac{\varsigma_{p}(z,y)}{\varsigma_p(\frac{1}{z},y)}.
\end{split}
\end{equation}

The subscript $B$ is motivated by the fact that $\varsigma^2_{B}(x,-x)$ is precisely the boundary dressing phases that have been previously introduced to study Wilson lines, and in particular, the cusp anomalous dimension \cite{Correa:2012hh,Drukker:2012de,Correa:2023lsm}.

We give additional relations for the dressing phases and provide their weak coupling expansions in Appendix~\ref{app:DressingPhases}.

\paragraph{Analytic properties of ${\mathbb Q}_a$.}
Another novel part of the ABBA equations is the functional Baxter equations at nodes $2$ and $6$. We denote the solutions to node $2$ as $\{\mathbb{Q}_2,\mathbb{Q}_{\tilde{2}}\}$ and the solutions to node $6$ as $\{\mathbb{Q}_6,\mathbb{Q}_{\tilde{6}}\}$. The solutions are distinguished by their asymptotic
\begin{equation}\la{asymptoticQ2}
\begin{split}
{\mathbb Q}_2 &\simeq u^{
\frac{+(\Delta-S_2)-J_2+J_3+K_4-1+i\sum_{j=1}^{N/2}(\theta_j-\theta_{N/2+j})}2}\;,\\
{\mathbb Q}_6 &\simeq u^{
\frac{+(\Delta+S_2)-J_2-J_3+K_4-1+i\sum_{j=1}^{N/2}(\theta_j-\theta_{N/2+j})}2}\;,
\end{split}
\end{equation}
and for the second pair 
\beqa\la{asymptoticDualQ2}
{\mathbb Q}_{\tilde 2}&\simeq& u^{
\frac{-(\Delta-S_2)-J_2+J_3+K_4-1+i\sum_{j=1}^{N/2}(\theta_j-\theta_{N/2+j})}2}\;,\\
{\mathbb Q}_{\tilde 6}&\simeq& u^{
\frac{-(\Delta+S_2)-J_2-J_3+K_4-1+i\sum_{j=1}^{N/2}(\theta_j-\theta_{N/2+j})}2}\;.
\eeqa
It is only through these asymptotics that the dependence on $\Delta$ and $S_2$ enters into the system. 
$\mathbb{Q}_a$ are further constrained by specific analytic properties. This is similar to the $\algsu(2)$ Heisenberg spin chain, where the analyticity constrains Q-functions to be polynomials. In our case, a polynomial constraint is too strong; we have to allow for poles at the positions
\beq\la{Q2poles}
\text{poles:}\quad \theta_{j}-\frac{i}{2}-i n\;\;,\;\;n\in \mathbb{N}_0\;\;,\;\;j=1,\dots,N/2\;.
\eeq
This pole structure follows from the underlying QSC as we derive in Appendix~\ref{app:ABBADerivation}.

In the spin chain language, this means that the auxiliary spin chains for nodes 2 and 6 have two sites in non-compact representations. All other sites remain compact, as can be seen from the absence of poles at $u_{1,k}-3i/2-i n/2$ and $u_{3,k}-3i/2-i n/2$, which would be present in the general solution of the finite difference Baxter equation. These poles vanish as a part of the quantisation condition for physical states. 

\paragraph{Ordering of massless roots.} 
Note that the Baxter functions $\mathbb{Q}_\alpha$ (for $\alpha=2,\tilde2,6,\tilde6$) depend implicitly on the ordering of the massless roots $\{\theta_i\}$, since the Baxter equation splits them into two equal subsets. The asymptotic behaviour in Eq.~\eqref{asymptoticQ2} likewise reflects this choice of ordering. This apparent asymmetry is not present in the full Quantum Spectral Curve—as we explain in Section~\ref{sec:ABBADerivation}—but rather arises from a convenient parametrisation. One can convert between different orderings by using the following permutation invariant object (a property we prove in Appendix~\ref{app:ABBADerivation})
\beq\la{Mcombo}
{\mathbb M}_\alpha=\frac{{\mathbb Q}_\alpha}{\Gamma_1^+}\;\;,\;\;\alpha=2,\;\tilde 2,\;6,\;\tilde 6\;,
\eeq
where we have introduced convenient notation
\beq\la{Gamma12}
\Gamma_1\equiv\prod_{i=1}^{ N/2}\Gamma(-iu+i\theta_i)\;,\qquad\quad\Gamma_2\equiv\prod_{i=1}^{ N/2}\Gamma(-iu+i\theta_i)\;.
\eeq
\paragraph{Cyclicity condition.}
Like in the case of the local operators, the solutions to the Bethe equations have to satisfy an additional cyclicity condition, which for local single-trace operators follows from the cyclicity of the trace. 

In our case, there is an additional contribution from the massless magnons
\beq
\la{cyclicity}
\prod_{i=1}^{M} \frac{x^+_{4,k}}{x^-_{4,k}} \prod_i^{N} z_i =1\;.
\eeq
Compared to the massless modes in AdS$_3$, the BFKL massless modes contribute only half to this relation.

\paragraph{Quantisation of the gluing constant.}

In order to enable the reader to solve the above equations, we need to determine the constant $\zeta$, which originates from the quantisation condition of the QSC Q-system.
A possible procedure to fix $\zeta$ is to use the cyclicity \eq{cyclicity}. In the case of the ABA, the condition \eq{cyclicity} appears when taking the product of all Bethe equations. This gives a non-trivial requirement, see \eq{zeromom} below, and can be used to fix $\zeta$.

\paragraph{The anomalous dimension $\gamma_S$.}

Once the above equations are solved, with 
\eq{cyclicity} taken into account, one can finally compute  $\gamma_S$ including both massive and massless contributions
\begin{equation}\label{eq:AnomalousSpinMain}
\gamma_S=\gamma_{S\bullet}+\gamma_{S\circ}\,,\qquad\gamma_{S\bullet}\equiv 2 i g \sum_{j=1}^{K_4}\(\frac{1}{x_{4,k}^+}-\frac{1}{x_{4,k}^-}\)\,,\qquad\gamma_{S\circ}\equiv i g\sum_{j=1}^N \(\frac{1}{z_j}-z_j\).
\end{equation}
Again, like in \eq{cyclicity}, the contribution of the massless modes is half compared to AdS$_3$, i.e. half compared to the massive contribution under the replacement $x^+ \to z,\;x^-\to 1/z$. Also, note that the first term gives a regular expansion in $g^2$ assuming that $u_{4,k} \simeq \mathcal{O}(g^0)$, whereas the second term is linear in $g$. We see the appearance of odd powers of $g$, which is a rare feature in ${\cal N}=4$ SYM, where most of the observables have a regular expansion in $g^2$ at weak coupling. This novel feature was first observed for the leading horizontal trajectories in \cite{Klabbers:2023zdz} for $J_1=3$ case.

\subsection{Rewriting Non-Compact Nodes}\label{subsec:SimplifyingNodes}
In this section, we rewrite the ABBA equations in a more convenient form for computations. The key observation is that the Baxter equations for nodes $2$ and $6$ produce non-polynomial Q-functions, but their ``transfer-matrix eigenvalues'' ${\mathbb T}$ and $\dot{\mathbb T}$ are polynomials and thus can be parameterised in an efficient way by just a few parameters.

\paragraph{Algebraic equations for ${\mathbb T}$.} Consider the Baxter equation for the node $2$, to simplify notation we introduce
\begin{equation}\label{eq:BetheQNotation}\begin{split}
    \betheQ_a = \prod_{k=1}^{K_a}(u-u_{a,k})\,,&\qquad\quad a=1,3,4,5,7\\
    \betheQ_{\theta_1} = \prod_{i=1}^{N/2}(u-\theta_i)\,,
    \qquad&\quad
    \betheQ_{\theta_2} = \prod_{i=N/2+1}^{N}(u-\theta_i)\,,
\end{split}\end{equation}
so that the Baxter equation associated to node $2$ reads
\beq\la{baxterQ2copy2}
{\mathbb Q}_2^{[+2]}{
{\mathbb Q}_{1}^-
{\mathbb Q}_{3}^-
{\mathbb Q}_{\theta_2}^+
}
+
{\mathbb Q}_2{\mathbb T}
+
{\mathbb Q}_2^{[-2]}
{\mathbb Q}_{1}^+
{\mathbb Q}_{3}^+
{\mathbb Q}_{\theta_1}^-
=0\;.
\eeq
Here we use the standard notations for shifting arguments $f^{[n]} = f\(u+\tfrac{\ii}{2}\)$.

First, the polynomial ${\mathbb T}$ can be restricted from the large $u$ asymptotic given in \eq{asymptoticQ2} and
\eq{asymptoticDualQ2}, which implies that the first $3$ coefficients in the large $u$-expansion of $\mathbb{T}$ are completely fixed. We find
\beq\la{Tintn}
{\mathbb T}=u^{N/2+K_1+K_3}
\(
-2+\frac{t_1}{u}+\frac{t_2}{u^2}+\dots
\)\,,
\eeq
where
\beqa\la{t1t2}
t_1&=&2\sum_{k=1}^{K_1} u_{1,k}+2\sum_{k=1}^{K_3} u_{3,k}+\sum_{k=1}^N\theta_k\;,\\ \nonumber
t_2&=&\frac{N-2-6(K_1+K_3)}{8}+\frac{(\Delta-S_2)^2}{4}-\frac{t_1^2}{4}+
\sum_{k=1}^{K_1} u^2_{1,k}+\sum_{k=1}^{K_3} u^2_{3,k}+\frac12\sum_{k=1}^N\theta^2_k\;.
\eeqa
As ${\mathbb T}$ is a polynomial, there are still
$N/2+K_1+K_3-2$ coefficients to fix: more precisely $t_{n},\;n=3,\dots,N/2+K_1+K_3$. Usually, these are fixed by first finding ${\mathbb Q}_2$ and imposing quantisation conditions. However, for $N=4$ one can fix ${\mathbb T}$ {\it completely } from a simple algebraic equation without finding ${\mathbb Q}_2$. Moreover, in the $N=2$ case, which is relevant for reflection areas, these equations become even more constraining and force one to set $\Delta$ to a particular value. We discuss this in detail in part~II~\cite{EkhammarGromovPreti:2025b}.

Let us show how to obtain $K_1+K_3$ additional algebraic constraints on ${\mathbb T}$. For that, consider the Baxter equation~\eqref{baxterQ2copy2} and evaluate it at $u=u_{1,k}\pm\tfrac{i}{2}$ or $u=u_{3,k}\pm\tfrac{i}{2}$. As ${\mathbb Q}_2$ does not have poles associated with these, slightly shifted, Bethe roots, one of the three terms vanishes 
\beqa\la{baxterQ2copy2_2}
{\mathbb T}^+
=-
\frac{{\mathbb Q}_2^{-}}{{\mathbb Q}^+_2}
{\mathbb Q}_{1}^{[+2]}
{\mathbb Q}_{3}^{[+2]}
{\mathbb Q}_{\theta_1}
\;,\qquad
{\mathbb T}^-=-
\frac{{\mathbb Q}_2^{+}}{{\mathbb Q}_2^-}{
{\mathbb Q}_{1}^{[-2]}
{\mathbb Q}_{3}^{[-2]}
{\mathbb Q}_{\theta_2}
}
\;,\qquad u=u_{1,k}\;{\rm or}\;u_{3,k}\;.
\eeqa
We notice that by multiplying these equations, we can get rid of ${\mathbb Q}_2$ to obtain
\beq
{\mathbb T}^+{\mathbb T}^- ={\mathbb Q}_{\theta}
{\mathbb Q}_{1}^{[+2]}
{\mathbb Q}_{1}^{[-2]}
{\mathbb Q}_{3}^{[+2]}
{\mathbb Q}_{3}^{[-2]}\;,\qquad u=u_{1,k}\;{\rm or}\;u_{3,k}\;.
\eeq
The above equations give precisely $K_1+K_3$ additional conditions on the coefficients in ${\mathbb T}$, allowing to completely constrain ${\mathbb T}$ with an algebraic equation in terms of the $u_{1,k},\;u_{3,k}$ and $\theta_k$, avoiding solving the complicated finite difference equation! The consideration above can be extended to $\dot {\mathbb T}$ and ${\mathbb Q}_6$ by obvious replacements e.g. $S_2\to -S_2$ and $J_3\to - J_3$. Furthermore, it is possible to rewrite the remaining auxiliary equations, excluding the complicated ${\mathbb Q}_2$ and ${\mathbb Q}_6$ functions, except for the massless node equation. Interestingly, at the leading order in $g$, the dependence on 
${\mathbb Q}_2$ and ${\mathbb Q}_6$ drop out from the massless equations too, allowing for an efficient way of finding, counting and classifying all solutions of ABBA.

\paragraph{Algebraic form of the auxiliary ABBA equations.}
\begin{eqfloat}[!htbp]
    \centering
\beq
\begin{aligned}
&\begin{tikzpicture}[baseline=(mid23.base),
node distance=1.35cm,
dynkinnode/.style={circle,draw,minimum size=6.8mm,inner sep=0pt}]
  \node[dynkinnode,label=left:{1}] (a1) {};
  \node[dynkinnode,label=left:{2}] [below=of a1] (a2) {};
  \node[dynkinnode,label=left:{3}] [below=of a2] (a3) {};
  \node[dynkinnode,fill=black!70,label=left:{4}] [below=of a3] (a4) {};
  \node[dynkinnode,label=left:{5}] [below=of a4] (a5) {};
  \node[dynkinnode,label=left:{6}] [below=of a5] (a6) {};
  \node[dynkinnode,label=left:{7}] [below=of a6] (a7) {};
  
  \node[inner sep=0pt, minimum size=0pt] (mid23) at ($(a4)!0.6!(a5)$) {};
  \node[dynkinnode,fill=gray!30,label=above:{massless}] [below=of a7] (extra) {};

  \draw[thick] (a1)--(a2)--(a3)--(a4)--(a5)--(a6)--(a7);
  % Additional disconnected node with light blue background
  % Cross marks on nodes 1,3,5,7
  \draw (a1.north west)--(a1.south east);
  \draw (a1.north east)--(a1.south west);
  \draw (a3.north west)--(a3.south east);
  \draw (a3.north east)--(a3.south west);
  \draw (a5.north west)--(a5.south east);
  \draw (a5.north east)--(a5.south west);
  \draw (a7.north west)--(a7.south east);
  \draw (a7.north east)--(a7.south west);
\end{tikzpicture}
&\quad
\begin{aligned}
% node 1
\nonumber 1&=\left.
\frac{\bB_{(-)}}{\bB_{(+)}}
\sqrt{\frac
{{\mathbb T}^-{\mathbb Q}_1^{[+2]}{\mathbb Q}_3^{[+2]}}
{{\mathbb T}^+{\mathbb Q}_1^{[-2]}{\mathbb Q}_3^{[-2]}} \frac{\bar\kappa}{\kappa}}\right|_{u=u_{1,k}}\\[19pt]
% node 2
{\mathbb T}^+{\mathbb T}^- &=\left.{\mathbb Q}_{\theta}
{\mathbb Q}_{1}^{[+2]}
{\mathbb Q}_{1}^{[-2]}
{\mathbb Q}_{3}^{[+2]}
{\mathbb Q}_{3}^{[-2]}\right|_{u=u_{1,k}\;{\rm or}\;u_{3,k}}\;\\[19pt]
% node 3
\nonumber 1&=\left.
\frac{\bR_{(-)}}{\bR_{(+)}}
\sqrt{\frac
{{\mathbb T}^-{\mathbb Q}_1^{[+2]}{\mathbb Q}_3^{[+2]}}
{{\mathbb T}^+{\mathbb Q}_1^{[-2]}{\mathbb Q}_3^{[-2]}} \frac{\kappa}{\bar\kappa}}\right|_{u=u_{3,k}}
\\[19pt]
% node 4
\(\frac{x^{+}_{4,k}}{x^{-}_{4,k}}\)^{L+2}&=
\left.-
\sig_{\bullet p}^2
\sig_{\bullet\bullet}^2
\sig_{\bullet\circ}^2
\frac{{\mathbb Q}_4^{[+2]}{\mathbb Q}_\theta^{+}}{{\mathbb Q}_4^{[-2]}{\mathbb Q}_\theta^{-}}
\frac{B^-_{1}R^-_{3} R_5^- B_7^-}{B^+_{1}R^+_{3} R_5^+ B_7^+}\right|_{u=u_{4,k}}
\\[17pt]
\nonumber 1&=\left.
\frac{\bR_{(-)}}{\bR_{(+)}}
\sqrt{\frac
{\dot{\mathbb T}^-{\mathbb Q}_7^{[+2]}{\mathbb Q}_5^{[+2]}}
{\dot{\mathbb T}^+{\mathbb Q}_7^{[-2]}{\mathbb Q}_5^{[-2]}} \frac{\kappa}{\bar\kappa}}\right|_{u=u_{5,k}}
\\[19pt]
\dot{\mathbb T}^+\dot{\mathbb T}^- &=\left.{\mathbb Q}_{\theta}
{\mathbb Q}_{7}^{[+2]}
{\mathbb Q}_{7}^{[-2]}
{\mathbb Q}_{5}^{[+2]}
{\mathbb Q}_{5}^{[-2]}\right|_{u=u_{7,k}\;{\rm or}\;u_{5,k}}\;
\\[19pt]\la{masslessgood}
\nonumber 1&=\left.
\frac{\bB_{(-)}}{\bB_{(+)}}
\sqrt{\frac
{{\mathbb T}^-{\mathbb Q}_7^{[+2]}{\mathbb Q}_5^{[+2]}}
{{\mathbb T}^+{\mathbb Q}_7^{[-2]}{\mathbb Q}_5^{[-2]}} \frac{\bar\kappa}{\kappa}}\right|_{u=u_{7,k}}
\\[10pt]
\la{middleGoodeq} z_{k}^{L+2}&=
\left.\zeta_P
\sig^2_{\circ p}
\sig^2_{\circ\bullet}
\sig^2_{\circ\circ}
\frac{{\mathbb Q}_4^+}{{\mathbb Q}_4^-}
\sqrt{
\frac{
{\mathbb Q}_2^-
{\mathbb Q}_{\tilde 6}^-
}{
\bar{\mathbb Q}_2^+
\bar{\mathbb Q}_{\tilde 6}^+}\frac{\bar{\Gamma}_1\Gamma_2}{\Gamma_1\bar{\Gamma}_2}
\frac{R_1 B_3B_5 R_7}{B_1R_3 R_5 B_7}
}\right|_{u=\theta_k}
\end{aligned}
\end{aligned}
\eeq
    \caption{Polynomial form of the ABBA equations - each instance of ${\mathbb Q}_2$ or ${\mathbb Q}_{\dot 6}$ in fact can be replaced by the polynomial ${\mathbb T}$ or $\dot {\mathbb T}$, except for the massless equation. This makes the equations way more tractable.
}\la{mostgeneralABBA}
\end{eqfloat}
In order to exclude ${\mathbb Q}_2$ and ${\mathbb Q}_6$ from the equations for the auxiliary roots, we compute the ratio appearing in \eq{baxterQ2copy2_2}
\beq\label{baxterQ2copy2_ratio}
\frac{{\mathbb Q}_2^-}{{\mathbb Q}_2^+}=
\sqrt{\frac{{\mathbb T}^+}{{\mathbb T}^-}
\frac{
{\mathbb Q}_{1}^{[-2]}
{\mathbb Q}_{3}^{[-2]}
{\mathbb Q}_{\theta_2}
}{
{\mathbb Q}_{1}^{[+2]}
{\mathbb Q}_{3}^{[+2]}
{\mathbb Q}_{\theta_1}
}}\;,\qquad u=u_{1,k}\;{\rm or}\;u_{3,k}\;.
\eeq
The l.h.s. of \eq{baxterQ2copy2_ratio} is precisely the combination entering into all massive
ABBA equations. Thus, one can simply substitute it with the combination of polynomials on the r.h.s. instead.

To present the algebraic form of ABBA in a more concise way, we introduce the standard notation for Zhukovsky polynomials, e.g. 
\begin{equation}\label{eq:MainRB}
\begin{split}
    B_a &= \prod_{k=1}^{K_a} \left(\frac{1}{x}-x_{a,k}\right)\,,
    \quad\quad\;\;
    R_a = \prod^{K_a}_{k=1}(x-x_{a,k})\,,
    \quad
    a=1,3,5,7
    \\
    \bB_{(\pm)} &= \prod_{k=1}^{K_4} \left(1-\frac{1}{x \,x_{4,k}^{\mp}}\right)\,,
    \quad
    \bR_{(\pm)} = \prod^{K_4}_{k=1}(x-\,x^{\mp}_{4,k})\,.
\end{split}
\end{equation}
Similarly, for the massless roots, we define the following quantities
\begin{equation}\label{eq:MainKappa}
\begin{split}
{\mathbb Q}_{\theta} = \prod_{i=1}^N(u-\theta_i)\,,\quad{\mathbb Q}_{\theta_1}=\prod_{i=1}^{N/2}&(u-\theta_i)\,,\quad{\mathbb Q}_{\theta_1}=\prod_{i=N/2+1}^{N}(u-\theta_{i})\;,
\\
\kappa_1 \!= \!\prod_{i=1}^{N/2}(x-z_i),
\quad
\kappa_2 \!= \!\!\!\!\prod_{i=N/2+1}^{N}\!\!\!\!(x-z_i),
&\quad
\bar{\kappa}_1 \!= \!\prod_{i=1}^{N/2}\!\(x-\frac{1}{z_i}\),
\quad
\bar{\kappa}_2 \!= \!\!\!\prod_{i=N/2+1}^{N}\!\!\!\(x-\frac{1}{z_i}\).
\end{split}
\end{equation}
We present the alternative form of the ABBA in \eqref{mostgeneralABBA}.
There, we also modified the massless equation by making use of the relation \eq{relationNeeded} from Appendix~\ref{App:Qrelations}.
\beq
\frac{
\mQ_{2}^-
}{
\bar{\mQ}_{\tilde 6}^+}\bigg|_{u=\theta_k}
=|\alpha_1|^2
\frac{\mQ_1\mQ_3}{\mQ_7\mQ_5}
\frac{
{\mQ}_{\tilde 6}^-}{
\bar\mQ_{2}^+}\bigg|_{u=\theta_k}\;,
\eeq
which also requires a re-definition of the constant factor $\zeta\to\zeta_P$ in the massless equation.
Moreover, we use the identification
\begin{equation}
\sig_{a\bullet}=\prod_{k=1}^{K_4}\sig_{a\bullet}(x,x_{4,k})\,,
\quad \sig_{a\circ}=\prod_{k=1}^{N}\sig_{a\circ}(x,z_{k})\,,
\quad
\sig_{ap}=\sig_{a p}(x,y_0)\,,
\quad
a=\bullet,\circ\,.
\end{equation}

\paragraph{Momentum conservation.}
The product of all ABBA is related to momentum conservation. Computing the product over all roots, we find
\begin{equation}
\begin{split}\la{zeromom}
\(\prod_{k=1}^N z_k \prod_{k=1}^{K_4} \frac{x_{4,k}^+}{x_{4,k}^-}\)^{L+2} &\!\!\!\!\!=\zeta^N_P \!\left(\prod_{k=1}^{N} \sig^2_{\circ p}(z_k,y_0) \right)\!\!\left(\prod_{k=1}^{K_4} \sig^2_{\bullet p}(x_{4,k},y_0) \right)
\!\sqrt{
\prod^{N}_{k=1}
\frac{
{\mathbb M}_2^-{\mathbb M}_{6}^-
}{\bar{\mathbb M}_2^+\bar{\mathbb M}_{\tilde 6}^+}}\bigg|_{u=\theta_k}\\
&\times \sqrt{\left(\prod_{a=1,3}\prod_{k=1}^{K_a}
-\frac{{\mathbb T}^-(u_{a,k})}{{\mathbb T}^+(u_{a,k})}
\right)
\left(\prod_{a=5,7}\prod_{k=1}^{K_a}
-\frac{{\dot{\mathbb{T}}}^-(u_{a,k})}{{\dot{\mathbb{T}}}^+(u_{a,k})}
\right)
}\;,
\end{split}
\end{equation}
where we used the unitarity for $\sig_{\bullet\bullet}^2$. Using momentum conservation~\eq{cyclicity}, the l.h.s. is $1$ and this gives an additional equation on $\zeta_P$.

\subsection{The Weak Coupling Limit}\label{subsec:WeakCouplingABBA}
Here we discuss ABBA to the leading order at weak coupling. We will assume a naive scaling, where all massive and auxiliary roots scale as $u_{a,k}\sim 1$, whereas the massless ones remain on the unit circle, i.e. $\theta_k\sim g$. As discussed below in Section~\ref{sec:SubleadingHT}, there are important exceptions to this scaling.

\paragraph{Weak coupling of massive equations.}
In the scaling $u_{a,k}\sim 1$,  $|z_k| \sim 1$ while $g\to 0$, we can simply perform the replacements $R\to {\mathbb Q}$ and $B\to 1$ in the auxiliary equations. Similarly $\kappa(u_{a,k})\sim \bar\kappa(u_{a,k})\sim u_{a,k}^N/g^n$
and ${\mathbb Q}_\theta\sim u^N$. We present the resulting expressions for the massive part of ABBA in \eqref{fig:WeakCouplingABBA}. As can be seen from this figure, the massless modes fully decouple; their only effect is to change the length as $L\to L-N$. 

\begin{eqfloat}[!htbp]
    \centering
\beq
\begin{aligned}
&\begin{tikzpicture}[baseline=(mid23.base),
node distance=1.35cm,
dynkinnode/.style={circle,draw,minimum size=6.8mm,inner sep=0pt}]
  \node[dynkinnode,label=left:{1}] (a1) {};
  \node[dynkinnode,label=left:{2}] [below=of a1] (a2) {};
  \node[dynkinnode,label=left:{3}] [below=of a2] (a3) {};
  \node[dynkinnode,fill=black!70,label=left:{4}] [below=of a3] (a4) {};

  \node [below=of a4] (a5) {};

  \node[inner sep=0pt, minimum size=0pt] (mid23) at ($(a2)!0.6!(a3)$) {};

  \draw[thick] (a1)--(a2)--(a3)--(a4);
  \draw[dashed] (a4)--(a5);
  % Additional disconnected node with light blue background
  % Cross marks on nodes 1,3,5,7
  \draw (a1.north west)--(a1.south east);
  \draw (a1.north east)--(a1.south west);
  \draw (a3.north west)--(a3.south east);
  \draw (a3.north east)--(a3.south west);
\end{tikzpicture}
&\quad
\begin{aligned}
% node 1
\nonumber 1&=\left.
\sqrt{\frac
{{\mathbb T}^-{\mathbb Q}_1^{[+2]}{\mathbb Q}_3^{[+2]}}
{{\mathbb T}^+{\mathbb Q}_1^{[-2]}{\mathbb Q}_3^{[-2]}}}\right|_{u=u_{1,k}}\\[17pt]
% node 2
{\mathbb T}^+{\mathbb T}^- &=\left.u^N
{\mathbb Q}_{1}^{[+2]}
{\mathbb Q}_{1}^{[-2]}
{\mathbb Q}_{3}^{[+2]}
{\mathbb Q}_{3}^{[-2]}\right|_{u=u_{1,k}\;{\rm or}\;u_{3,k}}\;.\\[17    pt]
% node 3
\nonumber 1&=\left.
\frac{{\mathbb Q}_4^-}{{\mathbb Q}_4^+}
\sqrt{\frac
{{\mathbb T}^-{\mathbb Q}_1^{[+2]}{\mathbb Q}_3^{[+2]}}
{{\mathbb T}^+{\mathbb Q}_1^{[-2]}{\mathbb Q}_3^{[-2]}}}\right|_{u=u_{3,k}}
\\[19pt]
% node 4
\(\frac{u_{4,k}+\tfrac{i}{2}}{u_{4,k}-\tfrac{i}{2}}\)^{L+2-N}&=
\left.
\frac{{\mathbb Q}_4^{[+2]}}{{\mathbb Q}_4^{[-2]}}
\frac{{\mathbb Q}^-_{3} {\mathbb Q}^-_{5}}{{\mathbb Q}^+_{3} {\mathbb Q}^+_{5}}\right|_{u=u_{4,k}}
\end{aligned}
\end{aligned}
\eeq
 \caption{Weak coupling of the massive ABBA equations in polynomial form.}\label{fig:WeakCouplingABBA}
\end{eqfloat}

\paragraph{Weak coupling scaling of the massless equations.}
In the massless equation itself, the terms depending on the massive roots can be removed because of the separation of the scale $\theta_a\sim g$, whereas the massive roots are assumed to scale as $u_a\sim 1$. Indeed, $\frac{R_a}{B_a}\to 1$ as $g\to 0$. The ratio of $\Gamma$-functions and the dressing phases also approaches $1$. The ratio $\frac{\betheQ^-_2 \betheQ_{\tilde{6}}^-}{\bar{\betheQ}^+_{2}\bar{\betheQ}_{\tilde{6}}^+}\big|_{\theta_k}$ does not necessarily go to $1$, but the result is independent of $\theta_k$ and hence can be reabsorbed into $\zeta_P$. We keep track of this redefinition by sending $\zeta_P\rightarrow \zeta'_P$. We are left with the following equations
\beq
\begin{aligned}
&\begin{tikzpicture}[baseline=(extra.base),
node distance=1.35cm,
dynkinnode/.style={circle,draw,minimum size=6.8mm,inner sep=0pt}]
  \node[dynkinnode,fill=gray!30,label=above:{massless}]  (extra) {};
\end{tikzpicture}
&\quad
\begin{aligned}
 z_{k}^{L+2}&=
\zeta_P'\;.
\end{aligned}
\end{aligned}
\la{masslessweakcoupling}
\eeq
Finally, to fix $\zeta'_P$ we can use \eq{cyclicity}, which at weak coupling becomes
\beq
\prod_{k=1}^{K_4}\frac{u_{4,k}+\tfrac i2}{u_{4,k}-\tfrac i2}\prod_{k=1}^N z_k = 1\;,
\eeq
which then from \eqref{mostgeneralABBA} implies
\beq\la{levelmatchweak}
(\zeta'_P)^N=
\prod_{k=1}^{K_4}\[\frac{u_{4,k}-\tfrac i2}{u_{4,k}+\tfrac i2}\]^{L+2} \;,
\eeq
so we conclude that at weak coupling the only interaction between the massive and massless equations occurs by means of the ``level matching" equation~\eq{levelmatchweak}. But otherwise, they remain fully independent. Note that for the parity invariant states, the r.h.s. \eq{levelmatchweak} simplifies to $1$ and we simply have $(\zeta'_P)^N=1$.

\section{The QSC Origin of the ABBA}\label{sec:ABBADerivation}

To derive the ABBA, the starting point is the non-perturbative QSC framework~\cite{Gromov:2013pga,Gromov:2014caa}. We have relegated a complete technical derivation to Appendix~\ref{app:ABBADerivation} complemented by a brief review of the QSC in Appendix~\ref{app:QSCGeneralities}. In this section, we will sketch how the various constituents of ABBA relate to QSC objects.  

\subsection{Lightning Review of QSC}
Let us give a quick review of the key QSC objects. The QSC is a collection of Q-functions that depend on one complex parameter $u$ called the spectral parameter. The most basic Q-functions are called $\bP_a,\bP^{a},\bQ_i,\bQ^{i}$, which have power-like asymptotics that encode the quantum numbers of the solution
\begin{equation}
    \bP_a \sim u^{-\tilde{M}_a}\,,
    \qquad
    \bQ_i \sim u^{\hat{M}_i-1}\,, 
    \qquad
    \bP^{a} \sim u^{\tilde{M}_a-1}
    \qquad
    \bQ^{i} \sim u^{-\hat{M}_i}\,,
\end{equation}
where $M$'s are some combinations of the quantum numbers given in \eq{MMt}.

The basic Q-functions are related by a set of finite difference equations according to
\begin{equation}\label{eq:FiniteDifferenceQai}\begin{split}
    Q&_{a|i}^+ -Q_{a|i}^- = \bP_{a} \bQ_{i}\,,
    \qquad\quad
    Q_{a|i}Q^{a|j} = -\delta_i^j\,,\\
    &\bP_{a} = -Q^{\pm}_{a|i}\,
    \bQ^{i}\,,
    \qquad\quad
    \bP^{a} = \left(Q^{a|i}\right)^{\pm}\bQ_i\,.
\end{split}\end{equation}
Finally, all Q-functions that feature in the QSC can be written as $Q_{A|I}$ where $A,I$ are antisymmetric multi-indices whose entries take values $1,\dots,4$. We identify $Q_{a|\es} = \bP_a$ $Q_{\es|i} = \bQ_i$, $\bP^{a} =-\frac{1}{6}\epsilon^{abcd}Q_{bcd|\fullset}$ and $\bQ^{i} = -\frac{1}{6}\epsilon^{ijkl}Q_{\fullset|jkl}$, where $\es$ is the empty set and $\fullset\equiv 1234$. All remaining Q-functions can be found from the following QQ-relations
\begin{equation}\label{eq:QQRelations}
\begin{split}
    \Wr(Q_{Aa|I},Q_{Ab|I}) = Q_{Aab|I}\,Q_{A|I}\,, \\ 
    \Wr(Q_{A|Ii},Q_{A|Ij}) = Q_{A|Iij}\,Q_{A|I}\,, \\ 
    \Wr(Q_{Aa|Ii},Q_{A|I}) = Q_{Aa|I}Q_{A|Ii}\,,
\end{split}
\end{equation} 
where the definition of the Wronskian is given by $\Wr(f,g)=f^+ g^--f^- g^+$.
Notice that the identities \eq{eq:FiniteDifferenceQai} ultimately follow from \eq{eq:QQRelations}. 

\begin{figure}[!t]
\begin{center}
    \begin{minipage}{0.3\textwidth}
    \begin{tikzpicture}<2->
    \node[] (a) at (-1.3,1.3) {$\bP_a$};
    \draw[thick] (-1,-1) rectangle (1,1);
    \node[circle,inner sep=1pt, fill=black] (p1) at (0.5,0) {};
    \node[circle,inner sep=1pt, fill=black] (p1) at (-0.5,0) {};
    \draw[blue] (-0.5,0)--(0.5,0);
    \end{tikzpicture}  
    \end{minipage}
    \begin{minipage}{0.3\textwidth}
    \begin{tikzpicture}<2->
    \node[] (a) at (-1.3,1.3) {$\bQ^{\downarrow}$};
    \draw[thick] (-1,-1) rectangle (1,1);
    \node[circle,inner sep=1pt, fill=black] (p1) at (0.5,0) {};
    \node[circle,inner sep=1pt, fill=black] (p1) at (-0.5,0) {};
    \draw[blue] (-0.5,0)--(0.5,0);
    \node[circle,inner sep=1pt, fill=black] (p1) at (0.5,-0.5) {};
    \node[circle,inner sep=1pt, fill=black] (p1) at (-0.5,-0.5) {};
    \draw[blue] (-0.5,-0.5)--(0.5,-0.5);
    \end{tikzpicture}  
    \end{minipage}
    \begin{minipage}{0.3\textwidth}
    \begin{tikzpicture}<2->
    \node[] (a) at (-1.3,1.3) {$\bQ^{\uparrow}$};
    \draw[thick] (-1,-1) rectangle (1,1);
    \node[circle,inner sep=1pt, fill=black] (p1) at (0.5,0) {};
    \node[circle,inner sep=1pt, fill=black] (p1) at (-0.5,0) {};
    \draw[blue] (-0.5,0)--(0.5,0);
    \node[circle,inner sep=1pt, fill=black] (p1) at (0.5,0.5) {};
    \node[circle,inner sep=1pt, fill=black] (p1) at (-0.5,0.5) {};
    \draw[blue] (-0.5,0.5)--(0.5,0.5);
    \end{tikzpicture}  
    \end{minipage}
\end{center}
    \caption{The analytic structure of the functions $\bP_{a},\bP^{a},\bQ^{\downarrow/\uparrow}_i$ and $(\bQ^{\downarrow/\uparrow})^i$}
    \label{fig:AnalyticStructurePQ}
\end{figure}
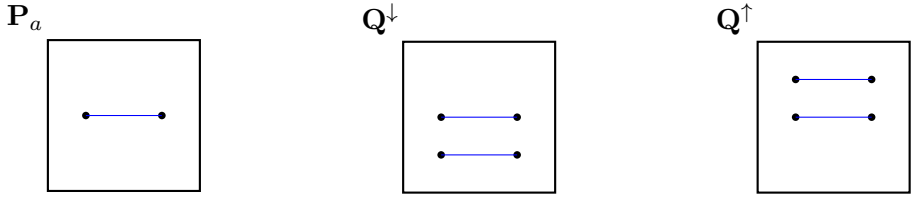

A crucial part of the QSC construction is to impose an analytic structure on the Q-functions. We demand that $\bP_a$ are analytic functions outside a cut $(-2g,2g)$ on the real axis. From \eqref{eq:FiniteDifferenceQai} it is possible to deduce that $Q_{a|i}$ and $\bQ_i$ must have an infinite number of cuts on the sheet we defined $\bP_a$. The finite difference equations do not uniquely specify how to choose where these cuts are. Rather, there are two distinguished choices: either we demand that the Q-functions are analytic in the upper-plane, written $Q^{\downarrow}$, or analytic in the lower half-plane, $Q^{\uparrow}$. We have illustrated this structure in Figure~\ref{fig:AnalyticStructurePQ}.

To quantise the system, we must glue together $\bQ^{\downarrow}$ and $\bQ^{\uparrow}$ along the cut on the real axis. Defining $\tilde{f}$ as the analytic continuation of any function around $2g$, the gluing condition is given by
\begin{equation}\label{eq:MainGluing}
\begin{split}
    &\tilde{\bQ}^{\downarrow,i} = G^{ij}\bQ^{\uparrow}_j\,,
    \qquad\quad
    \tilde{\bQ}^{\downarrow}_i = \bQ^{\uparrow,j}G_{ji}\,,
    \\
    & \!\!\!\!\!\!\!\!\!\!\!\!\!\!\!G^{ij} = \scalebox{0.8}{
    $\begin{pmatrix}
        \star & \star & \star & 0 \\
        \star & 0 & 0 & 0 \\
        \star & 0 & \star & \star \\
        0 & 0 & \star & 0
    \end{pmatrix}
    +
    e^{2\pi u}\begin{pmatrix}
        0 & 0 & \star & 0 \\
        0 & 0 & 0 & 0 \\
        \star & 0 & 0 & 0 \\
        0 & 0 & 0 & 0
    \end{pmatrix}
    +
    e^{-2\pi u}\begin{pmatrix}
        0 & 0 & \star & 0 \\
        0 & 0 & 0 & 0 \\
        \star & 0 & 0 & 0 \\
        0 & 0 & 0 & 0
    \end{pmatrix}$
    },
    \end{split}
\end{equation}
where we have schematically illustrated the structure of the gluing matrix, indicating its non-zero constant entries, see Appendix~\ref{app:QSCGeneralities} for a complete expression. This structure of the gluing matrix was deduced in~\cite{Alfimov:2014bwa,Gromov:2015wca}. 

Finally, we note that in practice it is often convenient to construct $\bQ^{\uparrow}$ from $\bQ^{\downarrow}$ using either a parity transformation $u \rightarrow -u$ (for the parity invariant states), or complex conjugation (for $\Delta$ and $g$ real). In those cases we have $\bQ^{\uparrow}(u) \propto \bQ^{\downarrow}(-u)$ and/or $\bQ^{\uparrow}(u) \propto \bar{\bQ}^{\downarrow}(u)$.

\paragraph{The $\bP\mu$-system.} An alternative formulation of the QSC, which goes back to the TBA~\cite{Gromov:2009bc,Bombardelli:2009ns,Arutyunov:2009ur} is the $\bP\mu$-system. A new key object is the antisymmetric matrix $\mu_{ab}$, which controls the monodromy of $\bP_a$ and is periodic on a section of the Riemann surface with long cuts. It satisfies the following relations
\begin{equation}\label{eq:PMu0}
    \tilde{\bP}_{a} = \mu_{ab}\bP^{b}\,,
    \qquad
    \tilde{\mu}_{ab}-\mu_{ab} = \bP_{a}\tilde{\bP}_b-\tilde{\bP}_{a}\bP_b\,,
    \qquad
    \tilde{\mu}_{ab} = \mu^{[2]}_{ab}\,.
\end{equation}
The matrix $\mu_{ab}$ can furthermore be related to the already introduced Q-functions as 
\begin{equation}\label{eq:MuAndQabij}
    \mu_{ab} = \frac{1}{2}Q^{-}_{ab|ij} \omega^{ij}\,,
    \qquad
    Q_{ab|ij} = \begin{vmatrix}
        Q_{a|i} & Q_{a|j} \\
        Q_{b|i} & Q_{b|j}
    \end{vmatrix}\,,
\end{equation}
with $\omega^{ij}$ an $\ii$-periodic matrix. The determinant expression on the right originates from the Wronskian relations \eqref{eq:QQRelations}.

The zeroes of $\mu_{12}$ play an important role in the asymptotic limit as they become the momentum-carrying roots. The key fact is that in the case of non-local operators $\mu_{12}$ is proportional to the gluing matrix and thus gets an exponential asymptotic at large $u$ growing as $e^{2\pi |u|}$ but not faster. Additional details are included in Appendix~\ref{app:QSCGeneralities}. In the following, it is shown that this exponential behaviour gives rise to the massless modes and the rate of growth is controlled by the number of massless roots.

\subsection{The Simplification for HT}\label{subsec:DerivingABBA}
So far, our discussion has been general and the statements made are valid at any coupling and any choice of quantum numbers. To derive the ABBA we now take a weak coupling limit such that $S\simeq -\Sigma$.

\paragraph{The momentum carrying roots.}
The key simplification which enables an analytic treatment of this regime is that we find
\begin{equation}
\mu_{ab} \simeq Q^{-}_{ab|13}\omega^{13}\;.
\end{equation}
Then, one can consider the ratio
\begin{equation}
\mathcal{F}=~\frac{\mu_{12}}{\mu_{12}^{[+2]}}\simeq \frac{Q^-_{12|13}}{Q^+_{12|13}}\;,
\end{equation}
which manifestly does not have cuts in the upper half-plane since $Q_{ab|13}$ is upper half-plane analytic. Moreover, using the reality of $\mu^+$ \eqref{eq:ConjugationMu}, one can conclude that no cuts can appear in the lower half-plane either. Thus, all cuts away from the real axis are suppressed in powers of the coupling. 
A detailed derivation is presented in Appendix \ref{app:ABBADerivation}, where we also show that 
$\mathcal{F}$ satisfies the following discontinuity relation
\beq\la{FFtilda}
\mathcal{F}\tilde{\mathcal{F}} = 1\,.
\eeq
Next, as in the case of the local operators, we define the momentum-carrying roots to be zeroes of $\mu_{12}^+$. The main difference as compared to the case of the local operators, is that $\mu_{12}$ now also has zeroes located directly on its branch cut. As shown below, these correspond precisely to the massless roots.

These crucial observations allow us to derive the ABBA. For example, the analyticity on the main sheet and the discontinuity condition \eq{FFtilda} imply that ${\cal F}$ is a rational function of the Zhukovsky variable $x$. Moreover, the poles and zeros originate from the roots of $\mu_{12}$,  we deduce the following representation
\begin{equation}\label{eq:Q1213Roots}
    {\cal F} =  \frac{\betheQ^-}{\betheQ^+}\, \left(\frac{\mathbf{B}_{(+)}}{\bf B_{(-)}}\right)^2\, \frac{\kappa}{\bar{\kappa}} + \mathcal{O}\left(g^\text{Wrapping Order}\right)\,,
\end{equation}
where we separated the roots/poles into those which are on the cut, namely $z_k$, and those that are away from the cut $x_{4,k}$, see \eqref{eq:MainRB} and \eqref{eq:MainKappa}.

It is now immediate to plug in the asymptotics $Q_{12|13} \simeq u^{1-J_1-S_1}$ to derive \eq{eq:AnomalousSpinMain} i.e.
\begin{equation}
\begin{split}
    S_1 &= -J_1+1-K_4+2 \ii g \sum_{i}^{K_4} \left(-\frac{1}{x_{4,i}^+}+\frac{1}{x_{4,i}^-}\right) + \ii g \sum_{i}^{N}\left(z_i-\frac{1}{z_i} \right) \\
    &= -J_1+1-K_4 - \gamma_{S}\,.
    \end{split}
\end{equation}
This expression implies $\Sigma = J_1-1+K_4$, in agreement with \eq{magnonnumbers}. Notice that each massive excitation increases $\Sigma$ by 1, whereas the massless modes $z_k$ just add a small perturbation of order $\sim g$.

\paragraph{Reconstructing $Q_{12|13}$.} Once the ratio ${\cal F}$ and the analyticity of $Q_{12|13}$ in the upper half-plane are known, one can solve \eqref{eq:Q1213Roots} explicitly to find
\begin{equation}\label{q1213sec42}
    Q_{12|13} \propto \betheQ_4\,\left(f^+\right)^2\, \prod_{n=0}^{\infty} \frac{\kappa^{[2n+1]}}{\bar{\kappa}^{[2n+1]}}\;,
\end{equation}
where we have introduced a new function $f$ satisfying
\begin{equation}\label{eq:MainF}
    \frac{f}{f^{[2]}} = \frac{\bfB_{(+)}}{\bfB_{(-)}}\,,
    \qquad\quad
    f = \prod_{n=0}^{\infty}\frac{\bfB_{(+)}^{[2n]}}{\bfB_{(-)}^{[2n]}}\,.
\end{equation}

\paragraph{Reconstructing $\mu_{12}$.} One can also reconstruct $\mu_{12}$, as described in detail in Appendix~\ref{app:ABBADerivation}, resulting in
\begin{equation}\label{mu12P0mt}
    \mu_{12} \propto p\,  \betheQ^-_4\,
    f \bar f^{--}
    \,
    \frac{\kappa}{x^{N/2}}\,\prod_{n=1}^{\infty} \frac{\kappa^{[2n]}\, \bar{\kappa}^{[-2n]}}{(x^{[2n]})^{N/2}(x^{[-2n]})^{N/2}}\;,
\end{equation}
where $p$ is a $2i$-analytic periodic factor. We now make the following observation: the infinite product in the r.h.s. of \eq{mu12P0mt} grows exponentially at infinity as $e^{\pi u N/2}$. Since the analytic prefactor $p$ is either a constant or also grows exponentially, we conclude that $N$ cannot exceed $4$. This establishes a connection between the number of the massless magnons and the exponential growth of the gluing matrix, as the exponential growth in $\mu$ can be traced back to the gluing matrix \eq{eq:MainGluing}.

\paragraph{The auxiliary roots.}
The ABBA construction of Section~\ref{sec:ABBA} features not only massive and massless momentum-carrying roots, but also a variety of auxiliary roots. These roots are encoded within a subset of the remaining Q-functions.  In principle, one can choose different subsets of $Q$-functions, each featuring different auxiliary roots and yielding different formulations of the ABBA. In the context of the standard ABA, this freedom corresponds to choosing different gradings of the Bethe ansatz. However, due to the non-compact nature of some of the auxiliary equations, there exists a natural and preferred choice of so-called nesting path:
\begin{equation}
    \left\{\,\bP_{1},\,Q_{1|1},\,Q_{1|13},\,Q_{12|13},\,Q_{123|13}\propto Q^{4|24},\,Q_{123|123}\propto Q^{4|4},\,Q_{123|1234}\propto\bP^{4} \,\right\}\,.
\end{equation}
We notice that this set of Q-functions is very similar to the one used for ABA for local operators, with an important difference that the index $2$ is interchanged with $3$. This is essentially equivalent to swapping the $\Delta$ and $S_1$ charges in the conformal group, which is of course expected as now $S_1$ plays the role of the energy whereas $\Delta$ is a parameter.

The analytic form of all $Q$-functions is derived in Appendix~\ref{app:ABBADerivation}, where we show that
\begin{equation}\label{eq:PABBA}
\begin{split}
    \bP_1 \propto x^{-L/2-1} R_1 B_3\, \sigma\,,
    \qquad
    Q_{1|13} \propto x^{L/2+N/2+1}\,  R_3 B_1\, \frac{f f^{[2]}}{\bar{\kappa}\sigma} \prod_{n=0}^{\infty} \frac{\kappa^{[2n]}}{\bar{\kappa}^{[2n]}}\;,\\
    \bP^4 \propto x^{-L/2-1} R_7 B_5\, \sigma\,,
    \qquad
    Q^{4|24} \propto x^{L/2+N/2+1}\, R_5 B_7\, \frac{f f^{[2]}}{\bar{\kappa}\sigma} \prod_{n=0}^{\infty} \frac{\kappa^{[2n]}}{\bar{\kappa}^{[2n]}}\;,
\end{split}
\end{equation}
which feature the auxiliary roots through the functions $R$ and $B$ defined in \eqref{eq:MainRB}.

It is in \eqref{eq:PABBA} that the dressing phases first appear. They essentially capture the deviation of $\bP$ from being a purely rational function of the Zhukovsky variable $x$.
We define the dressing phase as
\begin{equation}
\sigma = \sigma_p\,\sigma_{\bullet}\,\sigma_{\circ}\,,
\end{equation}
where $\sigma$s are written in terms of 
the previously introduced $\varsigma,\varsigma_{B}$ and $\varsigma_{p}$ as follows
\begin{align}\label{eq:PSigmaDef}
    \sigma_{\bullet} = \prod_{k=1}^{K_4} \varsigma_{\bullet}(x,x_{4,k})\,,
    \qquad
    \sigma_{\circ} = \prod_{k=1}^{N} \varsigma_{\circ}(x,z_k)\,\varsigma_{B}(x,z_k)\,,
    \qquad
    \sigma_{p} = \prod_{k=1}^{\frac{4-N}{2}} \varsigma_p(x,u_0)\,.
\end{align}

Finally, in general $Q_{a|i}$ cannot be put in a form as explicit as \eqref{eq:PABBA}. Instead, it is related to $\betheQ_2,\betheQ_{6}$ via the following relations
\begin{equation}
    Q_{1|1} = \betheQ_{2}\, f^+ \prod_{n=0}^{\infty} \left(\frac{\kappa_1}{\bar{\kappa}_2}\right)^{[2n+1]}\,,
    \quad
    Q^{4|4} = \betheQ_{6}\, f^+ \prod_{n=0}^{\infty} \left(\frac{\kappa_1}{\bar{\kappa}_2} \right)^{[2n+1]}\;.
\end{equation}
where $\betheQ_2$ is constrained by the Baxter equation \eqref{baxterQ2copy2} and $\betheQ_6$ by a similar equation obtained by sending $1\rightarrow 7,3\rightarrow 5$ in \eqref{baxterQ2copy2}.

\subsection{Deriving the ABBA}

In the previous paragraph, we identified the origin of all Bethe roots appearing in the ABBA — both momentum-carrying and auxiliary — as they emerge in the QSC in the weak coupling limit. What remains is to derive the equations satisfied by these roots.

\paragraph{Massive equations.}
Once the main Q-functions are determined, the derivation of the ABBA equations for the nodes $1,3,4,5,7$, as given in \eqref{mostgeneralABBA}, is a rather straightforward algebra exercise, following from the QQ-relations \eqref{eq:QQRelations}. Those calculations are presented in detail in Appendix~\ref{app:ABBADerivation}. Here, we demonstrate the procedure explicitly for node 4.

To find the massive middle node equation of the ABBA, we start from the QQ-relation
$$
    \text{Wr}(Q_{12|13},Q_{13|13}) \propto Q_{1|13}Q^{4|24}\,,
$$
which is a special case of \eqref{eq:QQRelations}.
Shifting and evaluating this equation at $u_{4,k}=0$ we find
\begin{equation}
    \frac{Q^{[2]}_{12|13}}{Q^{[-2]}_{12|13}}\Bigg|_{u = u_{4,k}} = -\frac{Q^+_{1|13}}{Q^-_{1|13}}\frac{\left(Q^{2|24}\right)^+}{\left(Q^{2|24}\right)^-}\Bigg|_{u = u_{4,k}}\;.
\end{equation}
Then, we can plug in the explicit expressions \eqref{q1213sec42} and \eqref{eq:PABBA} we obtained in Subsection~\ref{subsec:DerivingABBA}. This gives
\begin{equation}
   \left(\frac{x_{4,k}^+}{x_{4,k}^-}\right)^{L+2}= -\frac{\betheQ_4^{[2]}}{\betheQ_4^{[-2]}}\frac{\betheQ_{\theta}^{+}}{\betheQ_{\theta}^{-}} \left(\frac{\sigma^+}{\sigma^-}\right)^2\frac{B^-_1 R^-_3 R^-_5 B^-_7}{B^+_1 R^+_3 R^+_5 B^+_7}\bigg|_{u=u_{4,k}}\;,
\end{equation}
that is the middle node massive Bethe equation presented on \eqref{mostgeneralABBA}.

\paragraph{Massless equation.}

The derivation of massless equations is a bit more involved and is presented in Appendix~\ref{app:ABBADerivation}. Here we will be very brief.

For real parameters, one can relate $\bQ^{\uparrow}$ with $\bar{\bQ}^{\downarrow}$ and then the gluing equation \eqref{eq:MainGluing} implies
\begin{equation}\label{eq:GluingQMain}
    \bQ_{1} = \alpha_1\, \tilde{\bar{\bQ}}^{2}
    \quad,\quad
    \bQ_{3} = \alpha_3 \,\tilde{\bar{\bQ}}^{4}\;,
\end{equation}
where $\alpha_1$ and $\alpha_3$ are constants. Using the fundamental QQ-relations \eqref{eq:FiniteDifferenceQai} one can recast \eqref{eq:GluingQMain} in terms of $\bP$ and $Q_{a|i}$, which we parametrised in terms of ABBA data in the previous paragraphs. The tricky part in the derivation is that the asymptotic approximation fails in the vicinity of the branch cut. Nevertheless, as detailed in Appendix~\ref{app:ABBADerivation}, one can still extract information from the asymptotic regime to impose gluing conditions, which then results in the massless middle node equation as shown on \eqref{mostgeneralABBA}.

\section{The ABBA Minimal Sector: Solving HT$_0$}
\label{sec:minimal}

In this section, we show how the ABBA equations work in the simplest case,  which we refer to as the \textit{minimal sector}. This can be considered to be a ground state of the system. 
It is characterised by the absence of any massive Bethe roots, that is $K_4=0$, so the minimal sector is precisely what was denoted HT$_0$ in Section~\ref{sec:structure}. 
Whereas for local operators, the equivalent configuration would correspond to half-BPS operators $\tr(Z^L)$, which are protected and thus trivial from the spectrum point of view, the minimal sector corresponds to sets of non-local observables with non-trivial coupling dependence, governed by a dynamics in the massless sector. In particular, it includes the Pomeron eigenvalue for the special case of $J_1=2$~\cite{Alfimov:2014bwa,Gromov:2015wca,Alfimov:2018cms}. The $J_1=3$ case was also studied at weak coupling in the literature and some partial results are available in e.g.~\cite{Kotikov:2007cy,Beccaria:2007cn,Klabbers:2023zdz}.

A subset of this sector was previously considered in \cite{Ekhammar:2024kzp} for general $J_1$ within the ABBA framework, but for the parity invariant and LR invariant case. Here, we remove these restrictions and also provide a more detailed derivation.

We set $K_4=0$, so that \eqref{magnonnumbers} implies $-\left.S_1\right|_{g\to 0}\equiv \Sigma=J_1-1$. In addition, we remind the reader that for all HTs we should remove $\sigma_p$ and set $N=4$ in the ABBA equations from Section~\ref{sec:ABBA}. Note that setting $K_4=0$ implies that $K_{a}=0\,,\,a=1,3,5,7$ too. Indeed, \eqref{magnonnumbers} results in the following contraints
\begin{equation}
    K_1 + K_3 = J_3 - J_2 \geq 0\,,
    \quad
    K_5+K_7 = -J_3-J_2 \geq 0\,.
\end{equation}
If we assume that there are local operators on the Regge trajectories, then we also have $J_2\pm J_3\geq 0$ from \eq{inequalities}~\footnote{Of course there is an interesting possibility that there are Regge surfaces which has no local operators on them. We don't know if those exist, but if they do, we would like to call them ``Regge-ghosts".}. Thus, we conclude that $J_2=J_3=0$, which also implies that $L=J_1$ while $S_2$ is not constrained. To summarize
\begin{equation*}
\boxed{
\text{Minimal Sector (HT$_0$): }
\quad 
\Sigma = L-1\,,
\quad
N=4\,,
\quad
K_1=K_3=K_4=K_5=K_7=0}\;.
\end{equation*}

In the minimal sector, one needs only to consider the massless magnons and hence only the massless equation and the two Baxter equations out of the full system in \eqref{mostgeneralABBA}. In the following, we demonstrate an explicit solution of the ABBA in this setting.

\subsection{Solving the Baxter Equation}
In this section, we show that the Baxter equations for nodes $2$ and $6$ can be solved explicitly in terms of a hypergeometric function, granting us analytic control.
We focus on the equation for node $2$, the treatment of node $6$ is completely analogous upon the replacement $S_2\to -S_2$.

In the absence of massive roots, the Baxter equation \eqref{baxterQ2copy2} simplifies to
\beq\la{eq:BaxterMS}
{\mathbb Q}_a^{[+2]}{
{\mathbb Q}_{\theta_2}^+
}
+
{\mathbb Q}_a{\mathbb T}
+
{\mathbb Q}_a^{[-2]}
{\mathbb Q}_{\theta_1}^-
=0\;,
\qquad
a=2,\tilde{2}\,,
\eeq
which should be supplemented with asymptotics for $\betheQ_a$, \eq{asymptoticQ2} . In our current case, we have
\beq \label{eq:Q2AsymptoticsMinimalSector}
{\mathbb Q}_2\simeq u^{\alpha_{1,2}^+}\,,\qquad\;{\mathbb Q}_{\tilde 2}\simeq u^{\alpha_{1,2}^-}\,,
\eeq
with
\beq\la{DefAlpha}
\alpha^\pm_{a,b}=\frac{\pm (\Delta-S_2)-i(\sum_i\theta_i-2\theta_a-2\theta_b)-1 }{2}\;.
\eeq

We notice that the second degree polynomial ${\mathbb T}$ is completely fixed from the asymptotics of $\mathbb{Q}_a$ (see \eq{Tintn} and \eq{t1t2}) and it becomes
\beq
{\mathbb T}=-2 u^2
+u\sum_i^4 \theta_i
+\frac{1}{4} \left(\left(\Delta -S_2\right)^2-\left(\sum _i^4 \theta _i\right)^2+2 \sum_i^4 \theta_i^2+1\right)\;.
\eeq

A second-order finite-difference equation with polynomial second-order coefficients can be solved by the method described in e.g. \cite{Korchemsky_1995}. Let us briefly recall the main steps: one first performs a Mellin transform to convert the finite difference equation into a second-order differential equation
\beq
{\mathbb Q}(u)=
\int_0^1 (z-1)^{i u-i \theta _3-\frac{1}{2}} z^{-i u+i \theta _1-\frac{1}{2}} {\mathbb Q}_M(z) dz\;.
\eeq
For ${\mathbb Q}_M(z)$ we obtain the hypergeometric type equation
\beqa
{\mathbb Q}_M''(z)+{\mathbb Q}_M(z)  \frac{\alpha^-_{2,3} \alpha^+_{2,3}}{(z-1)z}+\left(\frac{i \theta _1-i \theta _2+1}{z}+\frac{-i \theta _3+i \theta _4+1}{z-1}\right) {\mathbb Q}_M'(z)=0\;.
\eeqa
Solving this equation and transforming it back, we find one of the solutions to be given by
\beqa
&&G(u)=
\frac{ {}_3F_2\left(\frac{1}{2}+i (u-\theta_3),-\alpha_{2,3}^+,-\alpha_{2,3}^-;1+i (\theta_1-\theta_3),1-i (\theta_3- \theta_4);1\right)}{
\frac{\Gamma \left(\frac{1}{2}-i (u- \theta_3)\right)}{
\Gamma \left(\frac{1}{2}-i (u-\theta _1)\right)
}2 e^{\frac{1}{4} \pi  \left(\theta_1+\theta_2-\theta_3-\theta_4\right)} 
\Gamma \left(1+i (\theta_1-\theta_3)\right) \Gamma \left(1-i (\theta_3-\theta_4)\right)}\;.
\label{explicitF}
\eeqa
Before we find the second solution, let us study the large $u$ asymptotics of this solution. Expanding at large $u$ the explicit expression \eq{explicitF}, we find
\beq\la{Glargeu}
G(u)\simeq 
\frac{\sqrt[4]{-1} e^{\frac{1}{4} i \pi  \left(\Delta -S_2\right)} \Gamma \left(S_2-\Delta \right) {u}^{\alpha _{1,2}^-}}{2 \Gamma \left(-\alpha _{3,4}^+\right) \Gamma \left(-\alpha _{2,3}^+\right) \Gamma \left(-\alpha _{1,3}^+\right)}+\frac{\sqrt[4]{-1} e^{-\frac{1}{4} i \pi  \left(\Delta -S_2\right)} \Gamma \left(\Delta -S_2\right) {u}^{\alpha _{1,2}^+}}{2 \Gamma \left(-\alpha _{3,4}^-\right) \Gamma \left(-\alpha _{2,3}^-\right) \Gamma \left(-\alpha _{1,3}^-\right)}\;.
\eeq
From the asymptotics, we see that the solution is invariant under $\theta_1\leftrightarrow\theta_2$ but not under $\theta_3\leftrightarrow\theta_4$.
As this is a symmetry of the initial Baxter equation, we can use this to generate the second linearly independent solution $G_{\theta_3\leftrightarrow\theta_4}(u)$.

By comparing the asymptotic \eq{Glargeu} with 
the required asymptotics of the ``pure" solutions ${\mathbb Q}_2$ and ${\mathbb Q}_{\tilde 2}$ \eq{eq:Q2AsymptoticsMinimalSector} we can identify a linear transformation leading to 
${\mathbb Q}_2$ in terms of $G(u)$ and $G_{\theta_3\leftrightarrow\theta_4}(u)$.
The expression is quite bulky, so we do not write it explicitly.

Before we proceed, we have to verify the poles are at the correct positions as required in \eq{Q2poles}. It is easy to see from its explicit form that $G(u)$ is regular in the upper half plane and has poles at $\theta_1-i/2-in$
and $\theta_2-i/2-in$ for $n=0,1,2,\dots$. Furthermore, at $u=\theta_3-i/2$ it vanishes while at  $u=\theta_4-i/2$ it is just a constant. For what follows, we also need to know the residues  $r_{\theta_1}$ and $r_{\theta_2}$ of $G(u)$ at $u=\theta_1-i/2$ and $u=\theta_2-i/2$ respectively. They can be found from the explicit expression~\eq{explicitF} quite straightforwardly\footnote{The poles appear as the divergence of the standard representation of the hypergeometric function in \eq{explicitF} as a series. Thus, the residues can be deduced from the tail of the series in the large index limit.}
\beqa\la{residue}
r_{\theta_1}&=&\frac{e^{-\frac{1}{4} \pi  \left(\theta _1+\theta _2-\theta _3-\theta _4\right)} \Gamma \left(i \left(\theta _2-\theta_1\right)\right) \sinh \left(\pi  \left(\theta _1-\theta_3\right)\right)}{2 \pi  \Gamma \left(\alpha _{2,4}^-+1\right) \Gamma \left(\alpha _{2,4}^++1\right)}\;,
\eeqa
and $r_{\theta_2}$ is obtained by replacing $\theta_1\leftrightarrow\theta_2$ in $r_{\theta_1}$.
These residues translate into simple analytic expressions for residues in ${\mathbb Q}_2$. 

\subsection{Massless Bethe Equations in the Minimal Sector}
As we managed to solve the Baxter equation explicitly in this sector, the only non-trivial equation left is the massless Bethe equation, which in the minimal sector reads
\begin{equation}
z_k^{L+2}=
\zeta_P\,\left[\frac{\betheQ_{2}^-}{\bar{\betheQ}_{\tilde 6}^+}\,\sqrt{\frac{\bar{\Gamma}_1\Gamma_2}{\Gamma_1\bar{\Gamma}_2}}\bigg|_{z=z_k}\right]\,
\prod_{j=1}^{4}\sig^2_{\circ\circ}(z_k,z_j)\;.
\end{equation}
Notice that the r.h.s. seems to be dependent on the way the set $\{\theta_i\}_{i=1}^4$ is split into two parts, as $\Gamma_1$ and $\Gamma_2$ contain the supplementary halves of all $\theta$'s. However, $\betheQ_{2}^-$ and $\betheQ_{6}^-$ also implicitly depend on the choice of the ordering of $\theta's$, as can be seen from the Baxter equation they solve~\eq{eq:BaxterMS}. As discussed in
Section~\ref{sec:ABBA} and proven in Appendix~\ref{app:ABBADerivation},  the r.h.s. should be symmetric under the permutation of $\theta_k$.
We use this fact to take $k=1$ and $k=2$, as in this case we know the value of the residues of ${\mathbb Q}_2$ and ${\mathbb Q}_{\tilde 6}$ explicitly from \eq{residue}. This allows us to get explicitly the following expression
\beq
\left.{\cal C}\frac{\betheQ_{2}^-}{\bar{\betheQ}_{\tilde{6}}^+}
\sqrt{
\frac{\bar\Gamma_1\Gamma_2}{\Gamma_1\bar\Gamma_2}
}\right|_{u\to\theta_k}\equiv \prod_{j\neq k}{\cal G}_{k,j}\,,
\eeq
where ${\cal C}$ is a constant
and
\beq
{\cal G}_{k,j}=\sqrt{
\frac{\Gamma \left(1-i \theta _k+i\theta _j\right) 
\Gamma \left(1+
\alpha_{jk}^-
\right) 
\Gamma \left(1+S_2+\alpha_{jk}^+\)
}
{\Gamma \left(1+i \theta _k-i\theta _j\right) 
\Gamma \left(-\alpha_{jk}^+
\right) 
\Gamma \left(S_2-\alpha_{jk}^-
\right)}
}\;,\la{Gmaineq}
\eeq
with $\alpha_{jk}^\pm$ defined in \eq{DefAlpha}.
This factor is not ``unitary" in the sense that ${\cal G}_{k,j}\neq 1/{\cal G}_{j,k}$; however, it does satisfy $\prod_{k\neq j} {\cal G}_{k,j}=1$.

To summarise, we get the following equation
\beq\la{Zmain}
z_k^{L+2}=\tilde{\zeta} \prod_{j\neq k}
{\cal G}_{k,j} 
\sig^2_{\circ\circ}(z_k,z_j)\;,
\eeq
where we have defined $\tilde{\zeta} = \zeta_P/\mathcal{C}$. Simply taking the product of the above equations for all massless roots and using cyclicity, unitarity of $\sigma_{\circ\circ}$ and the ``effective'' unitarity of $\mathcal{G}$ fixes $\tilde{\zeta} = e^{i\frac{\pi l}{2}},\;l=0,1,2,3$.

\subsection{Counting of Solutions and Low-Lying Spectrum}\label{subsec:CountingSolutionsExamples}
In this section, we count the number of solutions to \eqref{Zmain} and solve it for $L=2,3,4$. At weak coupling, the massless Bethe equation \eq{Zmain} simplify significantly as $\theta_j\sim g\to 0$ so that all non-trivial factors disappear. Thus we obtain
\beq\label{eq:MasslessTreeMS}
(z^{(0)}_k)^{L+2}= e^{i\pi l/2}\la{Zmain0}\;,
\eeq
with $z_k=z^{(0)}_k +\mathcal{O}(g)$. In addition, we have to impose the cyclicity condition \eq{cyclicity}, which in our case reads $\prod_{k=1}^{4} z_k=1$. We introduce a slightly undemocratic parametrisation, which automatically satisfies the cyclicity condition:
\beq\la{eq:kaDef}
z^{(0)}_4 = e^{-\frac{\pi i}{2}\frac{k_1+k_2+k_3}{L+2}}\,,
\qquad 
z^{(0)}_i=z^{(0)}_4 e^{2\pi i \frac{k_i}{L+2}}\,,
\qquad \quad
1\leq\! k_1\!<\! k_2\!<\!k_3\!\leq L+1\,.
\eeq
This parametrisation makes it apparent that
\begin{equation}\label{totnumber}
\texttt{Total number of states}=\frac{(L-1)L(L+1)}{6}\,,
\end{equation}
out of them $\lfloor{L^2/4}\rfloor$ are parity symmetric, as found in \cite{Ekhammar:2024neh}.

Once \eq{Zmain} is solved to leading order in $g$, one can expand it to find corrections order by order in the coupling. For $\gamma_S$ we find the first two orders to be
\begin{equation}\label{eq:gammaSg2}
\begin{split}
\gamma_S &=-2 g \left[
\sin \frac{\pi  k_{\text{tot}}}{2 (L+2)}+
\sum_{i=1}^3
\sin \frac{\pi  \left(k_{\text{tot}}-4 k_i\right)}{2 (L+2)}\right] \\
&-\frac{g^2\,\chi(\Delta,S_2)}{L+2}\bigg(\sum_{i=1}^{3}\left[2\cos{\frac{2\pi k_i}{L+2}}-3\cos\frac{\pi (k_{\text{tot}}-4k_i)}{L+2}+4\cos\frac{\pi (k_{\text{tot}}-2k_i)}{L+2}\right] \\
&\quad \qquad \qquad \qquad \quad -12+\sum_{i=1}^{3}\sum_{j=i+1}^{3}\cos\frac{2\pi(k_i-k_j)}{L+2} -3 \cos \frac{\pi  k_{\text{tot}}}{2 (L+2)}\bigg) + \mathcal{O}(g^3)\,,
\end{split}
\end{equation}
with $k_{\text{tot}}=k_1+k_2+k_3$ and
\begin{equation}\label{eq:ChiDefinition}
    \chi(\Delta,S_2) = \psi\(\frac{1-\Delta+S_2}{2}\) + \psi\(\frac{1+\Delta+S_2}{2}\) + 2 \gamma_E\,,
\end{equation}
the Pomeron eigenvalue with general conformal spin $S_2$, where $\psi(x)$ is the Digamma function and $\gamma_E$ the Euler-Mascheroni constant.
In the simple case of $L=2$, the only option for the parametrisation is $k_1=1,\;k_2=2,\;k_3=3$, and we obtain 
\beq
\gamma_S = 4\,g^2\,\chi(\Delta,S_2)+\mathcal{O}(g^4)\,,
\eeq
reproducing the result of \cite{Kotikov:2000pm}.

\begin{table}[!t]
$$
\begin{array}{c|ccc|cc|c|l}
 L  & k_1 & k_2 & k_3 & m_1 & m_2 & {\mathbb P} & \gamma_S \\ \hline
 2 & 1 & 2 & 3 & 1&  3& + & 4 \chi  g^2 \\ \hline
 3 & 1 & 2 & 3 &  &  &  -& 2 \chi  g^2+\left(\frac{4 \chi ^3}{3}+\frac{2 \pi ^2 \chi 
}{3}-\frac{20 \zeta _3}{3}-\frac{20 \chi ''}{3}\right) g^4 \\
 3 & 1 & 2 & 4 &  1&  3& +&-2 g+4 \chi  g^2+\frac{2 \pi ^2 g^3}{3}+\left(-\frac{4 \pi ^2 \chi \
}{3}-28 \zeta _3-24 \chi ''\right) g^4\\
 3 & 1 & 3 & 4 &  &  & - & 2 \chi  g^2+\left(\frac{4 \chi ^3}{3}+\frac{2 \pi ^2 \chi \
}{3}-\frac{20 \zeta _3}{3}-\frac{20 \chi ''}{3}\right) g^4 \\
 3 & 2 & 3 & 4 &  2&  4& +& 2 g+4 \chi  g^2-\frac{2 \pi ^2 g^3}{3}+\left(-\frac{4 \pi ^2 \chi 
}{3}-28 \zeta _3-24 \chi ''\right) g^4 \\  \hline
4 & 1 & 2 & 3 &  &  & -&\chi  g^2+\left(\frac{1}{2} \chi  \left(\chi ^2+\pi ^2\right)-2 
\left(\zeta _3+\chi ''\right)\right) g^4 \\
4 & 1 & 2 & 4 &  &  & -&-\sqrt{2} g+\frac{5 \chi  g^2}{2}+\frac{\left(4 \pi ^2-15 \chi ^2\right) g^3}{24 \sqrt{2}}+\frac{1}{12} \left(5 \chi ^3-3 \pi ^2 \chi \
-148 \zeta _3-130 \chi ''\right) g^4  \\
4 & 1 & 2 & 5 &  1&  3&  +&-2 \sqrt{3} g+\frac{10 \chi  g^2}{3}+\frac{\left(\chi ^2+2 \pi ^2\right) g^3}{\sqrt{3}}-\frac{4}{81} \left(8 \chi ^3+18 \pi ^2 \chi +531 
\zeta _3+369 \chi ''\right) g^4 \\
4 & 1 & 3 & 4 &  1&  5&  +&\frac{4 \chi  g^2}{3}-\frac{4}{81} \left(-16 \chi ^3-9 \pi ^2 \chi 
+72 \left(\zeta _3+\chi ''\right)\right) g^4 \\
4 & 1 & 3 & 5 &  &  & -&-\sqrt{2} g+\frac{5 \chi  g^2}{2}+\frac{\left(4 \pi ^2-15 \chi ^2\right) g^3}{24 \sqrt{2}}+\frac{1}{12} \left(5 \chi ^3-3 \pi ^2 \chi \
-148 \zeta _3-130 \chi ''\right) g^4 \\ 
 4 & 1 & 4 & 5 &  &  &  -&\chi  g^2+\left(\frac{1}{2} \chi  \left(\chi ^2+\pi ^2\right)-2 \left(\zeta _3+\chi ''\right)\right) g^4 \\
 4 & 2 & 3 & 4 &  &  & -&\sqrt{2} g+\frac{5 \chi  g^2}{2}+\frac{\left(15 \chi ^2-4 \pi ^2\right) g^3}{24 \sqrt{2}}+\frac{1}{12} \left(5 \chi ^3-3 \pi ^2 \chi \
-148 \zeta _3-130 \chi ''\right) g^4 \\
4 & 2 & 3 & 5 & 2& 4&  +&4 \chi  g^2+\left(-\frac{4 \pi ^2 \chi }{3}-32 \left(\zeta _3+\chi''\right)\right) g^4 \\
4 & 2 & 4 & 5 &  &  & -&\sqrt{2} g+\frac{5 \chi  g^2}{2}+\frac{\left(15 \chi ^2-4 \pi ^2\right) g^3}{24 \sqrt{2}}+\frac{1}{12} \left(5 \chi ^3-3 \pi ^2 \chi \
-148 \zeta _3-130 \chi ''\right) g^4 \\
 4  & 3 & 4 & 5 &  3&  5& + & 2 \sqrt{3} g+\frac{10 \chi  g^2}{3}-\frac{\left(\chi ^2+2 \pi ^2\right) g^3}{\sqrt{3}}-\frac{4}{81} \left(8 \chi ^3+18 \pi ^2 \chi +531 \zeta _3+369 \chi ''\right) g^4\\
  \hline
\end{array}
$$
\caption{We list all solutions for $L=2,3,4$ in the minimal sector.
We give the $k$-labels~\eq{eq:kaDef} for all states, and $m$-labels~\eq{eq:m1m2Def} for parity symmetric states. We have used short-hand notation $\chi=\chi(\Delta,S_2))$ defined in \eqref{eq:ChiDefinition} and derivatives as $\chi' = \frac{\partial}{\partial \Delta}\chi(\Delta,S_2)$ and $\chi''=\frac{\partial^2}{\partial \Delta^2}\chi(\Delta,S_2)$.}
\label{tab:tablesolutions}
\end{table}

\paragraph{Parity symmetric states.}
States exhibiting both parity and LR symmetry form a special subclass of the states discussed above. In particular, for such states we have $z_1=-\frac{1}{z_4}$ and $z_2=-\frac{1}{z_3}$. These states were previously studied in \cite{Ekhammar:2024neh}, where the following parametrisation was employed
\begin{equation}\label{eq:m1m2Def}
    z_1 \!= \!-\ii e^{\frac{\ii \pi}{L+2}m_1}
    ,
    \quad
    \!z_2 \!= \!-\ii e^{\frac{\ii \pi}{L+2}m_2},
    \quad
    \{k_1,k_2,k_3\}\! =\! \left\{\!m_1,\frac{m_1\!+\!m_2}{2},2+L+\frac{m_1\!-\!m_2}{2}\!\right\},
\end{equation}
with $m_i=1,\dots,L+1$, $m_1<m_2$ and $m_1+m_2$ restricted to be even. In this sector, \eqref{eq:gammaSg2} simplifies to \cite{Ekhammar:2024neh}
\begin{equation}\label{eq:GammaSymmetricSector}
    \gamma_S = -4g\sum_{k=1}^{2}\cos \frac{\pi \,m_k}{L+2}\,+\,\frac{16 g^2\,\chi(\Delta)}{L+2}\sum_{k=1}^{2}\sin^2\frac{\pi \, m_k}{L+2} \,+\, \mathcal{O}(g^3)\,.
\end{equation}

\paragraph{More orders and wrapping.}

\begin{figure}[!t]
    \centering 
    \includegraphics[width=\columnwidth]{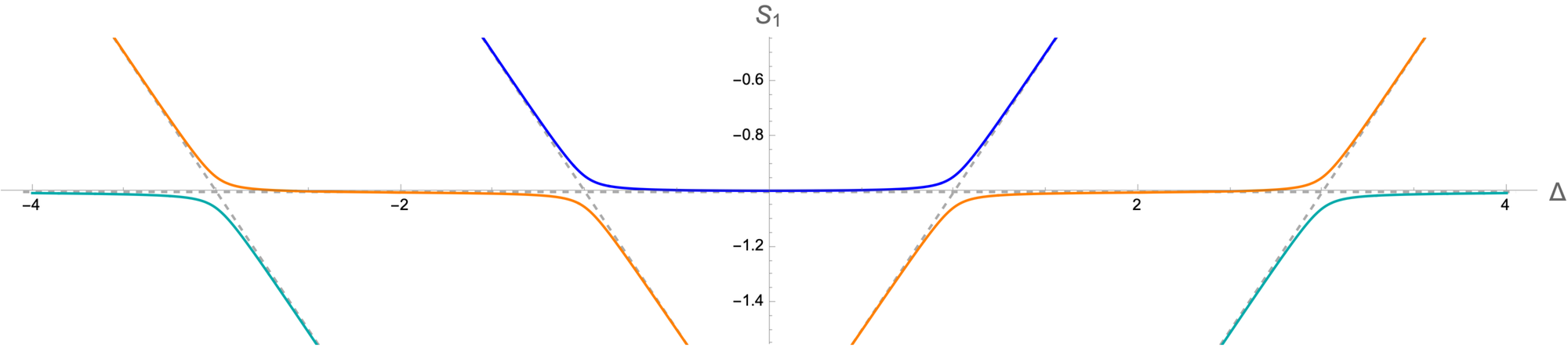}
    \includegraphics[width=\columnwidth]{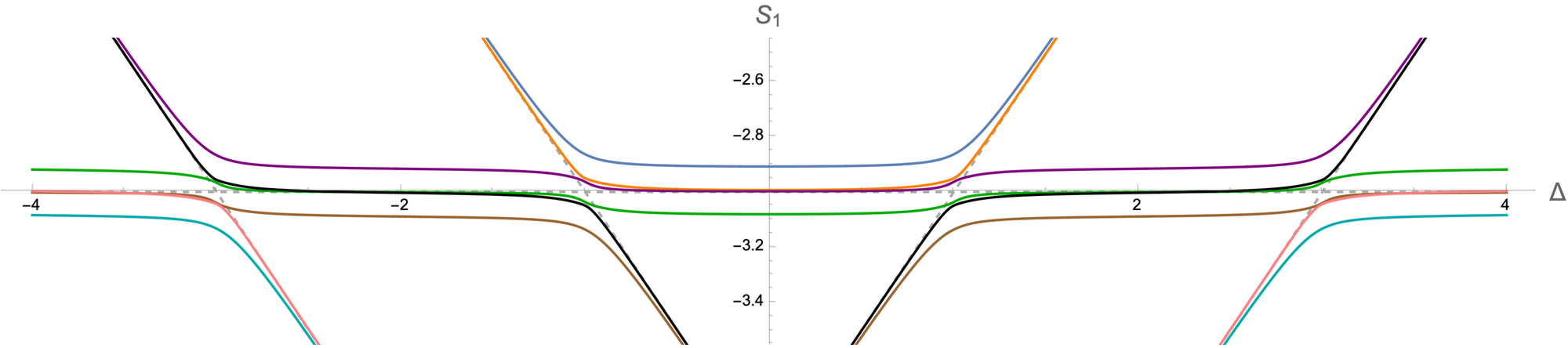}
    \caption{The leading horizontal trajectories ${\rm HT_0}$ for $L=2$ computed at $g=1/50$ and $L=4$ computed at $g=1/40$. We restrict attention to the parity even sector (for $L=2$ this is the only possible case). The $L=2$ case was first computed in \cite{Alfimov:2014bwa} while the $L=4$ one was presented for the first time in \cite{Ekhammar:2024neh}. The trajectories were computed numerically using the QSC, with each curve varying from 500 to 2000 data points.}
    \label{fig:L2L4HT0}
    \centering
    \includegraphics[width=0.49\columnwidth]{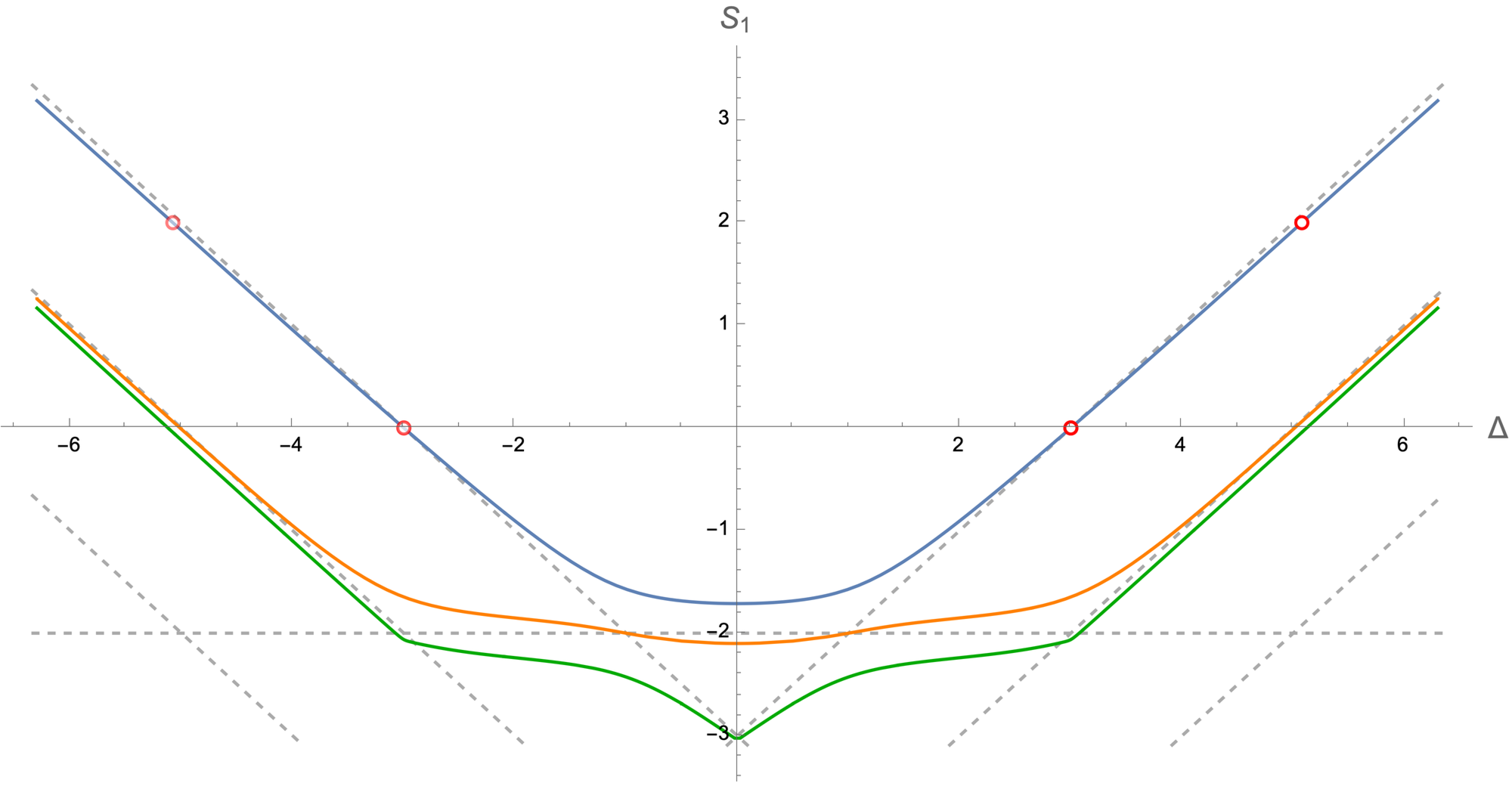}
    \includegraphics[width=0.49\columnwidth]{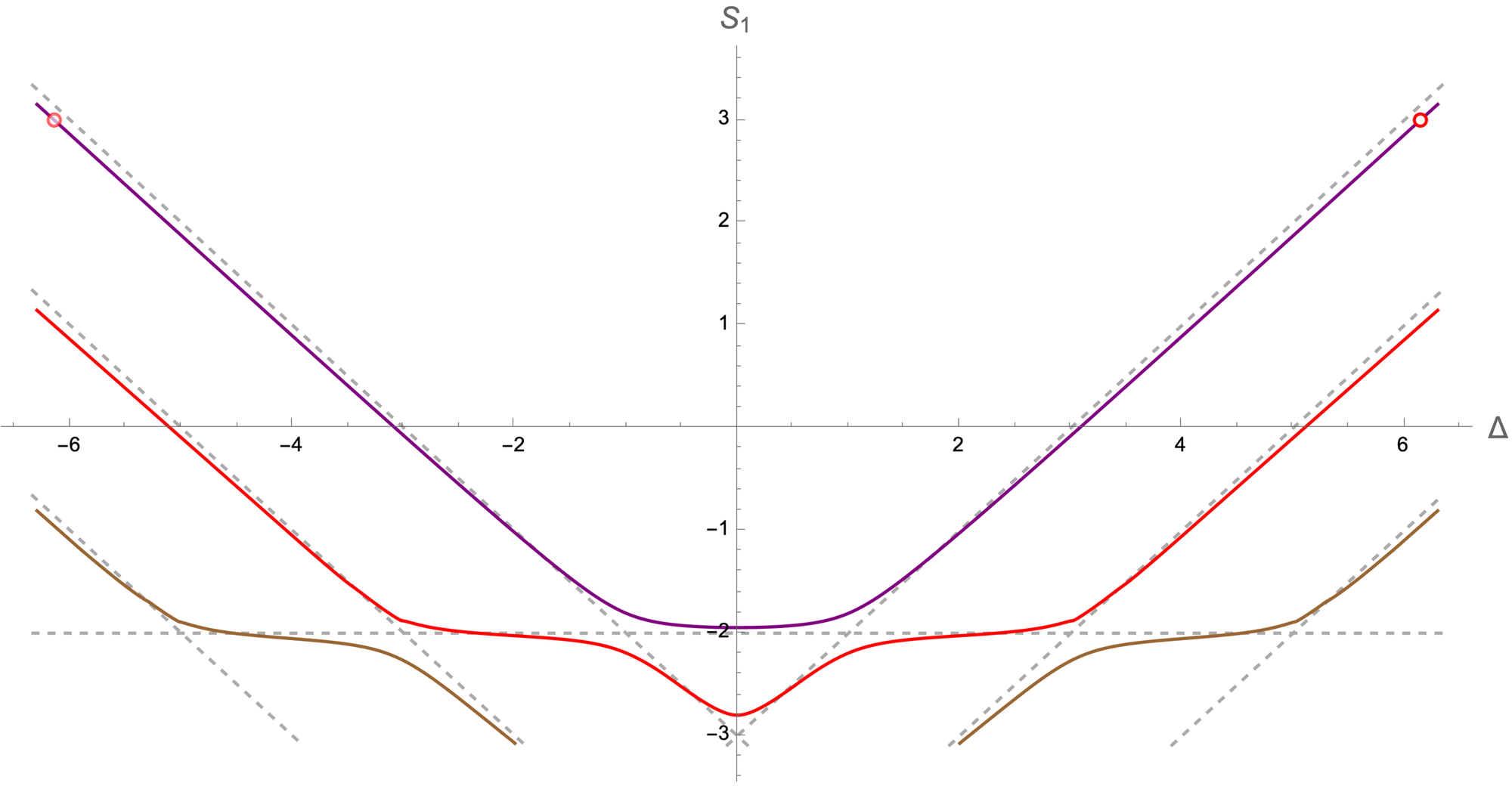}
    \caption{The leading horizontal trajectory ${\rm HT_0}$ for $L=3$ computed at $g=1/10$ for both the parity even (left, as computed in \cite{Klabbers:2023zdz}) and parity odd (right) sectors. The first corresponds to the analytic continuation to complex spin from operators at even integer spin values $S_1=0,2,4,...$, while the second originates from those with odd integer spin $S_1=3,5,...$. 
    The locations of the first few local operators along the leading trajectories, as well as their shadow counterparts, are marked with red circular markers. Due to the double degeneracy of odd-spin local operators, all parity-odd trajectories exhibit an exact two-fold degeneracy. The trajectories were computed numerically using the QSC, with each curve consisting of approximately 1000 data points.
    }
    \label{fig:L3HT0}
\end{figure}

Finally, as a last example of the predictive power of ABBA, we list the value of $\gamma_{S}$ up to $g^2$ for the unique $L=2$ state and up to $g^4$ for all states with $L=3,4$ in Table~\ref{tab:tablesolutions}. While we only display the result up to $g^{4}$, it is straightforward to solve the equations to the desired order (keeping in mind that the ABBA equations are only valid to $g^{L+1}$ order in this case). We have included a notebook in the arXiv submission that accomplishes this task. As can be seen from Table~\ref{tab:tablesolutions} all $\Delta$ and $S_2$ dependence in $\gamma_S$ is fully encoded in $\chi(\Delta,S_2)$, defined in \eqref{eq:ChiDefinition}, and its derivatives. This is a consequence of the fact that $\Delta$ and $S_2$ enter into the minimal sector through \eqref{Gmaineq}. We can see that the coefficient of $g^{m}$ has transcendentality $m-1$. Finally, notice that the spectrum exhibits degeneracies for all parity-odd states with a $g^2$ expansion. This degeneracy is lifted once we take further integrals of motions into account.

We emphasise that the result computed from ABBA eventually has to be supplemented with additional wrapping effects in order to reproduce the full answer. From numerical experiments, we found that the ABBA prediction is valid on its own to $g^{L+1}$, meaning that wrapping hits at $g^{L+2}$. 

\begin{figure}[!t]
    \centering   \includegraphics[width=0.8\columnwidth]{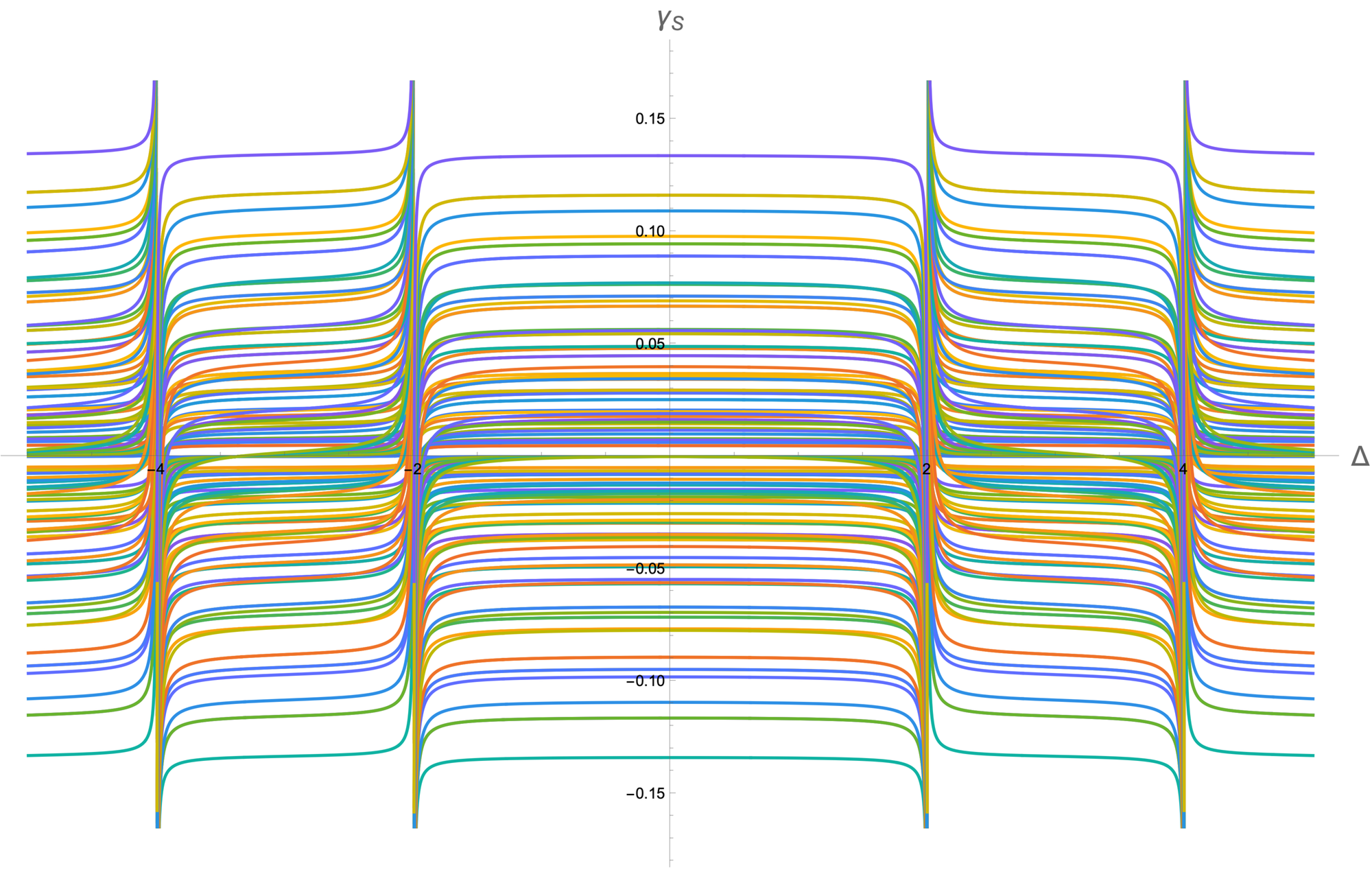}
    \caption{The spectrum of trajectories at HT$_0$ for $L=10$ and $S_2=1$ at $g=1/50$. The total number of states is 165, of which 25 are parity-symmetric and 140 are not. For simplicity, in this plot we just show ABBA solution up to $g^2$ term, even if in principle we can access analytical solution up to $g^{11}$.}
    \label{fig:highL}
\end{figure}

\paragraph{Visualisation of the minimal sector.} Having enumerated all possible trajectories in the minimal sector and solved it perturbatively, we proceeded to initialize QSC numerics, enabling us to plot the trajectories at finite coupling. We present the unique trajectory for $L=2$ and all parity-symmetric trajectories for $L=4$ in Figure~\ref{fig:L2L4HT0}. For all $L=3$ we plot all trajectories in Figure~\ref{fig:L3HT0}. In order to show the predictive power of ABBA equations for higher length, we also consider $L=10$ both for parities. We obtain the analytic prediction for all 165 states using the notebook attached to this work. We present all solutions in Figure~\ref{fig:highL}.

\section{Horizontal Trajectories with Massive Roots }\label{sec:SubleadingHT}

In general, HTs have massive and auxiliary roots in addition to massless ones.
In this section, we examine the simplest parity and left-right invariant trajectories containing two massive momentum-carrying roots, but the same method should apply in general. We will refer to these trajectories as ``sub-leading''.

The ability to analytically solve sub-leading trajectories provides us unique analytic control over trajectories also located deep down in the Chew-Frautschi plot. Reaching these trajectories numerically by starting from a local operator and traversing the relevant Riemann surface is in principle possible. However, it would be extremely inefficient and doomed to fail in a practical sense since the number of horizontal trajectories grows rapidly as \eqref{totnumber}. Thus, the ABBA framework is crucial to allow for an investigation of these trajectories.

In the following, we present how to solve the ABBA for sub-leading trajectories, focusing on the simplest but still highly non-trivial case. We have been able to find all trajectories for a subclass of HT$_2$, to be defined precisely below, for $L=2,4$. In addition, we performed intensive numerical tests to verify that no trajectory was missed.

\subsection{Type I Trajectories in the $\algsl(2)$ Sector}
The simplest possible trajectories to explore beyond the minimal sector are Type I trajectories containing local operators in the $\algsl_{2}$ sector. These are trajectories with the following quantum numbers
\begin{equation}\label{eq:QuantumNumbersHT2}
    J_2 = J_3 = S_2= 0\,,
    \qquad
    K_{4}=K_{3} = K_{5}= 2M\,,
    \qquad
    K_{1}=K_{7} = 0\,,
    \qquad
    N=4\,.
\end{equation}
We also impose parity invariance, giving
\begin{equation}\begin{split}\label{eq:ParityInvarianceHT2}
    z_{4} = -\frac{1}{z_1}\,,&
    \qquad
    z_{3} = -\frac{1}{z_2}\,,\\
    \{u_{3,i}\}_{i=1}^{K_3} = \{u_{3,i},-u_{3,i}\}_{i=1}^{M}\,,&
    \qquad
    \{u_{4,i}\}_{i}^{K_4} = \{u_{4,i},-u_{4,i}\}_{i=1}^{M}\,.
    \end{split}
\end{equation}

Let us summarise the ABBA in this sector. The Baxter equation is controlled by $\mathbb{T}$, which is explicitly given as \eqref{t1t2}
\begin{equation}\la{Tansatz}
    \mathbb{T} = -2 u^{2+2M} + u^{2M}\left(\frac{1-6M}{4}+\frac{\Delta^2}{4}+2\sum_{k=1}^{M} u^2_{3,k}+\sum_{k=1}^2\theta^2_k\right) + \sum_{m=1}^{M}t_{2m+2} u^{2M-2m}\,,
\end{equation}
to fix all the unknown coefficients $\{t_{2k+2}\}_{k=1}^{M}$ and $\{u_{3,k}\}_{k=1}^{M}$, we can follow the discussion in Section~\ref{subsec:SimplifyingNodes}. In order to avoid square-roots, we can pick a slightly different set of equations from those presented above. These new equations follows from \eqref{baxterQ2copy2_2} upon using the 3rd node ABBA equation in \eqref{fig:ABBAgeneral}
\begin{equation}\label{eq:AuxSub}
    \frac{\bfR_{(+)}}{\bfR_{(-)}}\frac{\bar{\kappa}_1}{\kappa_2}\betheQ_{\theta_2} \betheQ^{[-2]}_3 = -\mathbb{T}^-\big|_{u=u_{3,k}}\,,
    \qquad
    \frac{\bfR_{(-)}}{\bfR_{(+)}}\frac{\kappa_1}{\bar{\kappa}_2}\betheQ_{\theta_2} \betheQ^{[2]}_3 = -\mathbb{T}^+\big|_{u=u_{3,k}}\,,
\end{equation}
this gives $2M$ equations on $2M$ unknowns, i.e $u_{3,k}$ and $t_{k}$.

The remaining momentum carrying nodes are fixed from their Bethe equations that can be found in \eqref{mostgeneralABBA}, we repeat them here in our simplified sector for convenience. The massive roots $\{u_{4,k}\}_{k=1}^{M}$ satisfy
\begin{align}\label{eq:MiddleSub}
    \left(\frac{x^+_{4,k}}{x^-_{4,k}}\right)^{L+2} &= -\frac{\betheQ_4^{[2]}}{\betheQ_4^{[-2]}}\frac{\betheQ_{\theta}^{+}}{\betheQ_{\theta}^{-}} \left(\frac{R^-_3}{R^+_3}\right)^2\sig_{\bullet\bullet}^2\sig_{\bullet\circ}^2\bigg|_{u=u_{4,k}}\,,
\end{align}
while the massless roots are governed by 
\begin{equation}\label{eq:MasslessSub}
z_{k}^{L+2}=
\zeta_P\,
\sig^2_{\circ\bullet}
\sig^2_{\circ\circ}\,
\frac{{\mathbb Q}_4^+}{{\mathbb Q}_4^-}\,
\sqrt{
\frac{
{\mathbb Q}_2^-
{\mathbb Q}_{\tilde 2}^-
}{
\bar{\mathbb Q}_2^+
\bar{\mathbb Q}_{\tilde 2}^+}\frac{\bar{\Gamma}_1\Gamma_2}{\Gamma_1\bar{\Gamma}_2}
}\frac{B_3}{R_3}\bigg|_{u=\theta_k}\,.
\end{equation}
The set of equations \eqref{eq:AuxSub}, \eqref{eq:MiddleSub}, \eqref{eq:MasslessSub} together with the ansatz for $\mathbb T$ \eq{Tansatz} and the Baxter equation \eqref{baxterQ2copy2} now form a closed system of equations that can be readily solved numerically given an approximate starting point. We have implemented this algorithm in a notebook, see the ancillary files of the arXiv submission. We will discuss the numerically obtained values for the roots in Section~\ref{subsec:StartingUp}

In order to obtain a classification and counting of solutions, we will obtain an analytic solution to leading order in $g$ in the next section.

\subsection{The Weak Coupling Limit}
To enumerate solutions and find approximate starting points, we will now consider the $g\rightarrow0$ limit. We write $u_{a,k} \simeq u^{(0)}_{a,k} +\mathcal{O}(g)$, and add a subscript $(0)$ to indicate that all functions are to be consider at $g=0$, for example $\mathbb{T}_{(0)}=\mathbb{T}\big|_{g\rightarrow 0}$. Taking this limit for \eqref{eq:AuxSub} and \eqref{eq:MiddleSub} gives
\begin{align}
    \frac{\betheQ_{4,(0)}^{[2]}}{\betheQ_{4,(0)}^{[-2]}} \left(\frac{\betheQ^-_{3,(0)}}{\betheQ^+_{3,(0)}}\right)^{2} &= -\left(\frac{u^+}{u^-}\right)^{L-2}\bigg|_{u=u^{(0)}_{4,k}}\,, \label{eq:MomentumSubWeak}
    \\
    \frac{\betheQ^+_{4,(0)}}{\betheQ^-_{4,(0)}}\betheQ_{3,(0)}^{[-2]} u^{2} = -\mathbb{T}^-_{(0)}\bigg|_{u=u^{(0)}_{3,k}}\,,
    &\qquad
    \frac{\betheQ^-_{4,(0)}}{\betheQ^+_{4,(0)}}\betheQ_{3,(0)}^{[2]} u^{2} = -\mathbb{T}_{(0)}^+\bigg|_{u=u^{(0)}_{3,k}}\,,\label{eq:AuxSubWeak}
\end{align}
which importantly decouples fully from the massless part of the spectrum, a feature already discussed in Section~\ref{subsec:WeakCouplingABBA}. In the above we have $K_4+2K_3=3M$ equations for the equal number of variables $u_{4,k},\;u_{3,k},\;t_{2m+2}$.

Naively, one might expect that the massless part also decouples; this is true if all massive roots scale as $\mathcal{O}(g^0)$, which was the assumption in Section~\ref{subsec:WeakCouplingABBA}. 
In general, we found that this is not true. 
Instead, at least for the case $M=1$, we found that there are two options, either $u_{3,k}\sim 1$
or $u_{3,k}\sim g^2$, which will give rise to ``length-changing" effects. Indeed, assuming we have $M_0$ roots $u_{3,k}\sim g^2$ and all others $u_{3,k}\sim 1$, we have to take the limit of the r.h.s. of the massless equation more carefully to obtain
\begin{equation}
    \frac{B_3}{R_3}\,\underset{g\rightarrow 0}{\simeq}\,\frac{1}{x^{2M_0}}\,,
\end{equation}
and the massless equation becomes 
\begin{equation}\label{eq:MasslessSubWeak}
     \left(z^{(0)}_k\right)^{L_{\text{eff}}+2} = e^{\ii \pi l/2}\,,
    \qquad
    L_{\text{eff}} = L + 2 M_0 \,,
\end{equation}
where we use that all other terms can be absorbed into the overall phase, like was discussed in Section~\ref{subsec:WeakCouplingABBA}.
Since the equation \eq{eq:MasslessSubWeak} is exactly the same as that in the minimal sector \eq{eq:m1m2Def} with $L\to L_{\rm eff}$, we have
\begin{equation}\la{zeff}
    z^{(0)}_{k} = -\ii e^{\frac{\pi m_k}{\Leff+2}}\,,
    \quad
    k=1,2\,,
\end{equation}
with the range adjusted to be $1\le m_1\le m_2 \le L_{\rm eff}+1$ and $m_1+m_2$ is still even. This gives $\lfloor L_{\rm eff}^2/4 \rfloor$ solutions in total in the massless sector. We remind that for each solution in the massless sector, we have multiple solutions in the massive sector. In the following, we discuss some examples.

\begin{table}[!t]
    \centering
    \begin{tabular}{|c|c|c|c|c|}
        \hline
        $\solId$ & $u^{(0)}_4$ & $u^{(0)}_3=0$ & $L_{\rm eff}$ & mult. \\
        \hline
        \rule{0pt}{4ex}    
         1 & \scalebox{0.8}{$ \frac{\Delta ^6-55 \Delta ^4-\left(\sqrt{\left(\Delta ^2-49\right)^2 \left(\Delta ^4-2 \Delta
   ^2+17\right)}-307\right) \Delta ^2+5 \sqrt{\left(\Delta ^2-49\right)^2 \left(\Delta ^4-2
   \Delta ^2+17\right)}-637}{32 \left(\Delta ^2-49\right) \left(\Delta ^2-4\right)}$}  & No  & 2 & 1\\[1ex]
   \hline
   \rule{0pt}{4ex}
   2 & \scalebox{0.8}{$\frac{\Delta ^6-55 \Delta ^4+\left(\sqrt{\left(\Delta ^2-49\right)^2 \left(\Delta ^4-2 \Delta
   ^2+17\right)}-307\right) \Delta ^2+5 \sqrt{\left(\Delta ^2-49\right)^2 \left(\Delta ^4-2
   \Delta ^2+17\right)}-637}{32 \left(\Delta ^2-49\right) \left(\Delta ^2-4\right)}$}  & No & 2 & 1\\[1ex]
   \hline
   \rule{0pt}{4ex}
   3 & \scalebox{0.8}{$\frac{1}{\sqrt{12}}$}  & Yes & 4 & 4\\[1ex]
   \hline
    \end{tabular}
    \caption{Weak coupling solutions in the massive sector for $L=2$ sub-leading trajectories with the explicit values of $u^{0}_{4}$ with the corresponding $\solId$ identifier, which we use below. We also provide the $L_{\rm eff}$ and the number of solutions in the massless sector in the last column.}
    \label{tab:L2MassiveWeak}
\end{table}

For simplicity, let's restrict to $M=1$. In that case, we can write $u_3\equiv u_{3,1},\,u_{4}\equiv u_{4,1}$. Trajectories are labelled by $\{u_4,u_3,z_1,z_2\}$, which are solutions of \eqref{eq:MomentumSubWeak},\eqref{eq:AuxSubWeak} and \eqref{eq:MasslessSubWeak}. To streamline notation, we will introduce the following state-identification notation
\begin{equation}\la{stateiddef}
    \mathtt{State Id}: \{m_1,m_2,\solId\}_{\Leff}\,,
\end{equation}
where $\solId$ refers to particular solutions of the massive root equations.

Solving the weak-coupling equations for the massive roots \eqref{eq:MomentumSubWeak} and \eqref{eq:AuxSubWeak}, we found $3$ solutions for $L=2$: two of them with $u_3$ scaling as $1$ at $g\to 0$
and one solution with $u_3\sim g^2$. We list these solutions 
in the Table~\ref{tab:L2MassiveWeak}.

For $L=4$ we found $5$ solutions: $3$ with $u_3\sim 1$ and $2$ with $u_3\sim g^2$. The expressions with the exact dependence on $\Delta$ are quite bulky, so we presented the values of $u_4^{(0)}$ 
at a particular value $\Delta=1/2$
 in Table~\ref{tab:L4MassiveWeak}.

\begin{table}[!t]
    \centering
    \begin{tabular}{|c|c|c|c|c|}
        \hline
        $\solId$ & $u^{(0)}_4$ & $u^{(0)}_3=0$ & $L_{\rm eff}$ & mult.\\
        \hline
         1 & $\simeq 0.630037$  & No & 4 & 4\\
         2 &  $\simeq 0.331372$ & No& 4 & 4\\
         3 &  $\simeq 0.150865$ & No& 4 & 4\\
         \hline
         4 & $\simeq  0.688191 $ & Yes& 6 & 9\\
         5 & $ \simeq  0.16246 $ & Yes& 6 & 9\\  
         \hline
    \end{tabular}
    \caption{We list all solutions of \eqref{eq:MomentumSubWeak} and \eqref{eq:AuxSubWeak}  for the massive sector for $L=4$. For simplicity we pick a particular value $\Delta=\frac{1}{2}$ but general expressions are available in a form of a solution of a polynomial equation with the coefficients dependent on $\Delta$. We also provide the $L_{\rm eff}$ and the number of solutions in the massless sector in the last column.}
    \label{tab:L4MassiveWeak}
\end{table}

In conclusion, let us count states for HT$_2$, keeping in mind the restrictions \eqref{eq:QuantumNumbersHT2} and \eqref{eq:ParityInvarianceHT2}. To enumerate the solution, it is enough to consider the weak coupling limit where massive and massless equations decouple, the effect of the massless solutions is to make each massive solution  $\lfloor \Leff^2/2\rfloor$ degenerate.

Let us summarise what happens for $L=2,4$. For $L=2$, there are $2\times 1 + 1\times 4 = 6$ trajectories in total, and for $L=4$, there are $3\times 4 + 2\times 9 = 30$.
The massive sector only contributes to $\gamma_S$ at order $g^2$, it is the massless modes that give the leading contribution at order $g$. Consequently, for $L=4$ we have $4+9$ spectral lines at leading order. To next order, $g^2$, we observe a fine-structure splitting; each line in the first group splits into $3$, and each line in the second group into $2$. A very similar logic applies to $L=2$ where we find $1+4$ spectral lines at the LO and the first line splits into $2$ at NLO. In practice, some massless states have $g\to -g$ symmetry (for these states the set of $z_i$'s is invariant under the complex conjugation), which means that they collapse to zero at the leading order due to this additional symmetry. For instance, at $L=2$ there are $1+2$ states which have this additional symmetry, reducing the LO to $3$ lines with degeneracies $1+3+1$.

\subsection{The Full Q-System and Finite Coupling.}\label{subsec:StartingUp}

Upon solving the ABBA, we have in principle acquired enough information to reconstruct the entire Q-system in addition to those particular Q-functions introduced in Section~\ref{sec:ABBADerivation}. In the previous section we solved the ABBA to the leading order, but higher order corrections can be obtained too as ABBA remains exact all the way till the wrapping order.  The only complication is the finite difference Baxter equation which we solve numerically to obtain the higher order expansion from ABBA.

For practical purposes, the most important functions  to reconstruct perturbatively are $\bP_{a}$. In particular, they are key for the numerical algorithm of \cite{Gromov:2015wca}. To be precise, to solve the QSC numerically one starts from the ansatz
\begin{equation}\label{eq:PAnsatzNumerics}
    \bP_{a} = \sum_{n=-{\tilde{M}}_{a}}^{\infty} \frac{\hat{c}_{a,n}}{x^{n}}\,,
\end{equation}
where $\tilde{M}_a$ are given in \eqref{MMt}. The primary objective is thus to deduce $\hat{c}_{a,n}$ from ABBA. 

We will give the full details of how to accomplish this task in Appendix~\ref{app:StartingUpNumerics}. Let us here recap the main idea. First, we notice that the derivation of $\bP_1$ and $\mu_{12}$ from Section~\ref{sec:ABBADerivation} can be repeated without any change in general for the whole range of indexes, resulting in
\begin{equation}\label{eq:StartingUpNumericsAnsatz}
    \bP_a \propto x^{-L/2-1}\,\bp_{a}\, \sigma\,,
    \qquad\quad
    \mu^+_{ab} \propto \betheQ_{ab} f^+\bar{f}^- \prod_{n=1}^{\infty} \(\frac{\kappa}{x^{N/2}} \)^{[2n]}\(\frac{\bar{\kappa}}{x^{N/2}}\)^{[-2n]}\,,
\end{equation}
where $\bp_a$ are finite Laurent polynomials in $x$ and $\betheQ_{ab}$ polynomials in $u$. Comparing with \eqref{mu12P0mt} and \eqref{eq:PABBA} we find in particular that $\betheQ_{12} \propto \betheQ_4$ and $\bp_1 \propto R_1 B_3$. In general, one should also take a similar ansatz for $\bP^{a}$, but for the LR-symmetric states we consider in this section, we have $\bP^{a} \propto \bP_{5-a}$.  Importantly, we can parametrise $\bP_a$ and $\betheQ_{ab}$ with a finite number of coefficients. The only tricky point is to ensure that these coefficients scale properly with $g$, we work out the details in Appendix~\ref{app:StartingUpNumerics}. 

Since the $\bP\mu$-system \eqref{eq:PMu0} only involves $\bP$ and $\mu$, we can now plug our ansatz and solve the equations order by order in $g$. Interestingly, this procedure fixes the massive roots uniquely. We can use the higher-order data from ABBA in order to find corrections.

\begin{figure}[!t]
    \centering
    \includegraphics[width=\linewidth]{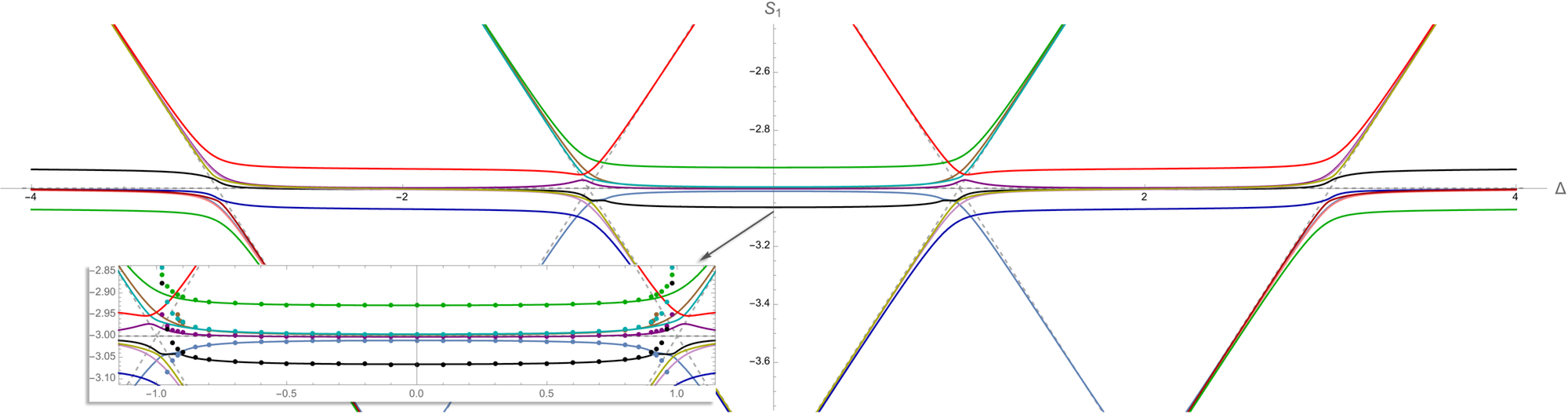}
    \caption{The $6$ states for HT$_2$ for $L=2$ plotted at $g=1/50$ and with $\Im \Delta = \frac{1}{100}$ to ease convergence of the numerics. Notice the interplay between diagonal and anti-diagonal trajectory and how the pattern depends on the particular branching area under consideration. 
    We also compare the QSC data with the ABBA perturbative prediction (denoted with dots). The ABBA results is very precise away from the Branching points. However, near the Branching Areas develop singularities, which can be resumed exactly as we explain in part II of this paper series, and then correctly describe the transitions of the trajectories. Each trajectory is constituted by a number between 400 and 1200 data points.}
    \label{fig:L2HT2}
\end{figure}

Once the $\bP\mu$-system has been solved to wrapping order, at which stage the ansatz \eqref{eq:StartingUpNumericsAnsatz} will fail, one proceeds to expand the dressing phases at weak coupling and compare with \eqref{eq:PAnsatzNumerics}. This gives an initial starting point for the numerics, which will be more and more reliable for higher $L$ since the wrapping order increases as the quantum numbers grow.

Applying this algorithm, we found approximate starting points that allowed us to start up numerics for all states for $L=2$ and $L=4$\footnote{We reserve a systematic scan of all quantum numbers, also involving $L=3$  to~\cite{UpcommingArpad}. }. We discuss some of the aspects of finite $g$ numerics in Section~\ref{sec:Intercepts} and finish this section by explicitly showing the resulting trajectories for $L=2$ and $L=4$ at $g=\frac{1}{50}$ in the range $\Delta\in (-4,4)$. 

\paragraph{$L=2$ trajectories.} As we found above, for $L=2$ there are a total of $6$ trajectories, see Table~\ref{tab:L2MassiveWeak} for a summary of these states. We showcase the interplay between these trajectories in Figure~\ref{fig:L2HT2}. 
Trajectories exhibit an intricate pattern  entering and leaving the HT$_2$ at the $4$ different branching areas featured. 
We  also zoomed in the central segment of the HT$_2$ showing the great agreement of the solutions of the ABBA equations (points) with the numerical data. Notice that at the boundary of the segment, namely in the vicinity of the branching areas, ABBA for HT breaks down and a different regime has to be considered. 
A in-depth study of these regions is one of the main objectives of part~II~\cite{EkhammarGromovPreti:2025b}.

\begin{figure}[!t]
    \centering
    \includegraphics[width=\linewidth]{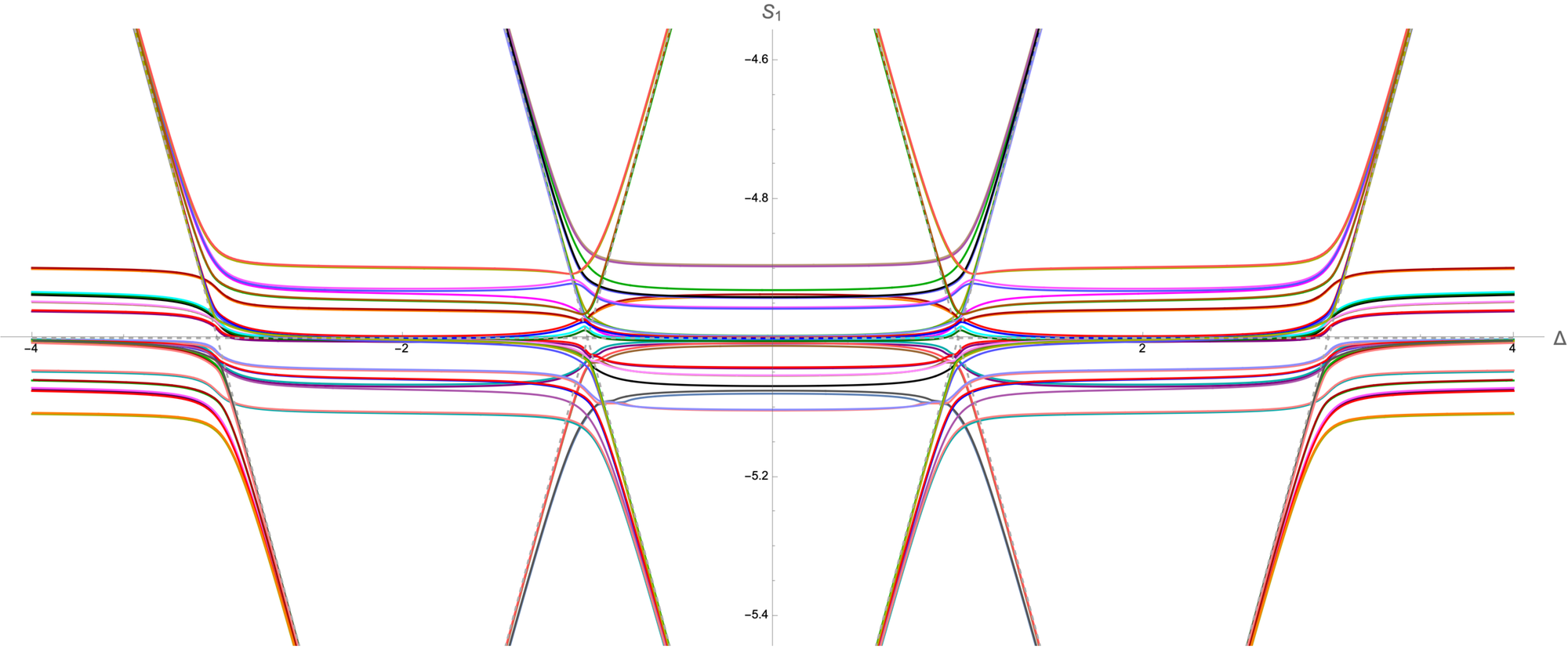}
    \caption{All $30$ states for HT$_2$ for $L=4$ plotted at $g=1/50$ with  $\Im \Delta = \frac{1}{100}$. Notice the interplay between diagonal and anti-diagonal trajectory and how the pattern depends on the particular branching area under consideration. Each trajectory is constituted by a number between 1000 and 5000 data points.}
    \label{fig:L4HT2}
\end{figure}

\paragraph{$L=4$ trajectories.} For $L=4$ there are a total of $30$ trajectories, see Table~\ref{tab:L4MassiveWeak}. We plot all these trajectories at $g=1/50$ and for $\Delta \in (-4,4)$ in Figure~\ref{fig:L4HT2}. 
We emphasise that finding all these trajectories starting from local operators would be a highly time-consuming task. Instead, with ABBA we can start in the BFKL regime, enabling us to find all trajectories in a reasonable timespan. 

\section{Intercepts}\label{sec:Intercepts}

A key quantity controlling the Regge behaviour is the so-called \textit{intercept}, defined as $S(\Delta)\big|_{S'(\Delta)=0}$. For a symmetric in $\Delta\to -\Delta$ trajectory the intercept is $S_1(\Delta=0)$, this is the quantity we consider in this section and will refer to as the intercept \footnote{We found a small number of cases where the trajectory is not an even function of $\Delta$; they may have an unusual position of the saddle point, i.e., where $S_1'(\Delta)=0$, but they are quite low in the spectrum of $S_1$. Therefore, in this section, we limit ourselves to the $\Delta=0$ case. In addition, for the even in $\Delta$ trajectories, there could be additional solutions of $S_1'(\Delta)=0$, different from $\Delta=0$. It would be interesting to study the critical points of the Regge-Riemann surfaces more systematically. }. In ${\cal N}=4$ SYM theory, the intercept controls the Regge limit of four-point functions \cite{Costa:2012cb} and is also linked to the high energy limit of scattering amplitudes in QCD through the maximal transcendentality principle~\cite{Kotikov:2010nd}. We briefly review known results for the intercept of the simplest trajectories before presenting new, more systematic results enabled by ABBA in the following subsection.

The most studied example is the Pomeron intercept, in our notation the $L=2$ minimal sector trajectory evaluated at $\Delta=0$. At weak coupling, the $\mathcal{N}=4$ Pomeron intercept was computed using the QSC up to three-loop order in \cite{Gromov:2015vua}, later confirmed in \cite{Velizhanin:2015xsa,Caron-Huot:2016tzz}, and extended to the case of $S_2 \neq 0$ in \cite{Caron-Huot:2016tzz,Alfimov:2018cms}. More recently, it was studied in \cite{Klabbers:2023zdz} for $L=3$, and in \cite{Ekhammar:2024neh} using ABBA, where the intercepts of trajectories with $L > 2$ for HT$_0$ were analysed both at weak coupling and non-perturbatively using numerical methods.

At strong coupling, the intercept for the leading trajectories containing $\mathfrak{sl}(2)$ sector local operators was first understood by leveraging the AdS/CFT correspondence \cite{Brower:2006ea,Cornalba:2007fs,Costa:2012cb}, enabling the use of string theory instead of the language of $\mathcal{N}=4$ SYM. By matching the intercept against the linear intercept of string theory in flat space and requiring that the trajectory passes through a BPS operator, it was found that
\begin{equation}\label{eq:InterceptStrongCoupling}
S_1(0) \simeq -\frac{L^2}{2\sqrt{\lambda}} + \mathcal{O}\(\frac{1}{\lambda}\)\,.
\end{equation}
These results were subsequently improved upon in \cite{Basso:2011rs,
Costa:2012cb,Kotikov:2013xu,Brower:2014wha,Gromov:2014bva}, culminating with \cite{Ekhammar:2024rfj}, where the first  $7$ non-trivial orders in $1/\sqrt\lambda$ were found analytically.

Despite the longstanding interest in the intercept as outlined above, our understanding of its general structure remains minimal, as all studies so far have focused only on the simplest cases. To build a truly complete picture, it is pertinent to systematically investigate all trajectories. In this section, we show that the ABBA is the right tool for this task by
going well beyond the leading Pomeron trajectory, both at weak, intermediate and strong coupling. 
We postpone a more systematic study to~\cite{UpcommingArpad}, and here
focus on $L=2,4$ case of sub-leading trajectories i.e. HT$_2$. We have already studied the ABBA for these trjaectories in Section~\ref{sec:SubleadingHT}, here we focus on the $\Delta=0$ case, showing some hints of simplifications at strong coupling and also some curious non-trivial behaviour of the intercepts at finite coupling. We also consider both parity even and odd intercepts for $L=3$ that, together with the data obtained in \cite{Klabbers:2023zdz}, complete the study of HT$_0$ in that case.

We start at weak coupling in Section~\ref{subsec:WeakCouplingIntercepts}, where we have the ABBA equations at our disposal. Thereafter, we will turn to finite coupling effects in Section~\ref{subsec:FiniteCouplingIntercepts}, relying on numerical QSC methods. Finally, we examine strong coupling in Section~\ref{subsec:StrongCouplingIntercepts} and find an emergent stringy spectrum.

\subsection{Weak Coupling Intercepts}\label{subsec:WeakCouplingIntercepts}
As we mentioned above, for demonstration purposes, we limit ourselves to the parity symmetric $\mathfrak{sl}(2)$ sector for HT${}_0$ and HT${}_2$ with R-charge $L=2$ and $L=4$ (where for $L=2$, HT${}_0$ is the Pomeron trajectory).

At weak coupling, the computation of the intercept up to wrapping order can be done using ABBA, as described in Section~\ref{sec:SubleadingHT}. We have collected all data for intercepts to $\mathcal{O}(g^3)$ for $L=2$ in Table~\ref{tab:SolutionsHT2L2}. For the first orders we have replaced the numerical estimate with exact numbers, which we verified to $10^{-20}$. We also provide the data for $L=3,4$ in the notebook found in the ancillary file of the arXiv submission.

As we learned in Section~\ref{sec:SubleadingHT}, for the
subleading trajectories HT${}_2$, the massless roots at the leading order are governed by the universal equation
\eq{eq:MasslessSubWeak}, which is the same as for the
purely massless sector i.e. HT${}_0$, with $L\to L_{\rm eff}$. This gives us an analytic expression for the intercept to order $g$ via \eq{eq:GammaSymmetricSector} with $L\to L_{\rm eff}$. Furthermore, we noticed that the minimal sector expression
\eq{eq:GammaSymmetricSector} correctly reproduces all $\log 2$ terms. This is because we know from Section~\ref{sec:SubleadingHT} that the massive and massless sectors decouple as $g\to 0$, and the leading order expressions for the massive roots do not contain $\log$'s or more complicated functions. The analytic expressions in Table~\ref{tab:SolutionsHT2L2} are deduced from the above argument and checked against ABBA numerics with $20$ digit precision.

\begin{table}[!t]
    \centering
    \begin{tabular}{|c|c|c|}
        \hline
        $\mathtt{StateId}$ & $S_1(0)+3$
        \\
        \hline
         $\{1,3,1\}_{2}$ & $g^2(8  \log 4-4 \left(5+\sqrt{17}\right) )+{\cal O}(g^4)$
         \\
         \hline
         $\{1,3,2\}_{2}$  & $g^2(8  \log 4-4 \left(5-\sqrt{17}\right) )+{\cal O}(g^4)$
         \\
         \hline
         $\{1,3,3\}_{4}$ & $\begin{aligned}
             &+2 \sqrt{3} g+\left(-2+\frac{20 \log 4}{3}\right) g^2 - 86.06097968 \, g^3 +{\cal O}(g^4)\end{aligned}$ 
             \\
         \hline
         $\{1,5,3\}_{4}$ & $g^2 \left(-8+\frac{8 \log 4}{3}\right)+{\cal O}(g^4)$
         \\  
          \hline
         $\{2,4,3\}_{4} $ & $8 g^2 \log 4+{\cal O}(g^4)$
            \\ 
          \hline
        $\{3,5,3\}_{4}$  & $\begin{aligned}
                          &-2 \sqrt{3} g+\left(-2+\frac{20 \log 4}{3}\right) g^2 + 86.06097968 \, g^3 +{\cal O}(g^4)\end{aligned}$ 
           \\
         \hline
    \end{tabular}
    \caption{The intercepts for HT$_2$ for $L=2$. The \texttt{StateId} is defined in \eq{stateiddef}.}
    \label{tab:SolutionsHT2L2}
\end{table}

\paragraph{Wrapping corrections.}

While the coefficients appearing in the examples above are relatively simple, we expect that more intricate combinations will emerge at higher orders. This is due to the necessity of solving the Baxter finite-difference equation at those orders, which can generate a wide variety of numbers. In contrast, within the minimal sector, the ABBA construction yields only derivatives of the Pomeron eigenvalue 
$\chi$ and factors of Riemann zetas $\zeta_n$, with increasing transcendentality as the loop order grows. Notably, for $L=2$ (the Pomeron), the $g^4$ term is the first order where wrapping effects are expected to invalidate the ABBA and lead to more complex structures. Indeed, from~\cite{Kotikov:2002ab}, we extract the following
\beqa\la{pomeronNLO}
\nonumber \left.S_1(0)\right|_{g^4}&=&\frac{8}{3}  \left(6 \pi ^3+192 \Im(\text{Li}_3(1-i))+96 C \log 2+12 \pi  \log^22-\pi ^2 \log 4-33 \zeta_3\right)\\
&\simeq& -84.078566807464919912 \,,
\eeqa
which contain the polylogarithm $\text{Li}_3$, and the Catalan constant $C$. 
While such numbers cannot be generated by the ABBA in the massless sector, they do appear in the computation of wrapping corrections in similar integrable quantum field theories, such as fishnet models (see, for instance,\cite{Cavaglia:2020hdb}). Moreover, certain similarities between fishnet theories and the BFKL regime have been noted, e.g., in \cite{Kazakov:2018qbr,Alfimov:2020obh}. It would be interesting to compute integrability-based L\"uscher corrections to reproduce, in particular, Eq.~\eqref{pomeronNLO}.

One could ask more generally at which order in $g$ the ABBA equations fail. Whereas for HT${}_0$ the answer is known to be $g^{L+2}$, from the numerical evidence in the case HT${}_2$ the situation is already more intriguing. We computed the intercept to order $g^{4}$ for $\{1,3,1\}_2$ and $\{2,4,3\}_4$ using both ABBA and QSC. For $\{1,3,1\}_2$ we found that the result from ABBA disagrees with that from QSC, albeit the correction is surprisingly small; $\gamma_{S,\text{ABBA}}/\gamma_{S,\text{QSC}}\big|_{\mathcal{O}(g^4)} \simeq 1.002$. For $\{2,4,3\}_4$ we however found a perfect match at  $\mathcal{O}(g^4)$, the first mismatch instead occurs at $\mathcal{O}(g^6)$, and is once again small. From these experiments, we thus expect that it is $\Leff$, rather than $L$ that sets the wrapping order, and that these effects kick in at $g^{\Leff+2}$. One could speculate that because the wrapping corrections are due to the massless modes, that is natural that the effective length in this sector controls the wrapping. It would be very interesting to understand this phenomenon properly both from an operatorial point of view from QFT side and from integrability by computing the L\"ucher correction.

\begin{figure}[!t]
    \centering   \includegraphics[width=0.49\columnwidth]{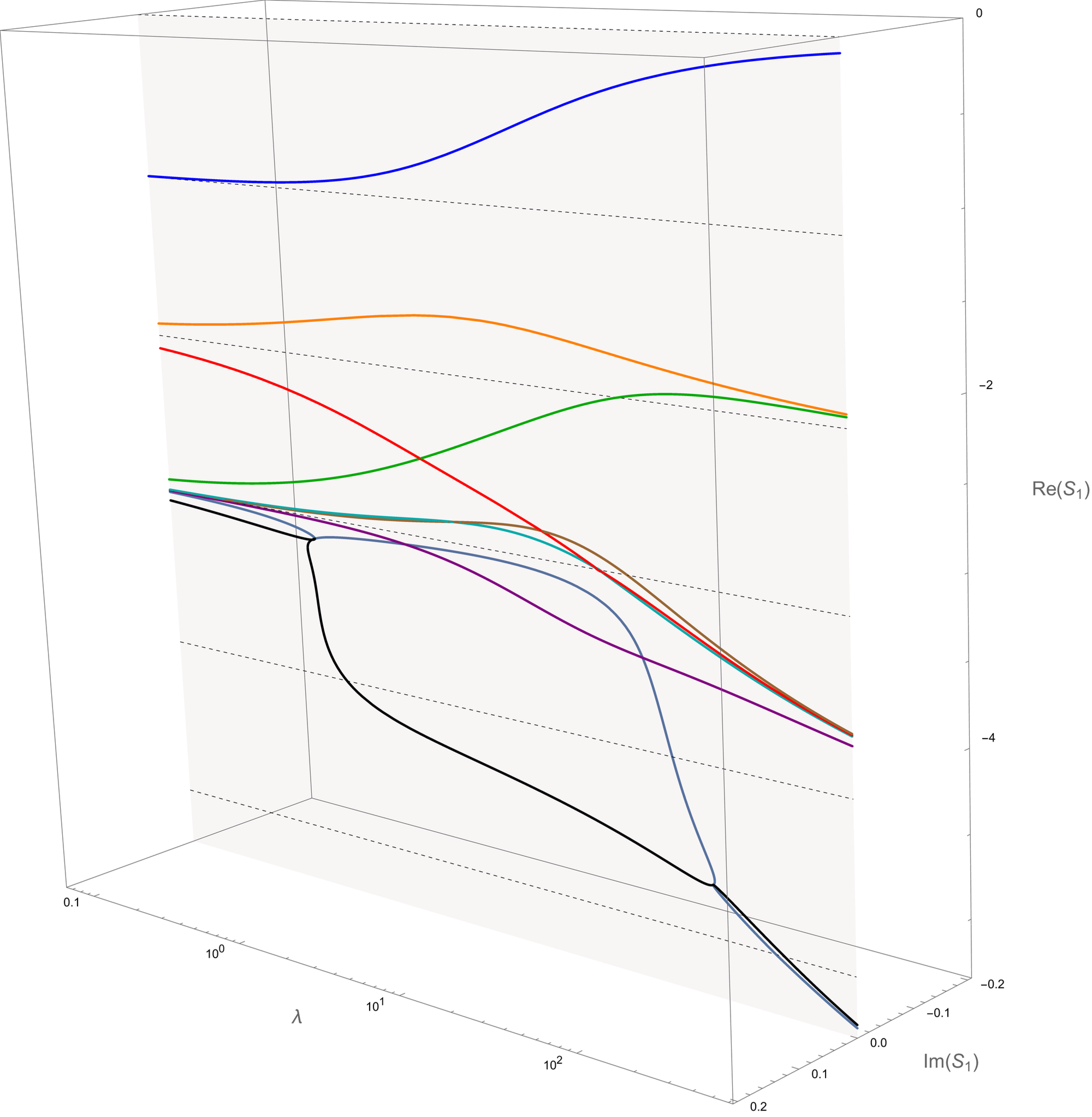}
    \includegraphics[width=0.49\columnwidth]{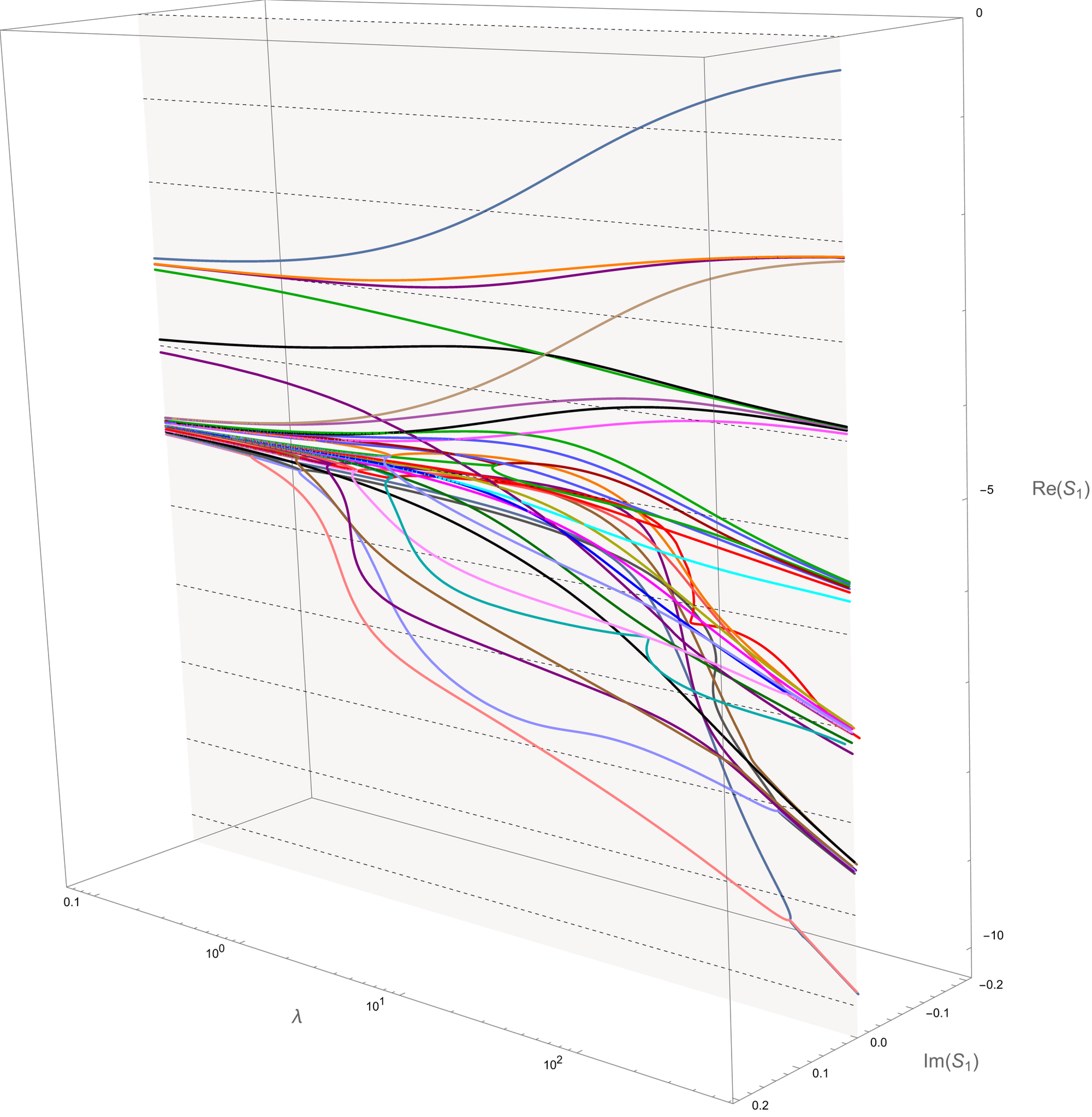}
    \caption{The figure shows the intercepts, defined as $S_1(\Delta=0)$, for the $\mathfrak{sl}(2)$ parity even Regge trajectories with $L=2$ (left) and $L=4$ (right). The curves starting from negative odd integers at $\lambda=0$ correspond to the HTs, while the ones starting at negative even integers correspond to RAs. The spectrum is not always real: trajectories collide and veer into the complex plane, forming conjugate pairs. However, at strong coupling, the imaginary parts become small and the intercepts approach even-integer values (for clarity, the imaginary component of $S_1$ has been magnified). The colour scheme is the one used for HTs in the previous sections.
}
    \label{fig:int3d}
\end{figure}

\subsection{Intercepts at Finite Coupling}\label{subsec:FiniteCouplingIntercepts}
Once we have solved the ABBA equations, we can construct perturbative solutions to the Q-system from which we can initialise numerics, for details see  Section~\ref{subsec:StartingUp}. As an example, we already presented plots of the HTs for $L=2$ and $L=4$ at relatively weak coupling $g=\frac{1}{50}$ in Figure~\ref{fig:L2HT2} and Figure~\ref{fig:L4HT2}. 

\begin{figure}[!t]
    \centering   \includegraphics[width=0.49\columnwidth]{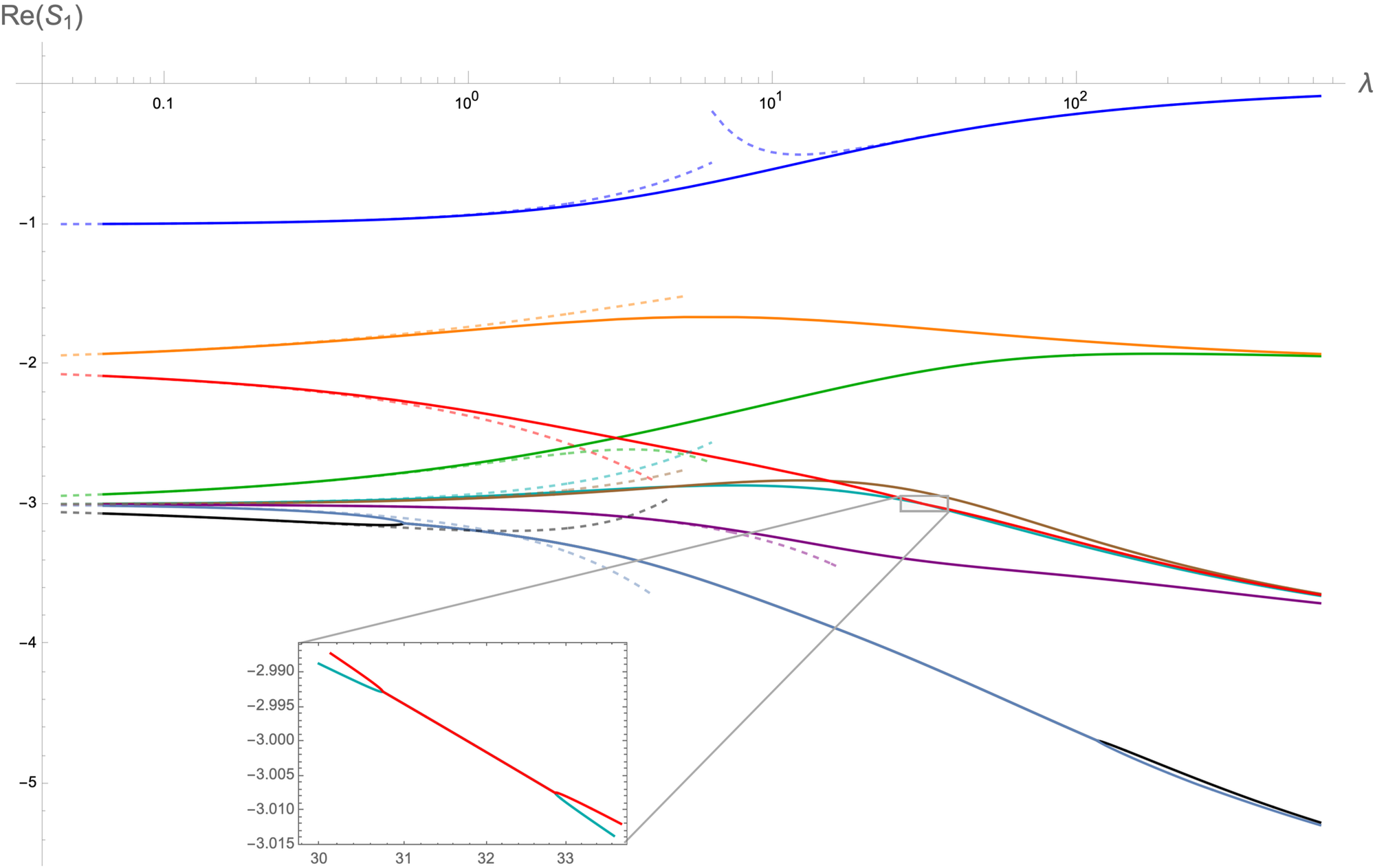}
   \includegraphics[width=0.49\columnwidth]{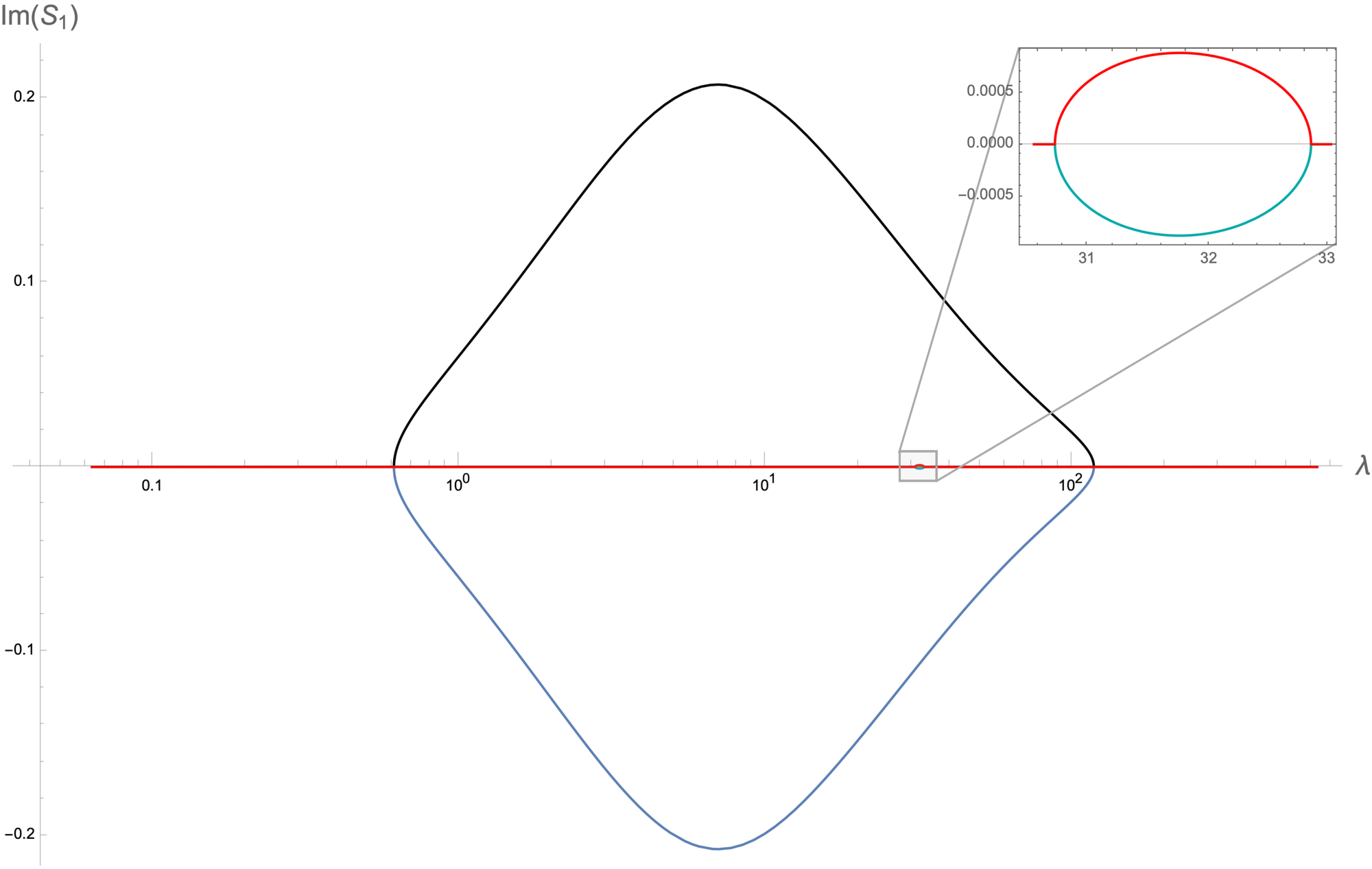} \includegraphics[width=0.49\columnwidth]{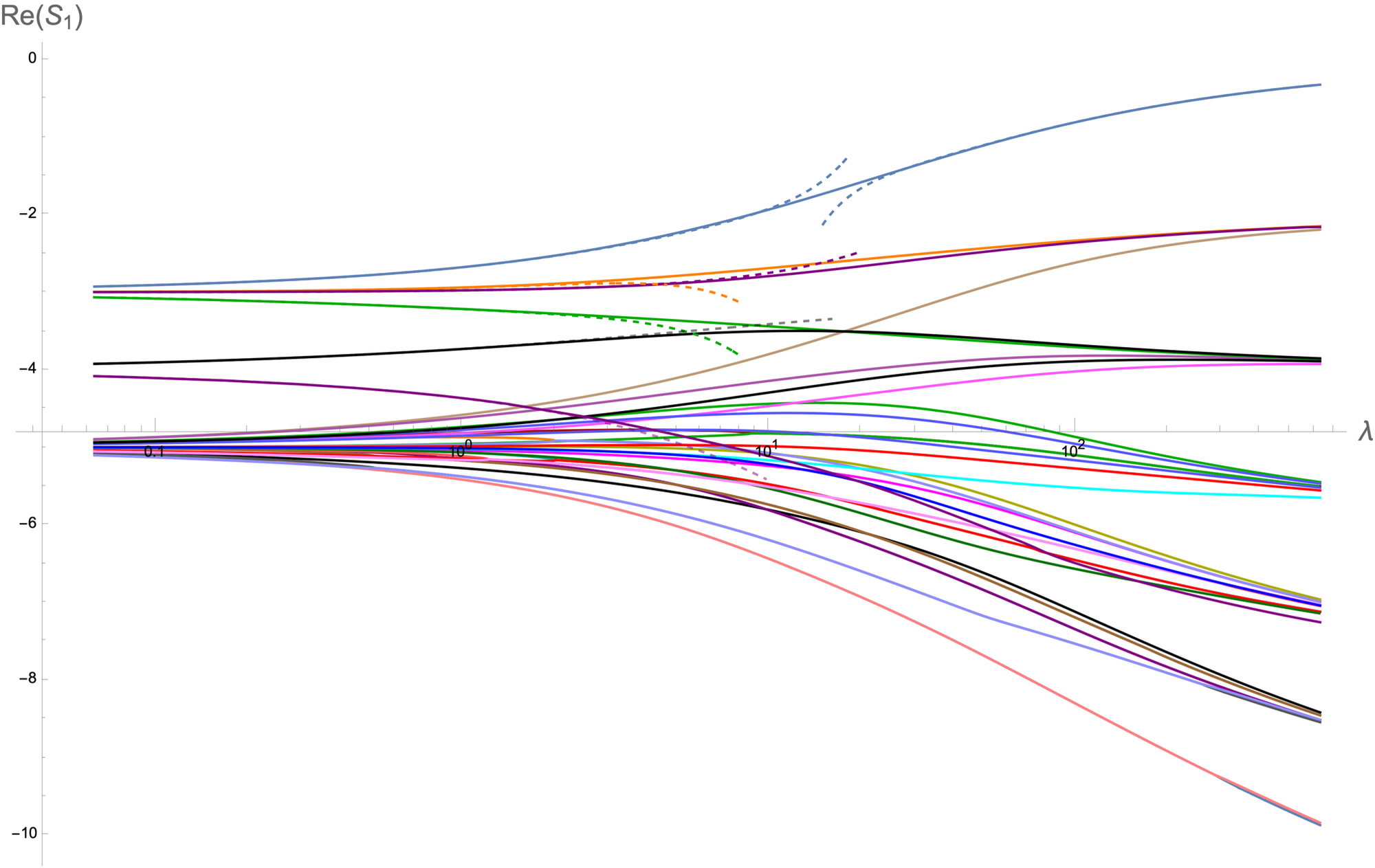}
     \includegraphics[width=0.49\columnwidth]{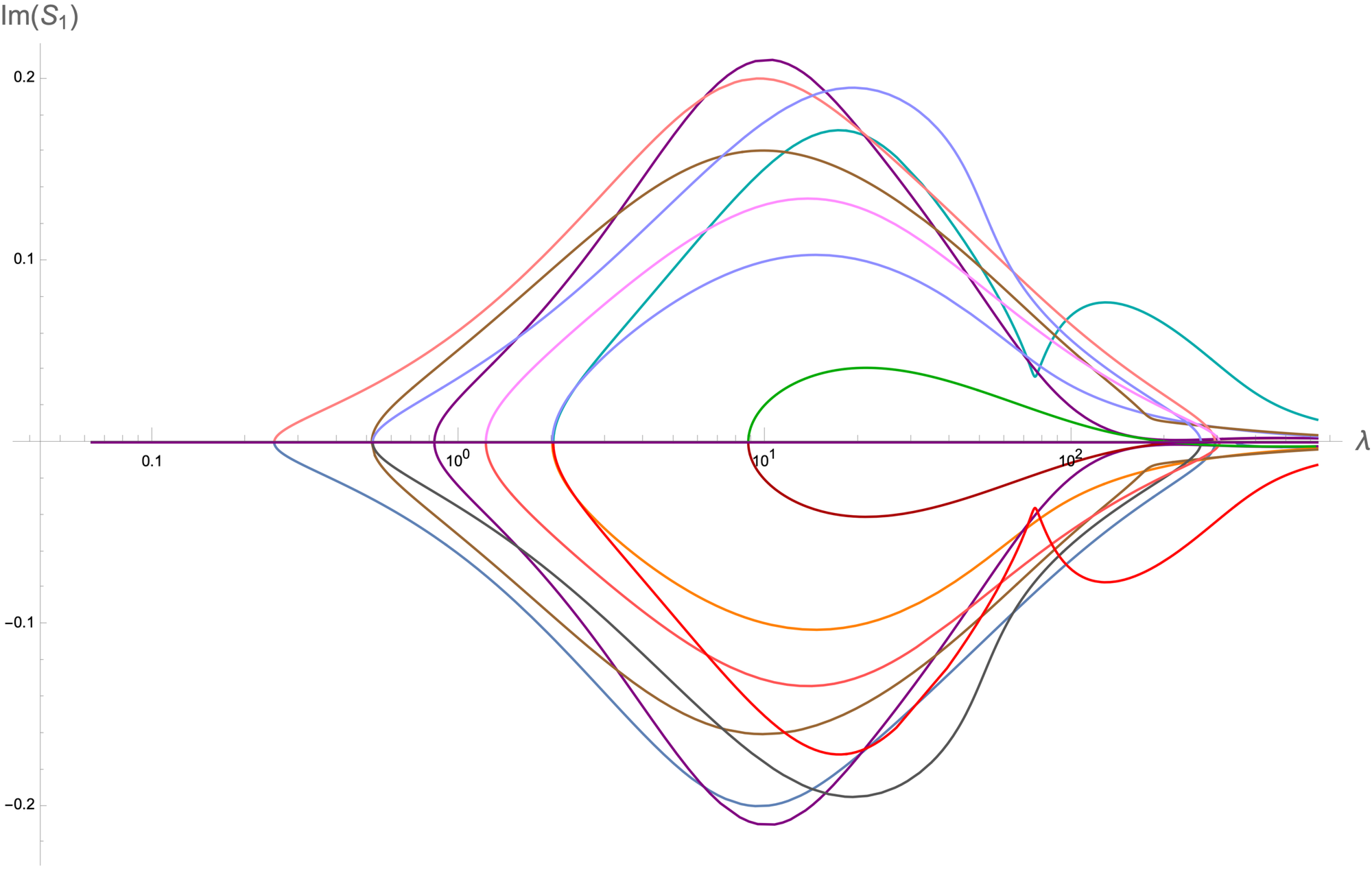}
    \caption{The real (left) and imaginary (right) part of the spectrum of intercepts for $L=2$ (top) $L=4$ (bottom) for $\Delta=10^{-7}+\ii/100$. Dashed lines correspond to the weak coupling ABBA predictions in Table~\ref{tab:SolutionsHT2L2} and to the known strong coupling expansion for the trajectory passing through the BPS point, that is the higher order version of \eqref{eq:InterceptStrongCoupling}. We anticipate the weak here the weak coupling curves for $RA_1^0$ that will be presented in part II \cite{EkhammarGromovPreti:2025b}. In $L=2$ plots, the framed insets represent the magnified collision between the red curve, the bottom contribution of the $\text{RA}_{1}^{0}$, and the cyan curve with \texttt{StateId} ($\{2,4,3\}_4$). We observe several collisions between different HT intercepts at finite values of the coupling. This translates to a non-zero realted imaginary part. However,as $g\rightarrow \infty$ the imaginary part becomes smaller again. The spectrum at strong coupling exhibits a ``stringy'' behaviour in the sense that it organises itself into discrete levels with integer difference, modelling the typical spectrum found from an oscillator-like algebra. Each intercept was computes with numerical QSC and it is constituted by up to 4000 points. The colour scheme follows the one chosen for the HTs in the previous sections.}
    \label{fig:intprojected}
\end{figure}

\begin{figure}[h]
    \centering   \includegraphics[width=0.8\columnwidth]{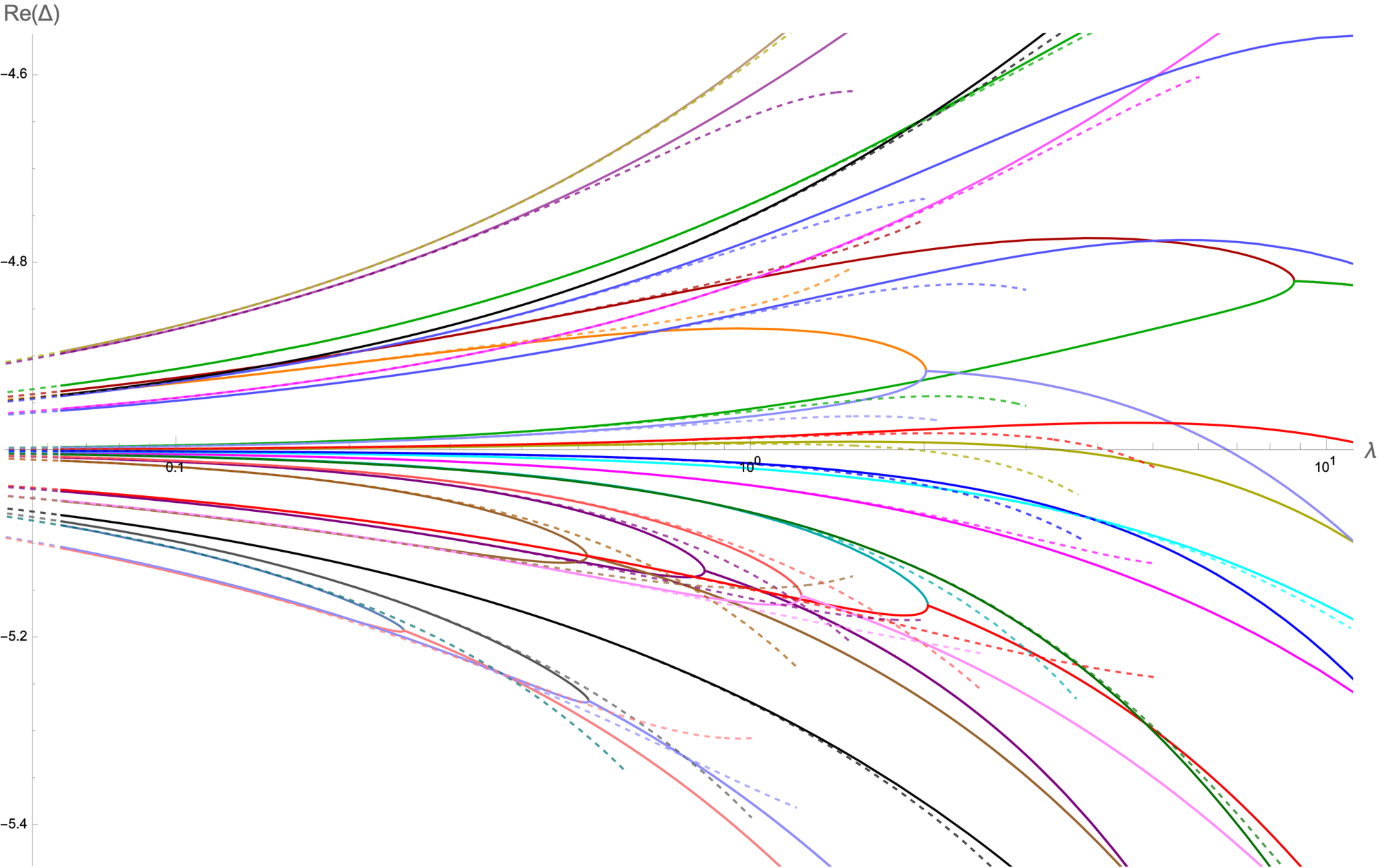}
    \caption{The transition between weak- to mid- coupling for the $L=4$ intercepts related to the HT$_2$ computed for $\Delta=10^{-7}+\ii/100$. Dashed lines represent the ABBA predictions at weak coupling. In this region of the coupling trajectories collide developing an imaginary part. In this plot we can count 8 collision between trajectories. The colour scheme for the states is the same used for the HT$_2$ in Figure~\ref{fig:L4HT2}. }
    \label{fig:zoom}
\end{figure}

In this section we explore what happens when we go out of the weak-coupling regime and the safety of ABBA using non-perturbative QSC. Interestingly, we observe new phenomena! While all data we found from ABBA gives real energies for all states and to all orders, at finite coupling trajectories can collide, whereafter they will develop an imaginary part. 
We provide data that illustrates this phenomenon for $L=2$ and $L=4$ in Figure~\ref{fig:int3d} where the full spectrum of intercept is presented.
We also separate real and imaginary part for clarity in Figure~\ref{fig:intprojected} while we zoom into the transition between weak- and mid-coupling in Figure~\ref{fig:zoom}. Similar collisions appear in the context of fishnet theories \cite{Gurdogan:2015csr}, where spectral line for local operators in certain multiplets veer into the complex plane forming conjugate pairs. See for instance plots in \cite{Gromov:2017cja}, where the non-reality of the spectrum is a consequence of non-unitarity. Analogous phenomena were also observed in the DGLAP regime in \cite{Korchemsky:2003rc} already at one loop. 
However, it was shown that for $\Delta= i\nu$ with $\nu\in {\mathbb R}$ the spectrum is real. It would be interesting to understand why the arguments presented in that paper do not work in the current setup \footnote{Strictly speaking the numerics presented in this section is for $\Delta=10^{-7}+\frac{\ii}{100}$ to make the numerical algorithm converge quicker.}.

In this paper, we have focused on horizontal trajectories, leaving the perturbative description of reflection areas RAs and branching areas BAs using ABBA for part~II~\cite{EkhammarGromovPreti:2025b}, see Section~\ref{sec:structure} for a reminder of our terminology. However, this division of regions on the Chew-Frautschi plot only makes sense in a weak coupling limit. At finite coupling, there is no natural way to distinguish them. Both regions at $\Delta=0$ would produce a saddle point at sufficiently small $g$, controlling the large energy asymptotic expansion. To see this more clearly, let us consider the following finite coupling example. Starting from HT$_2$ for $L=2$ we found that at finite coupling, a trajectory coming from RA$_1$ collides with one from HT$_2$. This enables one, in the vicinity of the collision, to continuously move from one solution to the other on the full Riemann surface. See the top part of Figure~\ref{fig:intprojected} for a focus of this mixing between perturbatively separated trajectories\footnote{While at sufficiently small $g$ all trajectories we consider in this section have $\Delta\to -\Delta$ symmetry, after the collision this may not be the case, and one should follow the $S_1'(\Delta)=0$ curve carefully. As this involves the next level of numerical precision, we postpone this consideration to~\cite{UpcommingArpad}.}.

While collisions and mixing between different regions perhaps can be anticipated, something a bit more surprising is that as we approach large coupling, the spectrum is starting to tend to be purely real again. One can hint at this in Figure~\ref{fig:int3d}, but to see it clearly we refer to Figure~\ref{fig:intprojected}, where we consider the intercept for all states and plot both the imaginary and real parts separately. Furthermore, the real part of the intercept starts grouping itself into a discrete set of levels at strong coupling. This apparent simplification, which is similar to the stringy strong coupling spectrum of the Wilson-Maldacena cusp with insertions \cite{Cavaglia:2021bnz}, invites us to study the strong coupling spectrum. We turn to this task in the next section.

\subsection{Strong Coupling Intercepts}\label{subsec:StrongCouplingIntercepts}
By employing the numerical QSC, we have gathered strong-coupling spectrum data not only for HT${}_0$ and HT${}_2$ but also for RA${}_1^0$.
We present our fits for $L=2$ and  $L=4$ in Table~\ref{tab:StrongCouplingFit}. As can be seen from these tables, the result for the intercept is strikingly simple at strong coupling. For $L=2$ and $L=4$ the leading term is always a negative even integer, and the $\mathcal{O}\left(\frac{1}{\sqrt{\lambda}}\right)$ coefficient an integer. Indeed, let $\delta = -S_1(0)\big|_{g\rightarrow \infty}$, then all our data fits the general formula
\begin{equation}\label{eq:StrongCouplingFit}
    S_1(0) = -\delta + \frac{1}{2\sqrt{\lambda}}\(\delta\(\delta+2\)-L^2\)+\mathcal{O}\(\frac{1}{\lambda}\)\,,
\end{equation}
which is a remarkably simple extension of the $\mathcal{N}=4$ Pomeron result \eqref{eq:InterceptStrongCoupling}. While \eq{eq:StrongCouplingFit} is deduced numerically we can be quite confident that the precision is sufficient to promote it to the exact formula.

\begin{table}[!t]
\resizebox{0.33\columnwidth}{!}{\begin{tabular}{|c|c|}
\hline
\multicolumn{2}{|c|}{$L=2$ intercepts at $g\rightarrow \infty$} \\
\hline
\texttt{StateId} & $S_1(0)$ \\
\hline
\{1,3\} & $0 - \frac{2}{\sqrt{\lambda}} - \frac{1}{\lambda}$ \\
RA$_{1}^{0+}$ & $-2 + \frac{2}{\sqrt{\lambda}} - \frac{3}{\lambda}$ \\
RA$_{1}^{0-}$ & $-4 + \frac{10}{\sqrt{\lambda}} - \frac{29}{\lambda}$ \\
\{1,3,1\}$_2$ & $-6 + \frac{22}{\sqrt{\lambda}} - \frac{121}{\lambda}$ \\
\{1,3,2\}$_2$ & $-4 + \frac{10}{\sqrt{\lambda}} - \frac{24.6}{\lambda}$ \\
\{1,3,3\}$_4$ & $-2 + \frac{2}{\sqrt{\lambda}} - \frac{12}{\lambda}$ \\
\{2,4,3\}$_4$ & $-4 + \frac{10}{\sqrt{\lambda}} - \frac{35}{\lambda}$ \\
\{1,5,3\}$_4$ & $-4 + \frac{10}{\sqrt{\lambda}} - \frac{78.3}{\lambda}$ \\
\{3,5,3\}$_4$ & $-6 + \frac{22}{\sqrt{\lambda}} - \frac{103.6}{\lambda}$ \\
\hline
\end{tabular}}
\resizebox{0.65\columnwidth}{!}{\begin{tabular}{|l|c|}
\hline
\multicolumn{2}{|c|}{$L=4$ intercepts at $g\rightarrow \infty$} \\
\hline
\texttt{StateIds} & $S_1(0)$ \\
\hline
 \{3,5\} & $0 - \frac{8}{\sqrt{\lambda}}$ \\
\hline
 \{2,4\}, \{1,5\}, \{1,3,4\}$_6$ & $-2 - \frac{4}{\sqrt{\lambda}}$ \\
\hline
 \makecell[l]{\{3,5\}, RA$_1^{0+}$, \{1,3,5\}$_6$, \{1,5,4\}$_6$, \{2,4,4\}$_6$} & $-4 + \frac{4}{\sqrt{\lambda}}$ \\
\hline
 \makecell[l]{\{1,3,2\}$_4$, \{1,3,3\}$_4$, \{3,5,4\}$_6$, \{3,7,4\}$_6$, \{1,5,5\}$_6$,\\ \{1,7,4\}$_6$, \{2,4,5\}$_6$, \{2,6,4\}$_6$} & $-6 + \frac{16}{\sqrt{\lambda}}$ \\
\hline
 \makecell[l]{RA$_1^{0-}$, \{1,3,1\}$_4$, \{2,4,2\}$_4$, \{1,5,2\}$_4$, \{1,5,3\}$_4$,\\ \{2,4,3\}$_4$, \{3,5,5\}$_6$, \{1,7,5\}$_6$, \{4,6,4\}$_6$, \{2,6,5\}$_6$} & $-8 + \frac{32}{\sqrt{\lambda}}$ \\
\hline
 \makecell[l]{\{1,5,1\}$_4$, \{2,4,1\}$_4$, \{3,5,2\}$_4$, \{3,5,3\}$_4$, \{3,7,5\}$_6$,\\ \{4,6,5\}$_6$, \{5,7,4\}$_6$} & $-10 + \frac{52}{\sqrt{\lambda}}$ \\
\hline
 \{3,5,1\}$_4$, \{5,7,5\}$_4$ & $-12 + \frac{76}{\sqrt{\lambda}}$ \\
\hline
\end{tabular}}
\caption{Numerically fitted intercept for all $L=2$ and $L=4$ states starting at HT$_0$, HT$_2$ and RA$_1$ at weak coupling. Superscript $\pm$ for RA refers to the concavity up or down of related trajectory (see Figure~\ref{fig:L2full}). Coefficients appearing in the expansions have been guessed from their numerical value with decreasing precision for higher powers of the coupling. Precision varies depending on the considered trajectory. However, we obtain at least 4 digits of precision for the $1/\sqrt{\lambda}$. For $L=2$ we look at one additional order where we have 2 digits for the whole numbers while for non-integers we just show 1 significant digit. Notice that while the $1/\sqrt{\lambda}$ is the same for all states sharing the leading order, the subleading $1/\lambda$ term differs for all states.}
\label{tab:StrongCouplingFit}
\end{table}

It would be fascinating to derive this formula using string theory in AdS$_5$, the second term lends itself to be interpreted as a Casimir element. A very similar expression can be found for local operators on the $\mathcal{N}=4$ Wilson loop at strong coupling. The Wilson loop at strong coupling is naturally connected to an AdS$_2$ string \cite{Giombi:2017cqn}, and in \cite{Ferrero:2023gnu} the origin of the subleading term was elucidated by constructing a dilation operator giving precisely the Casimir of $\mathfrak{osp}(4|2)$. Perhaps a similar structure could also explain our result.

\begin{figure}[!t]
    \centering   \includegraphics[width=0.485\columnwidth]{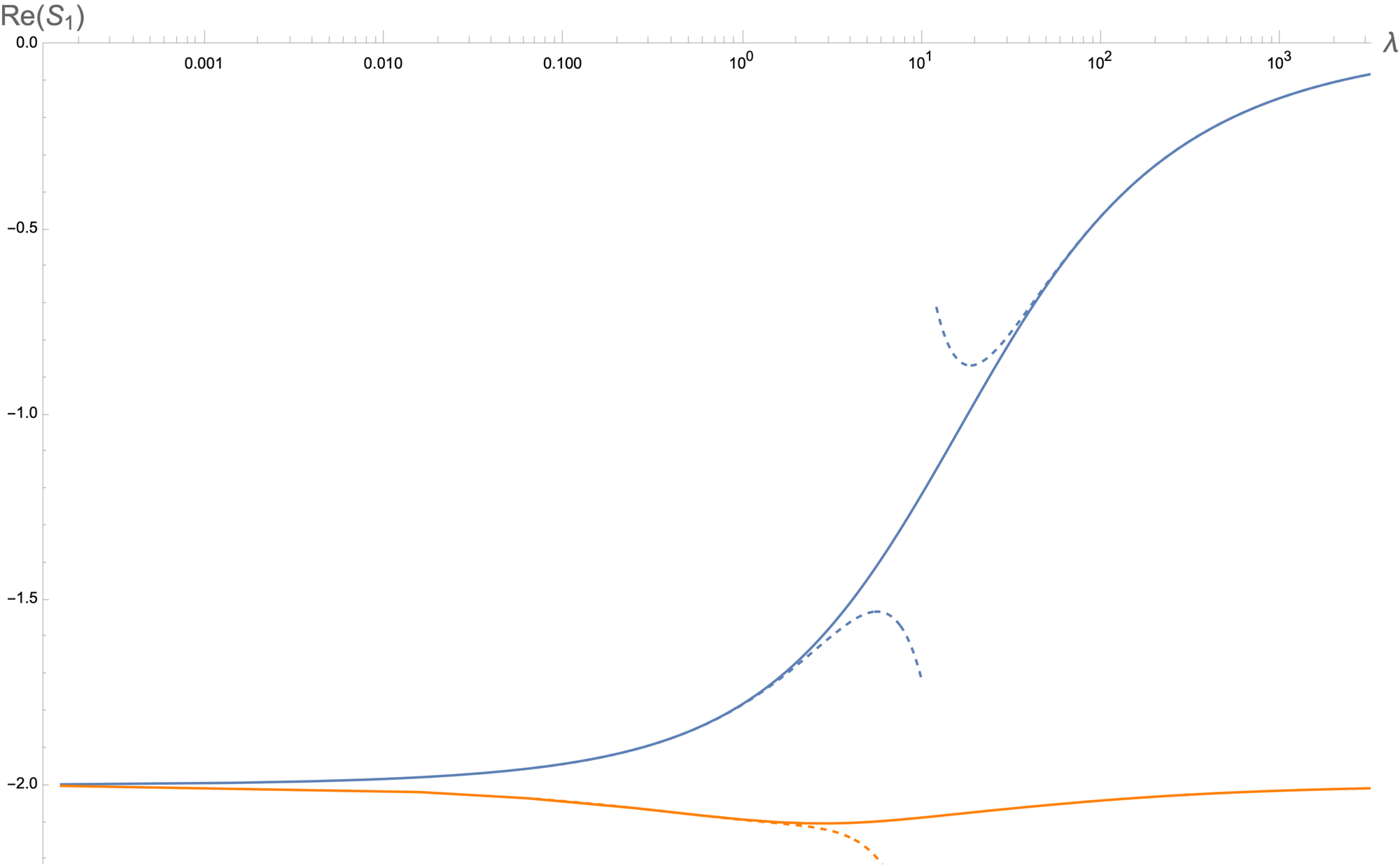}
    \includegraphics[width=0.50\columnwidth]{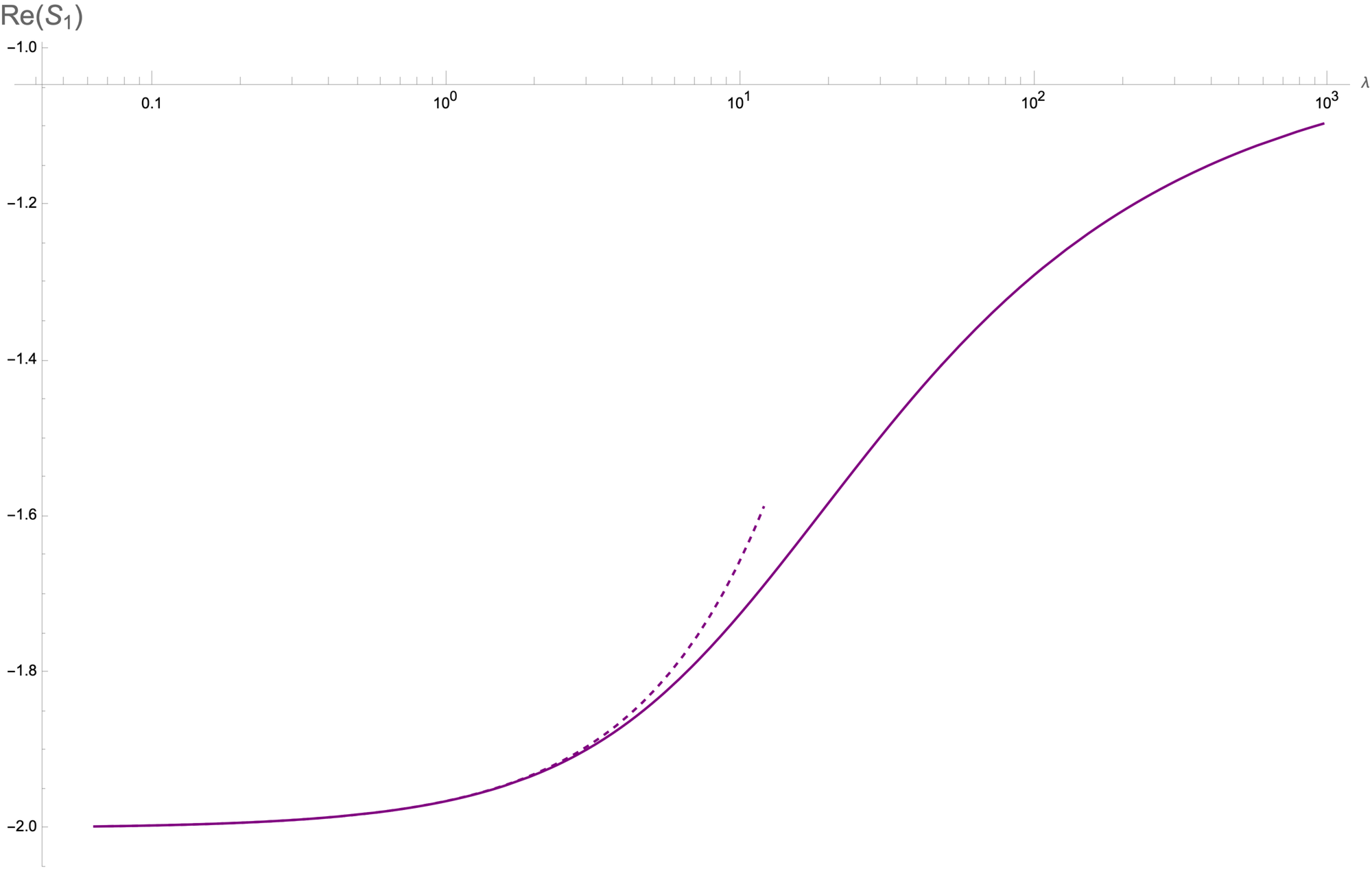}
    \caption{All $L=3$ intercepts for HT$_0$. On the left the 2 parity symmetric ones (blue was computed in \cite{Klabbers:2023zdz}), on the right the 2 coincident non-parity symmetric ones. Dashed lines correspond to the weak coupling ABBA predictions in Table~\ref{tab:tablesolutions} at $\Delta=0$ and $S_2=0$ and to the known strong coupling expansion for the trajectory passing through the BPS point, that is the higher order version of \eqref{eq:InterceptStrongCoupling}. Each intercept was computes
with numerical QSC and it is constituted by around 200 points. The colour scheme follows the one
chosen for the HT$_0$ in Figure \ref{fig:L3HT0}. }
    \label{fig:L3Intercepts}
\end{figure}

Given the fact that we deduced \eqref{eq:StrongCouplingFit} using only parity symmetric states for $L$ even, it is reasonable to worry that some feature might have been missed for more general cases. To probe odd $L$ and non-parity symmetric states, we also compute $L=3$ intercepts for HT$_0$ presented above in Figure~\ref{fig:L3HT0}. The curves are displayed in
Figure~\ref{fig:L3Intercepts} with a comparison with ABBA results and known strong coupling expansion for the leading curve. The latter was computed via numerical QSC in \cite{Klabbers:2023zdz}, while the remaining ones are novel results. Similarly to what we noticed for the $L=2,4$ cases, all curves approaches integer numbers at strong coupling. It is interesting to notice that, in contrast to all the cases analysed so far, the non-parity symmetric intercepts tend to an odd integer! This phenomenon resembles a similar behaviour that parity odd states have on the spectrum of local operator insertions on a Maldacena-Wilson line \cite{Cavaglia:2023mmu,Cavaglia:2024dkk}.
Fitting the orange parity symmetric intercept and the two degenerate purple non-parity symmetric ones appearing in Figure~\ref{fig:L3Intercepts}, we have
\begin{equation}\label{eq:NonParityStrong}
   S_{1}(0) = -2-\frac{1}{2\sqrt{\lambda}}+\frac{3}{4\lambda}+\frac{9}{8 \lambda^{3/2}}+\mathcal{O}\(\frac{1}{\lambda^2}\),\quad\text{and}\quad S_{1}(0) = -1-\frac{3}{\sqrt{\lambda}}+\mathcal{O}\(\frac{1}{\lambda}\),
\end{equation}
where the first refers to the parity symmetric state $\{k_1,k_2,k_3\}=
   \{2,3,4\}$, while the second to the two degenerate states $\{1,3,4\}$ and $\{1,2,3\}$ with $k_1,k_2,k_3$ our weak coupling labels introduced in Section~\ref{subsec:CountingSolutionsExamples}.
   We see that \eqref{eq:NonParityStrong} are both in perfect agreement with our proposal \eqref{eq:StrongCouplingFit}. Coefficients appearing in the expansion \eqref{eq:NonParityStrong} have been guessed from their numerical value with decreasing precision for higher powers of the coupling. In this case the last coefficient has 4 digits of precision. The remaining state automatically satisfy it since behaves at strong coupling as \eqref{eq:InterceptStrongCoupling} with $\delta=0$.

In this section, we have investigated intercepts at weak, intermediate and strong coupling. At weak coupling, we provided examples of how ABBA can provide highly accurate results up to wrapping order. For finite coupling, we showed numerically how trajectories can collide and mix, leading to complex values of the intercept and interplay between trajectories separated at weak coupling. Finally, at strong coupling, we found a stringy-like spectrum and managed to fit a universal formula, generalising the known expressions for trajectories passing through BPS operators. These are merely the first steps towards a more general understanding of intercepts in $\mathcal{N}=4$ SYM. To further our vision, we need to consider trajectories in more exotic sectors, for lower horizontal trajectories and find a coherent structure at strong coupling and also understand the role of other critical points, where $S_2'(\Delta)$ vanishes. We will continue this journey in \cite{UpcommingArpad}, as we clearly demonstrated the efficiency of the ABBA tools.

\section{Conclusions}\label{sec:Conclusions}
${\cal N}=4$ Super Yang-Mills theory is a remarkable theory within which we can explore such intriguing objects 
as Regge trajectories fully non-perturbatively. The integrability of $\mathcal{N}=4$ provides a unique opportunity to understand the interplay between DGLAP (diagonal) and 
BFKL (horizontal) trajectories, both with non-perturbative numerical tools and analytic methods. In this work, we have formulated a rich set of equations, the Asymptotic Baxter-Bethe Ansatz, which describes horizontal trajectories perturbatively, up to the wrapping order. Furthermore, as will be the topic of part~II~\cite{EkhammarGromovPreti:2025b} of this paper series, we will demonstrate that our equations allow for an analytic treatment of more challenging regimes, namely those where horizontal trajectories transition to diagonal ones and where the diagonal trajectory reflects to another diagonal trajectory. 

The structure of the ABBA equations closely resembles the Beisert-Eden-Staudacher equations for local operators, but with key novel ingredients. Firstly, some dressing phases share similarities with those that appear in the study of extended objects, such as Wilson loops with cusps or quark-antiquark potential \cite{Correa:2012hh,Drukker:2012de,Gromov:2016rrp}.
Secondly, and potentially even more intriguing, the ABBA equations contain massless modes, which were, until recently, only known to feature in AdS$_3$ holographic integrable systems. The form of ABBA seems to imply that the BFKL regime can be described by the standard long-range $\mathcal{N}=4$ $\mathfrak{psu}(2,2|4)$ spin chain, supplemented with additional integrable defects. It would be highly interesting to identify these 
defects precisely and derive the dilatation operator directly from QFT perturbation theory.
Especially noteworthy would be reproducing the expansion in integer orders of $g$ (manifesting the presence of massless modes) as opposed to the more standard $g^2$ expansion.

Some of the inspiration deduced from ${\cal N}=4$ should be naturally transferable to theories such as QSC.
Indeed, via the maximal transcendentality principle ${\cal N}=4$ should capture the most complicated (highest transcendentality)
part of the analogous observable in QCD, assuming the principle still applies for the observables we consider. Furthermore, as the BFKL regime is dominated by gluons, it is expected that at the leading order the 
BFKL regime of QCD and ${\cal N}=4$ should be very similar.
So in particular, exploring the possibility of finding trajectories in QCD with fractional powers of the coupling could have non-trivial phenomenological consequences.

Our results set the stage for a systematic exploration of Regge trajectories non-perturbatively, using numerical QSC. The analytic control at weak coupling granted by ABBA allows us to classify trajectories and find starting points for numerics, crucial to make the task of finding hundreds or more of trajectories feasible. We collected some preliminary numerical results, which indicate intriguing behaviour of the trajectories when the coupling is increased. It would be interesting to explore the strong coupling regime using quasi-classical string theory and the AdS/CFT correspondence.

As a concrete application of our equations, we studied intercepts $\Delta=0$ for a simple set of trajectories. We found that the intercepts have a non-trivial behaviour as we change the coupling constant (more sophisticated than that for the spectrum of local operators).
The intercepts display a whole range of behaviours, they can interchange, change their order, can intersect and become complex, but still have a very clean, quasi-classical-like strong coupling behaviour, indicating rich physics yet to be explored. The BFKL-regime at strong coupling was recently explored in \cite{Alday:2024xpq} using the Virasoro-Shapiro amplitude \cite{Alday_2023}, and 3-point functions were considered in \cite{Basso:2025mca} using the hexagon formalism. We hope our results will help push these advances beyond leading trajectories. Indeed, sub-leading trajectories have already proven useful at strong coupling \cite{Julius:2023hre,Julius:2024ewf}.

Finally, it still remains a mystery to find the analogue of the Odderon \cite{Lukaszuk:1973nt} in ${\cal N}=4$ SYM. The Odderon in QCD is of interest since it is responsible for the observed difference between hadron–hadron and hadron–antihadron scattering at high energies, and has only recently been measured experimentally \cite{Csorgo:2019ewn,D0:2020tig}. At the LO Odderon is expected to be a solution of length $3$ non-compact ${\mathfrak{sl}}(2,\mathbb C)$ spin chain~\cite{Lipatov:1993qn,Faddeev:1994zg}. In terms of the ABBA, the number of non-compact sites in the spin-chain is given by half of the number of massless modes, as one can see from the Baxter equation emerging in the auxiliary nodes. With our prescription of analytic continuation from local operators, which is expected to be compatible with light-ray operators at weak coupling, the number of massless modes is at most $4$, making it unlikely that any of the trajectories will manifest a non-compact ${\mathfrak{sl}}(2,\mathbb C)$-style spin chain with $3$ or more non-compact sites.

\section*{Acknowledgements}
We are grateful to participants of the ``Bootstrap Meets Integrability'' workshop in S\~{a}o Paulo 2024. We are also grateful for helpful discussions with   
J., B.~Gabai, R.~Klabbers, S.~Komatsu, 
G.~Korchemsky, P.~Kravchuk, J.~Mann,  V.~Gorbenko,  E.~Sobko, N.~Sokolova, I.~Szécsényi, P.~Vieira and A.~Zhiboedov.

The work of N.G. was
supported by the European Research Council (ERC) under the European Union’s Horizon 2020 research and innovation program – 60 – (grant agreement No. 865075) EXACTC. N.G.’s research is supported in part by the Science Technology \& Facilities Council (STFC) under the grants ST/P000258/1 and ST/X000753/1. The work of S.E. was conducted with funding awarded
by the Swedish Research Council grant VR 2024-00598. The work of M.P. was supported by Marie Skłodowska-Curie Global Fellowship
(HORIZON-MSCA-2022-PF-01) BOOTSTRABILITY-101109934 and by the INFN SFT specific initiative.

\appendix

\section{QSC Generalities}\label{app:QSCGeneralities}

The main tool underlying the results of this paper is the non-perturbative Quantum Spectral Curve (QSC) \cite{Gromov:2013pga,Gromov:2014caa}. Originally developed to study local operators, it was adapted to Regge trajectories in \cite{Alfimov:2014bwa,Gromov:2015wca} and an extensive investigation of the Riemann surface containing twist-2 operators with conformal spin was presented in \cite{Alfimov:2018cms}, building on numerical methods developed in \cite{Gromov:2015wca}.

In this appendix, we recall the structure of the QSC and highlight the main differences between Regge trajectories as compared to local operators. Our presentation will be brief; more pedagogical introductions can be found in \cite{Gromov:2015wca,Kazakov:2018ugh,Levkovich-Maslyuk:2019awk} and a useful summary is given in \cite{Gromov:2023hzc}.

\subsection{The Q-Functions of the QSC}

The basic building blocks of the QSC are Q-functions, functions of a complex parameter $u$ called the spectral parameter. There are $16$ basic Q-functions, they are denoted as $\bP_{a}(u), \bP^{a}(u), \bQ_{i}(u), \bQ^{i}(u), a,i=1,\dots,4$ and satisfy $\bP_a(u)\bP^{a}(u)=\bQ_i(u) \bQ^i(u)=0$. We will often drop the explicit dependence on the spectral parameters. 

The basic Q-functions can be ``rotated'' into each other with the help of additional Q-functions $Q_{a|i}$ in the following way
\begin{equation}\label{eq:QPRotations}
    \bP_{a} = -Q^{\pm}_{a|i}\bQ^{i}\,,
    \quad
    \bQ_{i} = -Q^{\pm}_{a|i} \bP^{a}\,,
    \quad
    \bQ^{i} = \bP_a\left(Q^{a|i}\right)^{\pm}\,,
    \quad
    \bP^{a} = \left(Q^{a|i}\right)^{\pm}\bQ_i\,,
\end{equation}
where we use standard shift notation $f^{\pm} = f(u+\frac{\ii}{2}), f^{\pm \pm} = f(u\pm \ii)$ and more generally $f^{[n]} = f(u\pm \frac{\ii n}{2})$. The Q-functions $Q_{a|i}$ and $Q^{a|i}$ are constrained by
\begin{equation}\label{eq:QaiRelations}
    Q_{a|i}^+ - Q_{a|i}^- = \bP_a \bQ_i\,,
    \quad
    Q_{a|i} Q^{a|j} = -\delta^{j}_i\,,
    \quad
    Q_{a|i}Q^{b|i} = -\delta^{b}_a\,.
\end{equation}

\paragraph{Asymptotics.}
The asymptotics of $\bP$ and $\bQ$ encodes the $\mathfrak{psu}(2,2|4)$ quantum numbers of the trajectory under consideration \cite{Gromov:2014caa}
\begin{equation}\label{eq:AsymptoticsQ}
    \bP_{a} \simeq A_{a}\,u^{-\tilde{M}_{a}}\,,
    \quad
    \bP^{a} \simeq A^{a}\,u^{\tilde{M}_a-1}\,,
    \quad
    \bQ_{i} \simeq B _{i}\,u^{\hat{M}_i-1}\,,
    \quad
    \bQ^{i} \simeq B^{i}\,u^{-\hat{M}_i}\,,
\end{equation}
with 
\begin{align}\la{MMt}
    \tilde{M}_{a} = \frac{1}{2} \{J_1+J_2-J_3+2,J_1-J_2+J_3,-J_1+J_2+J_3+2,-J_1-J_2-J_3\}\,, \\
    \hat{M}_i = \frac{1}{2}\{\Delta-S_1-S_2+2,\Delta+S_1+S_2,-\Delta-S_1+S_2+2,-\Delta+S_1-S_2\}\,.
\end{align}
Upon combining these polynomial asymptotics with \eqref{eq:QPRotations} and \eqref{eq:QaiRelations} we find that the prefactors $A_{a},B_i$ must satisfy
\begin{equation}
    A_{a}A^{a} = \ii \frac{\prod_{i=1}^{4}(\tilde{M}_a-\hat{M}_i)}{\prod^{4}_{b\neq a}\left(\tilde{M}_a-\tilde{M}_b \right)}\,,
    \quad
    B_{i}B^{i} = -\ii \frac{\prod_{a=1}^{4}(\hat{M}_i-\tilde{M}_a)}{\prod^{4}_{j\neq i}\left(\hat{M}_i-\hat{M}_j \right)}\,,
\end{equation}
with no sums over $a$ and $i$.

The asymptotics of the remaining Q-functions can be deduced from QQ-relations, see \eqref{eq:QQRelations}. We list the most important examples for this paper in Table~\ref{tab:Qasymptotics}.
\begin{table}[!t]
    \centering
    \begin{tabular}{|c|c|c|}
        \hline
        Q-function & Asymptotics & ${\mathfrak sl}(2)\, (J_1=L, J_2=J_3=S_2=0)$\\
        \hline
        $\bP_{1}$ & $\frac{-J_1-J_2+J_3-2}{2}$& $-\frac{L+2}{2}$ \\
        $Q_{1|1}$ & $\frac{\Delta-J_1-J_2+J_3-S_1-S_2}{2}$& $\frac{\Delta-L-S}{2}$\\
        $Q_{1|13}$ & $\frac{-J_1-J_2+J_3-2S_1}{2}$ & $-\frac{L+2S}{2}$ \\
        $Q_{12|13}\propto Q^{34|24}$ & $1-J_1-S_1$& $1-L-S$ \\ 
        $Q_{123|13} \propto Q^{4|24}$ & $\frac{-J_1-J_2-J_3-2S_1}{2}$& $-\frac{L+2S}{2}$ \\
        $Q_{123|123} \propto Q^{4|4}$ & $\frac{\Delta-J_1-J_2-J_3-S_1+S_2}{2}$& $\frac{\Delta-L-S}{2}$  \\
        $Q_{123|1234}\propto Q^{4|\es}$ & $\frac{-J_1-J_2-J_3-2}{2}$& $-\frac{L+2}{2}$
        \\
        \hline
    \end{tabular}
    \caption{Asymptotics for a selection of Q-functions.
    }
    \label{tab:Qasymptotics}
\end{table}

\paragraph{$\Lambda$-symmetry.}
To be precise, $\hat{M}_{a}$ and $\hat{M}_i$ are only defined up to certain shifts of relative integers. This follows from the fact that the QSC is invariant under the following transformation
\beq\la{LambdaSym}
\bP_a\to x^{+\Lambda} \bP_a\;\;,\;\;
\bP^a\to x^{-\Lambda} \bP^a\;\;,\;\;
\bQ^i\to x^{+\Lambda} \bQ^i\;\;,\;\;
\bQ_i\to x^{-\Lambda} \bQ_i\;\;,\;\;Q_{a|i}\to Q_{a|i}\;,
\eeq
see \cite{Gromov:2014caa} for more details.

\paragraph{Analytic Properties.}
To find the spectrum of $\mathcal{N}=4$ the algebraic relations described above must be supplemented with analytic constraints on the Q-functions. We require that $\bP_{a}$ and $\bP^{a}$ are analytic functions outside a cut $(-2g,2g)$ on the real axis. Compatibility with the Q-system then requires $Q_{a|i}$ and $\bQ_i$ to have a ladder of cuts, at maximum we can pick them to be either upper half-plane analytic (UHPA) functions or lower half-plane analytic (LHPA) functions. We will write $Q^{\downarrow}$ for UHPA Q-functions and $Q^{\uparrow}$ for LHPA functions, the arrow indicate the directions of the cuts. We summarize the analytic structure of $\bP_{a},\bQ^{\downarrow/\uparrow}_i$ in Figure~\ref{fig:AnalyticStructurePQ}. Since $\bQ_i^{\uparrow/\downarrow}$ all satisfy the same Bethe equations we have
\begin{equation}\label{eq:OmegaBQ}
    \bQ^{\uparrow}_i = \(\Omega^{\bQ}\)_i{}^j\,\bQ^{\downarrow}_j\,,
    \qquad
    \left(\Omega^{\bQ,+}\right)_i{}^j = -Q^{\uparrow}_{a|i}\,Q^{\downarrow,a|j}\,.
\end{equation}

\subsection{Symmetry Properties}\label{sec:SymmetryPropertiesQSC}

\paragraph{Complex conjugation.}

When all quantum numbers are real, we can demand that the Q-system respect complex conjugation.  Since $\bP$ are functions with one short cut, complex conjugation will map $\bP$ onto itself (up to an overall constant factor). For $\bQ$, complex conjugation is slightly more complicated since it necessarily maps UHPA functions to LHPA functions. We can and will pick conventions such that
\begin{equation}
\begin{split}
    &\bar{\bP}_{a} = \mathbb{c}_{a}{}^{b}\bP_{b}\,,
    \quad
    \bar{\bP}^{a} = -\bP^{b}\mathbb{c}_b{}^{a}
    \quad
    \overline{\bQ^{\downarrow}_i} = \mathbb{c}_{i}{}^{j}\bQ^{\uparrow}_j\,,
    \quad
    \overline{\bQ^{\downarrow,i}} = -\bQ^{\uparrow,j}\mathbb{c}_{j}{}^{i}\,,
    \quad
    \\
    &\mathbb{c}_{i}{}^{j}=\mathbb{c}_{a}{}^{b} =\text{diag}(-1,-1,1,1)\,. 
\end{split}
\end{equation}

Apart from complex conjugation, there are two more symmetries that will play a role in our analysis: LR-symmetry and parity. Contrary to complex conjugation, these two properties only hold true for a subset of states in $\mathcal{N}=4$ SYM. 

\paragraph{Parity symmetry.}
Parity symmetry is similar to complex conjugation in nature as it preserves $\bP$ but maps $\bQ^{\downarrow}$ to $\bQ^{\uparrow}$. It takes the form
\begin{equation}
    \bP_{a}(-u) = (-1)^{-\tilde{M}_{a}}\,\bP_{a}(u)\,,
    \quad
    \bQ^{\downarrow}_{i}(-u) = (-1)^{\hat{M}_i-1}\bQ^{\uparrow}_{i}\,.
\end{equation}

\paragraph{LR-symmetry.} This symmetry implies a relation between $\bP_{a}$ and $\bP^{a}$ as well as $\bQ_{i}$ and $\bQ^{i}$. Using a linear transformation, this symmetry can be phrased as
\begin{equation}
    \bP^{a} = \chi^{ab}\bP_{b}\,,
    \quad
    \bQ^{i} = \chi^{ij} \bQ_{j}\,,
    \quad
    \chi = \begin{pmatrix}
        0 & 0 & 0 & -1 \\
        0 & 0 & 1 & 0 \\
        0 & -1 & 0 & 0 \\
        1 & 0 & 0 & 0
    \end{pmatrix}\;.
\end{equation}

\paragraph{Types of states.} The solutions of the Q-system can be classified as type I-IV depending on what symmetries they exhibit, see \cite{Gromov:2023hzc}. We summarize these conventions in Table~\ref{tab:TypeOfStates}.
\begin{table}[]
    \centering
    \begin{tabular}{c|c|c}
         Type & LR-Symmetry & Parity    \\
         \hline
        I & $\checkmark$ & $\checkmark$ \\
        II & $\checkmark$ & $\cross$ \\
        III & $\cross$ & $\checkmark$ \\
        IV & $\cross$ & $\cross$
    \end{tabular}
    \caption{Classification of solutions to the Q-system according to symmetries}
    \label{tab:TypeOfStates}
\end{table}

\subsection{Gluing}
The conditions imposed on the Q-system so far are not yet enough to fully quantise the system. To do so, we must also prescribe that $\bQ_i$ are functions with long cuts. Explicitly, this means that we are instructed to glue together $\bQ^{\uparrow}$ with $\bQ^{\downarrow}$ along the cut on the real axis. We can equally well describe this as prescribing what happens upon analytically continuing $\bQ_{i}$ around the branch points at $\pm 2g$. Writing $\tilde{f}$ for the analytic continuation of a function $f$ around $u=2g$ we have
\begin{equation}\label{eq:GluingQU}
    \tilde{\bQ}^{\downarrow}_{i} = \bQ^{\uparrow,j}G_{ji}\,,
    \quad
    \tilde{\bQ}^{\downarrow,i} = G^{ij}\bQ^{\uparrow}_j\,,
\end{equation}
where the matrix $G$ is known as the gluing matrix. In order to ensure that the new Q-functions obtained after analytic continuation are still compatible with QQ-relations, it needs to be $\ii$ periodic; following~\cite{Gromov:2015wca,Alfimov:2018cms}, we further require that it has no branch cuts.

In this work, we will use a specific gluing matrix introduced in \cite{Alfimov:2018cms}. It is the choice of this matrix that specifies the analytic continuation away from integral spin $S_1$ while keeping $S_{2}$ integer (it is also possible to analytically continue in both $S_1$ and $S_2$ as described in \cite{Alfimov:2018cms} by further modifying the gluing matrix). Using the relation between $\bQ^{\downarrow}$ and $\bQ^{\uparrow}$ introduced above we can rephrase the gluing equation \eqref{eq:GluingQU} as
\begin{equation}\label{eq:TildeQComplex}
    \tilde{\bQ}_i = \bar{\bQ}^{j}\,M_{ji}\,,
    \quad
    \tilde{\bQ}^i = M^{ij}\,\bar{\bQ}_j\,,
\end{equation}
where $M^{ij}$ is given as
\begin{equation}
\begin{split}
    M^{ij} &=\begin{pmatrix}
        \alpha^{11} & 0 & \alpha^{13} & 0 \\
        0 & 0 & 0 & 0 \\
        \bar{\alpha}^{13} & 0 & \alpha^{33} & 0 \\
        0 & 0 & 0 & 0
    \end{pmatrix}
    + e^{2\pi u} \begin{pmatrix}
        0 & 0 & \beta^{13} & 0 \\
        0 & 0 & 0 & 0 \\
        \bar{\beta}^{13} & 0 & 0 & 0 \\
        0 & 0 & 0 & 0
    \end{pmatrix}
     + e^{-2\pi u}\begin{pmatrix}
        0 & 0 & \gamma^{13} & 0 \\
        0 & 0 & 0 & 0 \\
        \bar{\gamma}^{13} & 0 & 0 & 0 \\
        0 & 0 & 0 & 0
    \end{pmatrix}  \\
    &+\begin{pmatrix}
        0 & \alpha^{12} & 0 & 0 \\
        \bar{\alpha}^{12} & 0 & 0 & 0 \\
        0 & 0 & 0 & \alpha^{34} \\
        0 & 0 & \bar{\alpha}^{34} & 0
    \end{pmatrix}\,,
\end{split}
\end{equation}
with $\alpha^{11},\alpha^{33} \in \mathbb{R}$ and we have an explicit relation $\gamma^{13} = -e^{\ii \pi(S_2-\Delta)}\, \beta^{13}$ \cite{Alfimov:2018cms}.

\subsection{The $\bP\mu$- and $\bQ\omega$-System}
Let us finally connect the above discussion to the more traditional presentation of the QSC through the so-called $\bP\mu$ and $\bQ\omega$ systems. The $\omega$ matrix is obtained by yet another rewriting of \eqref{eq:GluingQU} as
\begin{equation}
    \tilde{\bQ}_{i} = \omega_{ij}\bQ^{j}\,,
    \quad
    \tilde{\omega}_{ij}-\omega_{ij} = \bQ_i \tilde\bQ_j - \tilde\bQ_i \bQ_j\,,
    \quad
    \omega^{ij} = G^{ik}\, \(\Omega^{\bQ}\)_{k}{}^{j}\,,
\end{equation}
where $\omega^{ij}$ is the inverse of $\omega_{ij}$, explicitly
\beq\la{omdual}
\omega^{ij}=-\frac{1}{2}\frac{1}{\text{Pf}(\omega)}\epsilon^{ijkl}\omega_{kl}\;,
\;\;\;
\text{Pf}(\omega) = \omega_{12}\omega_{34}-\omega_{13}\omega_{24}+\omega_{14}\omega_{23}\,.
\eeq
and $\epsilon$ is the fully antisymmetric tensor normalised so that $\epsilon^{1234}=1$. In particular  we find $\omega^{13}=\omega_{24}$.
Using then the Q-system, and in particular \eqref{eq:QPRotations}, one can deduce the $\bP\mu$-system
\begin{equation}\label{eq:PMu}
    \tilde{\bP}_{a} = \mu_{ab}\bP^{b}\,,
    \quad
    \tilde{\mu}_{ab}-\mu_{ab} = \bP_{a}\tilde{\bP}_b-\tilde{\bP}_{a}\bP_b\,,
    \quad
    \tilde{\mu}_{ab} = \mu^{[2]}_{ab}\,,
    \quad
    \mu_{ab} = \frac{1}{2}Q^{-}_{ab|ij}\omega^{ij}\,.
\end{equation}
Furthermore, we have
\begin{equation}
    \mu^{ab} = -\frac{1}{2}\frac{1}{\text{Pf}(\mu)}\epsilon^{abcd}\mu_{cd}\,,
    \quad
    \text{Pf}(\mu) = \mu_{12}\mu_{34} - \mu_{13}\mu_{24} + \mu_{14}\mu_{23}\,.
\end{equation}

A convenient rewriting of \eqref{eq:PMu} is \begin{equation}\label{eq:tildePgen}
    \tilde{\bP}_a = \frac{1}{2}Q_{a|ij} \omega^{ij}\,,
    \quad
    \tilde{\bP}^{a} = \frac{1}{2}Q^{a|ij}\,\omega_{ij}\;.
\end{equation}
To see this identity one starts from \eqref{eq:MuAndQabij} and uses  \eqref{eq:QaiRelations} as well as \eqref{eq:QQRelations}. 

Other useful identities for $\mu$ and $\omega$ are in terms of $Q_{a|i}$ and its complex conjugate $\bar Q_{a|i}$
\begin{align}\label{eq:MuGluingComplex}
    \mu_{ab} &= -Q^{-}_{a|i} M^{ij}\left(\bar{Q}_{c|j}\right)^- \mathbb{c}_b{}^c\;,\\
    \omega^{ij} &= M^{ik} \bar{Q}^-_{c|k}\,\mathbb{c}_{b}{}^{c}\, \left(Q^{b|j}\right)^-\,.
\end{align}
The reality properties of $\bP$ and $\bQ$ discussed in Section~\ref{sec:SymmetryPropertiesQSC} also propagate to $\mu$. Using \eqref{eq:MuGluingComplex}, and antisymmetry of $\mu$, we find that in a gauge in which $\mathbb{c}$ is diagonal that
\begin{equation}\label{eq:ConjugationMu}
    \overline{\mu^+_{ab}} = -\mathbb{c}_a^a \mathbb{c}_b^b\, \mu^+_{ab}\;\;\Leftrightarrow\;\;
    \overline{\mu_{ab}} = -\mathbb{c}_a^a \mathbb{c}_b^b\mu^{[+2]}_{ab}
    \,.
\end{equation}

\section{The Full ABBA Derivation}\label{app:ABBADerivation}
In this appendix, we provide the full derivation of the ABBA equations presented in Section~\ref{sec:ABBA}, filling in the details omitted in Section~\ref{sec:ABBADerivation}. 

To describe the weak coupling limit that gives rise to the ABBA it is convenient to start from the $\bP\mu$-system, see Appendix~\ref{app:QSCGeneralities} and in particular \eqref{eq:PMu}. The main simplification at weak coupling, in the BFKL regime, 
is that the term with  $\omega^{13}$ dominates the sum 
in the last equation of \eq{eq:PMu} and
we can approximate $\mu_{ab}$ with only one term
\begin{equation}\label{eq:MuABBALimit}
    \mu_{ab} = \frac{1}{2}Q^-_{ab|ij} \omega^{ij} \simeq_{g \rightarrow 0 } Q^-_{ab|13}\omega^{13} + \mathcal{O}\left(g^\text{Wrapping Order}\right)\,.
\end{equation}
In order to determine the wrapping order precisely one should consider so-called L\"usher correction (see~\cite{Janik:2007wt,Bajnok:2009vm,Janik:2010kd}). In the cases we considered, the wrapping corrections started normally at $g^{J_1+2}$ order.

We emphasise that the limit picks out $\omega^{13}$ precisely because we are in the BFKL regime. In the case of the local operators, a similar phenomenon would occur, but the term with $\omega^{12}$ would dominate.
Those two options are related by interchanging  $S_1\leftrightarrow\Delta$, hinting at the fact that $-S_1$ now plays the role of the energy.

We have extensively verified \eqref{eq:MuABBALimit} from numerical solutions and found that it holds in all cases considered, including both horizontal trajectories and reflection areas. We will present an even more thorough numerical analysis in \cite{UpcommingArpad}. In the remainder of this appendix, we assume the validity of \eqref{eq:MuABBALimit} and proceed to show that it is a consistent assumption leading to an analytic, closed set of equations; the ABBA equations.

\subsection{Finding $\mu_{12},\ \omega^{13}$ and $Q_{12|13}$}\label{subapp:FindingMuOmega}
To solve the QSC in the regime \eqref{eq:MuABBALimit}, we follow the general strategy developed in \cite{Gromov:2014caa}, where it was used to study the spectral problem of local operators in the ABA limit. Similar calculations have also been carried out in AdS$_4$ and AdS$_3$ \cite{Bombardelli_2017,Cavaglia:2021eqr,Ekhammar:2021pys,Ekhammar:2024kzp}.

Let us start by defining
\begin{equation}\la{calF}
\mathcal{F} = \frac{\mu_{12}}{\mu_{12}^{[2]}} = \frac{\mu_{12}}{\tilde{\mu}_{12}} \simeq \frac{Q^-_{12|13}}{Q^+_{12|13}}\,,
\end{equation}
where the first equality follows from \eq{eq:PMu} and the last equality uses \eqref{eq:MuABBALimit}. Since $Q^-_{ab|ij}$ by construction are UHPA this implies that ${\cal F}$ is analytic above the real axis. Furthermore,  $\bar{\cal F}=\pm 1/{\cal F}$, a consequence of \eqref{eq:ConjugationMu}, and thus ${\cal F}$ is analytic in the lower half plane. To summarize, ${\cal F}$ is a function with poles and zeros and only one cut at $(-2g,2g)$.

Let us denote the zeros of $\mu_{12}(u+i/2)$ which are not on the branch cut by $\{u_{4,i}\}_{i=1}^{K_4}$. For zeros on the branch cut we use the more convenient Zhukovsky parametrisation, which allows us to keep track of the two different sides of the branch cut. We denote the zeros as $\{z_i\}_{i=1}^{N}$, i.e $\mu_{12}(x=z_i) = 0$. Finally, we assemble these roots into the following functions
\begin{align}
    \betheQ_4 = \prod_{i}^{K_4}(u-u_{4,i})\,,
    \quad
    \kappa = \prod_{i=1}^{N}(x-z_i)\,,
    \quad
    \bar{\kappa} = \prod_{i=1}^{N}\left(x-\frac{1}{z_i}\right)\,.
\end{align}
With this notation in hand, we can explicitly remove all zeroes and poles in  $\mathcal{F}$, which prompts us to define a new function $\mathcal{G}$
\begin{equation}\label{eq:CalGDef}
    \mathcal{G} = \mathcal{F} \frac{\betheQ^+}{\betheQ^-} \frac{\bar{\kappa}}{\kappa}\;.
\end{equation}
Let us list all properties of $\mathcal{G}$:
\begin{enumerate}
    \item $\mathcal{G}$ is, up to wrapping order, a function with a single cut at $(-2g,2g)$.
    \item ${\cal G}$ does not have poles or zeroes in the strip around the cut, as all of them are absorbed into $\kappa$ and ${\mathbb Q}$ already.
    \item Under complex conjugation: $\bar {\cal G}=1/{\cal G}$, as follows from \eq{eq:ConjugationMu}.
    \item $\mathcal{G}$ has asymptotic behaviour $\mathcal{G}(\infty) = 1$, as follows from the definition \eq{calF}.
    \item Under analytic continuation around $2g$ we find as a consequence of  \eqref{calF} 
    \begin{equation}\la{eq:GtG}
    \tilde{\mathcal{G}}\, \mathcal{G} = \left(\frac{\betheQ^+}{\betheQ^-}\right)^2 \frac{1}{\prod^{N}_i z_i^2}\;,
    \end{equation}
    where to fix the sign one can use that $\tilde {\cal G}(2g)={\cal G}(2g)$.
\end{enumerate}

A solution to \eqref{eq:GtG}, which satisfies all the requirements listed above, is
\begin{equation}\label{eq:calG}
    \mathcal{G} = \left(\prod_{k=1}^{N}\frac{1}{z_k}\right)
    \prod_{k=1}^{K_4} \frac{x^-_{4,k}}{x^+_{4,k}}
    \left(\frac{\bB_{(+)}}{\bB_{(-)}}\right)^2\;,
\end{equation}
where $\bB_{(\pm)}$ is defined in \eq{eq:MainRB}.
with the additional constraint
\begin{equation}\la{cyclicity2}
    1 = \prod_{i=1}^{K_4} \frac{x^+_{4,k}}{x^-_{4,k}} \prod_i^{N} z_i\,,
\end{equation}
needed to ensure $\mathcal{G}(\infty) = 1$. We will refer to \eqref{cyclicity2} as the ``cyclicity'' condition due to its similarity with the cyclicity condition familiar from local operators.

Let us verify that \eqref{eq:calG} 
does satisfy all the above requirements, and furthermore, is the unique solution. $1)-2)$ follows immediately from the definition of $\mathcal{G}$, see \eqref{eq:CalGDef}. To see $3)$ we note that $|z_k|=1$ and hence $\bar z_k=1/z_k$, and thus $\bar{\cal G}=1/{\cal G}$. The asymptotics constraint in $4)$ is immediate from the definition of $\bfB$. Regarding $5)$, it is easy to verify that \eqref{eq:GtG} is satisfied as $\left(\prod_{k=1}^{K_4} x_{4,k}^\mp\right)\;\bB_{\pm}\tilde\bB_{\pm}= (-1)^{K_4} {\mathbb Q}^\pm_4 $. Finally, to establish uniqueness we can consider the homogeneous equation $\tilde {\cal G}_0 \, {\cal G}_0 = 1$.  ${\cal G}_0$ is regular and has no zeros for $|x|>1$, thus the same is true about $\tilde{\cal G}_0$, meaning that it is a polynomial of $x$. Since ${\cal G}_0 \rightarrow_{u\rightarrow \infty} 1$, we conclude that ${\cal G}_0 \equiv 1$.

\paragraph{Fixing $\mu_{12}$.}
From \eqref{eq:calG} and using the cyclicity condition \eqref{cyclicity2} we find
\begin{equation}\label{eq:MuMuEq}
    \frac{\mu_{12}}{\mu^{[2]}_{12}} = \frac{\betheQ^-}{\betheQ^+}\, \left(\frac{\mathbf{B}_{(+)}}{\bf B_{(-)}}\right)^2\, \frac{\kappa}{\bar{\kappa}}\;.
\end{equation}
Next, we introduce
\beq
f \propto \prod_{n=0}^\infty
\frac{\bfB^{[+2n]}_{(+)}}{\bfB^{[+2n]}_{(-)}}\;,
\quad
\bar f \propto  \prod_{n=0}^\infty
\frac{\bfB^{[-2n]}_{(-)}}{\bfB^{[-2n]}_{(+)}}\,,
\quad
\frac{f}{f^{++}}=
\frac{\bfB_{(+)}}{\bfB_{(-)}}\,.
\eeq
In this notation, we can find the solution \eq{eq:MuMuEq} up to a periodic prefactor ${\cal P}$ as
\begin{equation}\la{mu12P}
    \mu_{12} \propto\mathcal{P}\,  \betheQ^-_4\,
    f \bar f^{--}
    \,\kappa\,\prod_{n=1}^{\infty} \kappa^{[2n]}\, \bar{\kappa}^{[-2n]}\;.
\end{equation}
More precisely, \eqref{eq:MuMuEq} is satisfied if  $\mathcal{P}$ is a periodic function
\beq
\mathcal{P}(u+i)=\mathcal{P}(u)\;.
\eeq

In order to fix the periodic function ${\cal P}$, consider the following two relations, following immediately from
\eq{mu12P}
\begin{align}
    &\tilde{\mu}_{12} = \frac{\tilde{\mathcal{P}}}{\mathcal{P}}\,\frac{\bfR_{(+)}}{\bfR_{(-)}}\,\frac{\bfB_{(-)}}{\bfB_{(+)}}\frac{\tilde{\kappa}}{\kappa}\,\left(\prod_{k=1}^{K_4}\frac{x^+_{4,k}}{x^-_{4,k}}\right)\, \mu_{12}\,,
    \quad
    \mu^{[2]}_{12} = (-1)^{N}\frac{\betheQ^+_4}{\betheQ^-_4} \left(\frac{\bfB_{(-)}}{\bfB_{(+)}}\right)^2 \, \frac{\bar{\kappa}}{\kappa}\,\mu_{12}\,.
\end{align}
As we should have $\tilde{\mu}_{12}=\mu^{[2]}_{12}$, we find
\begin{equation}
    \frac{\tilde{\mathcal{P}}}{\mathcal{P}} = \frac{\betheQ^+_4}{\betheQ^-_4}\frac{\bfB_{(-)}}{\bfB_{(+)}}\frac{\bfR_{(-)}}{\bfR_{(+)}} \frac{\bar{\kappa}}{\tilde{\kappa}} = (-1)^{N}\left(-x\right)^{N}\left(\prod_{k=1}^{K_4}\frac{x^-_{4,k}}{x^+_{4,k}}\right)\frac{1}{\prod^{N}_{k=1} z_k} = x^{N}\,,
\end{equation}
and thus we should have, formally
\begin{equation}\la{Pcal}
    \mathcal{P} \propto p\prod_{n=-\infty}^{\infty}\frac{1}{(x^{[2n]})^{N/2}}\,,
\end{equation}  
where $p$ is an analytic (anti)periodic function. For $N$ odd, we get additional cuts, so in what follows we assume $N$ to be even.

Having fixed ${\cal P}$, we finally find
\begin{equation}\la{mu12P0}
    \boxed{\mu_{12} \propto p\,  \betheQ^-_4\,
    f \bar f^{--}
    \,
    \frac{\kappa}{x^{N/2}}\,\prod_{n=1}^{\infty} \frac{\kappa^{[2n]}\, \bar{\kappa}^{[-2n]}}{(x^{[2n]})^{N/2}(x^{[-2n]})^{N/2}}}\;.
\end{equation}

\paragraph{Recovering ${\cal Q}_{12|13}$.}
In order to recover ${\cal Q}_{12|13}$ we plug $\mu_{12} = Q^-_{12|13}\omega^{13}$ into \eqref{eq:MuMuEq}, which gives
\beq\label{eq:Q1213Eq}
\frac{{Q}_{12|13}^-}{{ Q}_{12|13}^+}=
\frac{\betheQ^-_4}{\betheQ^+_4}\, \left(\frac{\mathbf{B}_{(+)}}{\bf B_{(-)}}\right)^2\, \frac{\kappa}{\bar{\kappa}}\;.
\eeq
We have to keep in mind that, unlike $\mu_{12}$, 
${\cal Q}_{12|13}^-$ is an upper-half-plane analytic function
with power-like asymptotics. The unique solution to \eqref{eq:Q1213Eq}, up to proportionality coefficients,  with this analytic structure is 
\begin{equation}\la{Q1213}
\boxed{ Q_{12|13} \propto \betheQ_4 \,(f^+)^{2} \, \prod_{n=0}^{\infty} \frac{\kappa^{[2n+1]}}{\bar{\kappa}^{[2n+1]}}}\,.
\end{equation}
\paragraph{Recovering $\omega^{13}$.}
Finally, we can use \eq{eq:MuABBALimit} to recover 
$\omega^{13}$ from $\mu_{12}$ \eq{mu12P0} and ${\cal Q}_{12|13}$ \eq{Q1213}. We get the following result:
\begin{equation}\la{omegares}
\boxed{
    \omega^{13} \propto p\frac{\bar{f}^{[-2]}}{f}\prod_{n=-\infty}^{\infty}\frac{\bar{\kappa}^{[2n]}}{(x^{[2n]})^{N/2}}
    }\;.
\end{equation}

\paragraph{Number of Massless Modes and the Function $p$}

As mentioned in the main text, a fundamental assumption in our numerical investigations is that the gluing matrix has exponential asymptotic $G \sim e^{2\pi u}$, see \eqref{eq:MainGluing}. This, through \eqref{eq:MuGluingComplex}, implies that also $\mu_{12}$ and $\omega^{13}$ will have the same asymptotics. Let us assume $p \sim e^{\pi u n_p}$, then from our explicit expressions \eqref{mu12P0} and \eqref{omegares}
\begin{equation}
    \mu_{12} \sim e^{\pi u \left(\frac{N}{2}+n_p\right)}\,.
\end{equation}

In our numerical investigations, we have so far found that there are $4$ massless roots on each HT, forcing
\begin{equation}
    \boxed{N=4\,, \quad p\equiv 1\,, \quad \text{Horisontal Trajectory}}\,.
\end{equation}

On the contrary, for the RA, which will be treated in detail in \cite{EkhammarGromovPreti:2025b}, we found numerically that $N=2$ and thus in this regime we need to turn on $p$. In this case $p$ by construction is an anti-periodic function with asymptotic $p\sim e^{\pi u}$ and so
\begin{equation}
    \boxed{N=2\,, \quad p\equiv \cosh{\pi(u-u_0)}\,, \quad \text{Reflection Area}}\,.
\end{equation}
It is not impossible that there could be other values of $N$ relevant for the spectrum of $\mathcal{N}=4$. We will investigate this question for more general Regge trajectories in upcoming work \cite{UpcommingArpad}. However, since we at the moment lack evidence for any other values of significance except for $N=4,2$ we will restrict our attention to these cases when discussing the explicit form of $p$ in this paper.

\subsection{Finding $\bP_\alpha$ and ${ Q}_{\alpha|12}$}\label{subsec:PaQalpha}

In the previous subsection we found $\mu_{12}, \omega^{13}$ and $Q_{12|13}$.
In this subsection we will find $\bP_a$ and ${\cal Q}_{a|13}$, which are needed for ABBA.
We start from the following ansatz
\begin{equation}\label{eq:PABBADef0}
\bP_{a} =g_a \,\sigma_p \, \sigma_{\circ}\,\sigma_{\bullet}
\;\;,\;\;
\bP^{a} =g^a \,\sigma_p \, \sigma_{\circ}\,\sigma_{\bullet}\;,
\end{equation}
where we introduced the functions $\sigma_{p},\sigma_{\bullet}$ and $\sigma_{\circ}$ which are define to satisfy
\begin{equation}\label{eq:CrossingABBADerivation}
\tilde\sigma_p\,\sigma_p \propto p\,,
\quad
\tilde\sigma_\bullet\;\sigma_\bullet \propto f^{[+2]}\bar f^{[-2]}\,,
\quad
\tilde\sigma_\circ\;\sigma_\circ \propto \prod_{n=1}^\infty \frac{\kappa^{[+2n]}\bar\kappa^{[-2n]}}{
(x^{[+2n]})^{N/2}
(x^{[-2n]})^{N/2}
}\;,
\end{equation}
and constrained to be real, analytic outside a short cut at $(-2g,2g)$ and with no zeroes on their defining sheet and tending to $1$ at infinity. We will explicitly solve \eqref{eq:CrossingABBADerivation} with those additional conditions in Appendix~\ref{app:DressingPhases}. 
The function $g_\alpha$ has the same properties as $\bP_a$, i.e. it is real (up to a sign) and has a single cut on the main sheet.

Let us show that the function $g_\alpha(u)$ should be a rational function of $x$.
We use \eq{eq:tildePgen} and neglect all contributions except the terms containing $\omega^{13}$, so that up to the wrapping correction we get
\begin{equation}\label{eq:tildePABBA}
    \tilde{\bP}_a \propto Q_{a|13} \omega^{13}\,,
    \quad
    \tilde{\bP}^{a} \propto Q^{a|24}\,\omega_{24}\,,
\end{equation}
we recall that $\omega^{13} =\omega_{24}$, see \eqref{omdual}. At the same time, from \eq{eq:PABBADef0} we have
\beq
\frac{\tilde{\bP}_a}{\omega^{13}}=\frac{\tilde g_a\tilde \sigma_p\tilde \sigma_\circ\tilde \sigma_\bullet}{\omega^{13}}
\propto
\tilde g_a\[
\frac{x^{N/2} f f^{[+2]}}{
\sigma_p \sigma_\circ \sigma_\bullet
{\bar{\kappa}}
}
 \prod_{n=1}^\infty \frac{\kappa^{[+2n]}}{\bar\kappa^{[+2n]}}\]\;.
\eeq
From \eq{eq:tildePABBA} we know that the r.h.s. should be UHPA.
We see that, indeed, all terms in the r.h.s. 
inside the square bracket are UHPA, and thus $\tilde g_a$ should also be UHPA. As $g_a$, and thus $\tilde g_a$ are real functions (up to a sign) we conclude that $g_a$ should be a polynomial of $x$ times a suitable power of $x$.
We split the zeroes in this polynomial into those outside the unit circle and inside the unit circle to define
\beq
g_1 = x^{-L/2-1}R_1 B_3\;,
\eeq 
with $R$ and $B$ defined in \eq{eq:MainRB}. 
By applying the same chain of arguments to $\bP^4$ we conclude that
\begin{equation}\label{eq:PABBADef}
    \boxed{\bP_{1} = x^{-L/2-1}\,\sigma_p \, \sigma_{\circ}\,\sigma_{\bullet} \, R_{1} B_{3}\,,
    \quad
    \bP^{4} = x^{-L'/2-1}\,\sigma_p \, \sigma_{\circ}\,\sigma_{\bullet} \,R_{7} B_{5}\,,}
\end{equation}
We can use the $\Lambda$-symmetry of QSC \eq{LambdaSym} to set $L'=L$. Note, however, that by doing so the power of $x$ may become half-integer, which creates an additional fictitious sign cut in $\bP$, formally violating the analyticity axioms; however, it is just a matter of convenience to set $L'=L$ and is the standard conversion, which we follow in the rest of the paper.
 
Upon returning to \eqref{eq:tildePABBA} after having found \eq{eq:PABBADef} we can identify $Q_{a|13}$ and $Q^{a|24}$ 
\begin{equation}\label{eq:Q113ABBADef}
\boxed{\begin{split}
Q_{1|13} &\propto
x^{L/2+1+N/2}
B_{1}R_{3}
\frac{f f^{[2]}}
{\sigma_p\sigma_{\circ}\sigma_\bullet\bar\kappa}
\prod_{n=1}^{\infty}\frac{\kappa^{[2n]}}{\bar{\kappa}^{[2n]}}\,,
\\
Q^{4|24} &\propto
x^{L/2+1+N/2}
R_{5}B_{7}
\frac{f f^{[2]}}
{\sigma_p\sigma_{\circ}\sigma_\bullet\bar\kappa}
\prod_{n=1}^{\infty}\frac{\kappa^{[2n]}}{\bar{\kappa}^{[2n]}}\,.
\end{split}}
\end{equation}

\subsection{Energy and Quantum Numbers}
From Table~\ref{tab:Qasymptotics} we know that
$Q_{1|13}\sim u^\frac{-J_1-J_2+J_3-2S_1}{2}$, this allow us to ``measure'' the ``energy'' $S_1$ by checking the large $u$ asymptotic of \eq{eq:Q113ABBADef}. We find
\beq
u^{L/2+1-N/2+K_3+\gamma_\circ+\gamma_\bullet}\;\;,\;\;\gamma_\circ = ig \sum_{k=1}^N\(\frac{1}{z_k}-z_k\)\;\;,\;\;\gamma_\bullet = 2i g \sum_{k=1}^{K_4}\(\frac{1}{x^+_k}-\frac{1}{x^-_k}\)\,.
\eeq
Thus we see the ``anomalous'' part of $S_1$ is given by $-\gamma_\circ-\gamma_\bullet$. For the integer part we get
\beq
K_3=\frac{1}{2}\left(-J_1-J_2+J_3+2\Sigma-L-2+N\right)\,,
\eeq
in agreement with \eq{magnonnumbers}.

More generally we find the following asymptotic
\begin{equation}
\begin{split}
    &\{\bP_1,Q_{1|13},Q_{12|13},Q^{4|24},\bP^{4}\}\\
    &\simeq u^{\{-L/2-1+K_1,L/2+1-N/2+K_3+\gamma,K_4+\gamma,-L/2-1+K_7\}}\\
    &=u^{\frac{1}{2}\{-J_1-J_2+J_3-2,-J_1-J_2+J_3-2S_1,2\(1-J_1-S_1\),-J_1-J_2-J_3-2S_1,-J_1-J_2-J_3-2\}}\,,
\end{split}
\end{equation}
where the second line follows from Table~\ref{tab:Qasymptotics}, setting these expressions equally gives precisely \eqref{magnonnumbers}.

\subsection{Finding $Q_{1|1}$ and $Q^{4|4}$}
The Q-functions $Q_{a|i}$ turn out to be more complicated in our construction than in the case of local operators. The reason is that their asymptotics contain $\Delta$, which is an arbitrary parameter.

Indeed, instead of parameterizing $Q_{a|i}$ in terms of some roots, we will derive a Baxter equation that fixes it.
Before that, we can still simplify them a bit by stripping away an overall prefactor to ensure that the remaining piece is a meromorphic function with poles at specific locations.

Whereas $Q_{a|i}^-$ are analytic in the upper half plane, they have infinitely many cuts in the lower half plane (including on the real axis). At the same time, using \eqref{eq:MuABBALimit} in \eqref{eq:MuGluingComplex} we have 
\begin{equation}
     Q^-_{a|1}\omega^{13} \propto M^{3l} \bar{Q}^-_{c|l}\mathbb{c}_{a}{}^{c}\,,
     \quad
     Q^-_{a|3}\omega^{13} \propto M^{1l} \bar{Q}^-_{c|l}\mathbb{c}_{a}{}^{c}\,.
\end{equation}
In other words, when multiplying $Q_{a|\mu}^-$ with $\mu=1,3$ by $\omega^{13}$ we should get a function analytic in the UHP including the real axis. Let us write it more explicitly,
using the form of $\omega^{13}$ from \eq{omegares} (and periodicity of $\omega^{13}$) we have
\beq
{Q}_{1|\mu}\omega^{13+}={Q}_{1|\mu}\omega^{13-}\propto
{Q}_{1|\mu}\;
p^-\;\frac{\bar{f}^-}{f^{+}}\prod_{n=-\infty}^{\infty}\frac{\bar{\kappa}^{[2n-1]}}{(x^{[2n-1]})^{N/2}}\;\;,\;\;\mu=1,3\;,
\eeq
where the r.h.s. should be LHP analytic. 
By removing the terms which are explicitly LHP analytic we conclude that
\beq
{Q}_{a|\mu}\frac{1}{f^{+}}
\prod_{n=0}^\infty \frac{\bar\kappa^{[2n+1]}}{(x^{[2n+1]})^{N/2}}\;,
\eeq
is LHPA, at the same time all factors are explicitly UHPA, so, naively this function is analytic in the whole plane. However, the infinite product does not converge even up to an infinite u-independent factor.
One way to make sense of the product is to add additional factors, which go as $1/u^2$ and thus we introduce the following combination
\beq\la{Q11}
{\mathbb M}_{1|\mu}\equiv {Q}_{1|\mu}\frac{{\bf F}}{f^{+}}\;,
\eeq
with the regularised factor
\beq
{\bf F}\propto
\frac{1}{\prod\limits_{i=1}^{N/2}\Gamma\(\frac{1}{2}-i(u-\alpha_i)\)}\prod_{n=0}^\infty \(\frac{\bar\kappa}{x^{N/2}\prod\limits_{i=1}^{N/2}(u-\alpha_i)}\)^{[2n+1]}\;.
\eeq
It is easy to see that the ratio of ${\bf F}$'s with different $\alpha$'s is just a constant, meaning that (up to an irrelevant proportionality coefficient) ${\bf F}$ does not actually depend on $\alpha$, which we can choose the way we want. A convenient choice is to take $\alpha_i$ to be a subset of all $\theta$'s, without reducing the generality we can pick it to be $\{\theta_{i}:\;i=1,\dots,N/2\}$ (we recall that $\theta_i\equiv g(z_i+1/z_i)$). We then denote
\beq
\kappa_1 = \prod_{i=1}^{N/2}(x-z_{i})\;,\qquad\kappa_2 = \kappa/\kappa_1\,,
\eeq
and similarly for $\bar\kappa_1$ and $\bar\kappa_2$. In these notations we have
\beq
{\bf F}\propto\frac{1}{\Gamma^+_1}\prod_{n=0}^\infty \(\frac{\bar\kappa_2}{
\kappa_1
}\)^{[2n+1]}
\;\;,\;\;\Gamma_1\equiv \prod_{i=1}^{N/2}
\Gamma\left(-iu+i\theta_{i}
\right)
\;,
\eeq
where the r.h.s. is independent of the way the splitting between $\kappa_1$ and $\kappa_2$ is done. Note that the price to pay for the analyticity is that the function $\bf F$ grows exponentially at infinity, ${\bf F}\sim u^{i u}$.

We managed to simplify $Q_{a|\mu}$ to a simpler analytic function ${\mathbb M}_{a|\mu}$.
For local operators $\mathbb{M}$ would be a polynomial, here however, as the asymptotic of ${\mathbb M}_{1|\mu}$ is exponential ${\mathbb M}_{1|\mu}$ must have infinitely many roots.
Our next goal is to find an equation for the analytic part ${\mathbb M}_{\alpha|\mu}$. For that, consider the exact
QQ-relation \eq{eq:QQRelations}
\beq\la{Q1113}
{Q}_{1|1}^+{Q}_{1|3}^-
-
{Q}_{1|1}^-{Q}_{1|3}^+
=
{\bf P}_1 {Q}_{1|13}
\;,
\eeq
after plugging the expression \eq{Q11} and \eq{eq:Q113ABBADef}, \eq{eq:PABBADef}, 
which after canceling common factors,
becomes
\beq\la{wronskianM}
{\mathbb M}_{1|1}^+
{\mathbb M}_{1|3}^-
-
{\mathbb M}_{1|1}^-
{\mathbb M}_{1|3}^+
\propto
\frac{{\mathbb Q}_{1}{\mathbb Q}_{3}}
{\Gamma^{++}}\;\;,\;\;
\Gamma\equiv\prod_{i=1}^{N}
\Gamma\left(-iu+i\theta_{i}
\right)\;.
\eeq
Note that the factor ${\mathbb Q}_{1}{\mathbb Q}_{3}$ is a finite degree polynomial, whereas $1/\Gamma^{++}$ introduces an infinite number of zeros into the r.h.s.
at $u=\theta_i+n,\;n=1,2,\dots$. Thus we conclude
\beq\la{recursion}
\frac{{\mathbb M}_{1|1}^+}{{\mathbb M}_{1|1}^-}
=
\frac{{\mathbb M}_{1|3}^+}{{\mathbb M}_{1|3}^-}\;\;,\;\;u=\theta_i+n\;\;,\;\;n=1,2,\dots\;.
\eeq
Similarly, define the following combination
\beq\la{Tdef}
{\cal T}\equiv -\left({Q}_{1|1}^{++}{Q}_{1|3}^{--}
-
{Q}_{\alpha|1}^{--}{Q}_{\alpha|3}^{++}\right)
\eeq
after stripping the same type of factors we define
\beq
{{\mathbb T}}\equiv
\frac{{\cal T}}{f^- f^{[+3]}}
\frac{\bar\kappa^-}{\kappa^+}
\prod_{n=1}^\infty 
\frac{\bar\kappa^{[2n+1]}}
{\kappa^{[2n+1]}}\;.
\eeq
From \eq{Tdef} and \eq{Q11} we can see that it satisfies 
\beq
 -
{\mathbb M}_{\alpha|1}^{++}{\mathbb M}_{\alpha|3}^{--}
+
{\mathbb M}_{\alpha|1}^{--}{\mathbb M}_{\alpha|3}^{++}
= \frac{{\mathbb T}}{\Gamma^+}\;.
\eeq
Note that the l.h.s. vanishes at poles of $\Gamma^+$ due to \eq{recursion}. Thus ${\mathbb T}$ is an analytic function in the whole complex plane. Furthermore, from \eq{Tdef}, it is easy to see that it has the power-like asymptotic
\beq
{\mathbb T}\sim u^{N-J_2+J_3+K_4-2}\,,
\eeq
thus we conclude that ${\mathbb T}$ is a polynomial!

Finally, we consider the trivial determinant
\beq
\left|
\begin{array}{ccc}
{\mathbb M}_{1|\mu}^{[-2]} & {\mathbb M}_{1|\mu} & {\mathbb M}_{1|\mu}^{[+2]}\\
{\mathbb M}_{1|1}^{[-2]} & {\mathbb M}_{1|1} & {\mathbb M}_{1|1}^{[+2]}\\
{\mathbb M}_{1|3}^{[-2]} & {\mathbb M}_{1|3} & {\mathbb M}_{1|3}^{[+2]}
\end{array}
\right|=0\,,
\eeq
and upon expanding it along the first row
we find the following Baxter equation
\beq\la{Mbaxter}
{\mathbb M}_{1|\mu}^{[+2]}{
{\mathbb Q}_{1}^-
{\mathbb Q}_{3}^-
{\mathbb Q}_\theta^+
}
+
{\mathbb M}_{1|\mu}{\mathbb T}
+
{\mathbb M}_{1|\mu}^{[-2]}
{\mathbb Q}_{1}^+
{\mathbb Q}_{3}^+
=0\;\;,\;\;\mu=1,3\;.
\eeq
\paragraph{Introducing ${\mathbb Q}_{2}$ and
${\mathbb Q}_{\tilde 2}$.}
Whereas ${\mathbb M}_{1|\mu}$ are analytic in the whole complex plane, they have very messy asymptotics. It is convenient to redefine ${\mathbb M}_{1|\mu}$ slightly to obtain a set of meromorphic functions $\mathbb{Q}_{1|\mu}$, which we use in the main text.
We define
\beq\la{QtoM}
{\mathbb Q}_{1|\mu} \propto {\mathbb M}_{1|\mu} \Gamma_1^+
={Q}_{1|\mu}
\frac{1}{f^{+}}\prod_{n=0}^\infty \(\frac{\bar\kappa_2}{
\kappa_1
}\)^{[2n+1]}\;,
\eeq
or
\beq\la{QtoMDef}
\boxed{
{Q}_{1|\mu}
\propto
{\mathbb Q}_{1|\mu}
{f^{+}}\prod_{n=0}^\infty \(\frac{
\kappa_1
}{\bar\kappa_2}\)^{[2n+1]}}
\;.
\eeq

In the main text we will use slightly different notation to emphasize the role of $\betheQ_{1|\mu}$ in the full ABBA, we introduce
\begin{equation}
    \betheQ_{2} = \betheQ_{1|1}\,,
    \quad
    \betheQ_{\tilde{2}} = \betheQ_{1|3}\,.
\end{equation}
One can see that the additional factor in the r.h.s. give simple power-like asymptotic
\beq
{\mathbb Q}_{1|\mu}\sim
{Q}_{1|\beta}
u^{-\gamma_\bullet/2+\omega_{\circ_1}}\,,
\quad
\omega_{\circ_k} = i g\sum_{i=1}^{N/2}
\(z_{i+(k-1)N/2}-\frac{1}{z_{k-(k-2)N/2}}\)\,.
\eeq
Since
\beq
{Q}_{1|1}\sim 
u^\frac{+\Delta-S_2 -\omega -J_2+J_3+K_4-1}{2}\;\;,\;\;
{Q}_{1|3}\sim
u^\frac{-\Delta+S_2 -\omega -J_2+J_3+K_4-1}{2} \,,
\eeq
we see that ${\mathbb Q}_{1|\mu}$ indeed behaves power-like at infinity. More precisely:
\beqa
{\mathbb Q}_2\equiv {\mathbb Q}_{1|1} \sim u^{\frac{+(\Delta-S_2)+\omega_{\circ_1}-\omega_{\circ_2} -J_2+J_3+K_4-1}{2}}
= u^{\frac{+(\Delta-S_2)-J_2+J_3+K_4-1+i(\theta_1+\theta_2-\theta_3-\theta_4)}{2}}\;,\\
{\mathbb Q}_{\tilde 2}\equiv {\mathbb Q}_{1|3} \sim u^{\frac{-(\Delta-S_2)+\omega_{\circ_1}-\omega_{\circ_2} -J_2+J_3+K_4-1}{2}}
= u^{\frac{-(\Delta-S_2)-J_2+J_3+K_4-1+i(\theta_1+\theta_2-\theta_3-\theta_4)}{2}}\;.
\eeqa
As a consequence of \eq{Mbaxter} we have
\beq\la{baxterQ2appendix}
{\mathbb Q}_\alpha^{[+2]}{
{\mathbb Q}_{1}^-
{\mathbb Q}_{3}^-
{\mathbb Q}_{\theta_2}^+
}
+
{\mathbb Q}_\alpha{\mathbb T}
+
{\mathbb Q}_\alpha^{[-2]}
{\mathbb Q}_{1}^+
{\mathbb Q}_{3}^+
{\mathbb Q}_{\theta_1}^-
=0\;\;,\;\;\alpha=2,\tilde 2\;.
\eeq
The price to pay for the nice asymptotic is twofold: Firstly, as one can see from \eq{baxterQ2appendix}, ${\mathbb Q}_\alpha$ depends on the choice of the ordering of $\theta$'s; Secondly, it is no longer analytic but has poles at $u=\theta_i-i/2-i n\;,\;n=0,1,\dots$ for the first subset of $\theta$'s i.e. for $i=1,\dots,N/2$.

Note that as ${\mathbb M}$ was indifferent to the ordering of $\theta$ we can always use \eq{QtoM} to relate ${\mathbb Q}$ for different sets. As we can see, to translate between different orderings, we simply add different factors of $\Gamma$'s.

\paragraph{$Q^{4|4}$ and $Q^{4|2}$} 
Let us also note what happens for $Q^{4|\dot{\mu}}\,, \dot{\mu}=2,4$. For these functions, one can simply repeat the argumentation above with minimal modifications. We will be content with showing how to find a similar Baxter equation as in \eqref{baxterQ2appendix}. We start from
\begin{equation}
    \left(Q^{4|2}\right)^+\left(Q^{4|4}\right)^- - \left(Q^{4|2}\right)^- \left(Q^{4|4}\right)^+ = \bP^{4}Q^{4|24}
\end{equation}
instead of \eqref{Q1113}. As we can see from \eqref{eq:PABBADef} and \eqref{eq:Q113ABBADef} the only difference between $\bP_1 Q_{1|13}$ and  $\bP^4 Q^{4|24}$ is that we should interchange the auxiliary roots $1\leftrightarrow 7$ and $3\leftrightarrow 5$. We thus introduce 
\begin{equation}\la{QtoMDef6}
\boxed{
{Q}^{4|\mu}
\propto
{\mathbb Q}^{4|\mu}
{f^{+}}\prod_{n=0}^\infty \(\frac{
\kappa_1
}{\bar\kappa_2}\)^{[2n+1]}}
\;,
\end{equation}
and the following notation
\begin{align}
    \mathbb{Q}_6 &\equiv \betheQ^{4|4} \simeq u^{\frac{(\Delta+S_2)-J_2-J_3+K_4+\omega_{\circ_1}-\omega_{\circ_2}-1}{2}}\,, \\
    \mathbb{Q}_{\tilde{6}} &= \betheQ^{4|2} \simeq u^{\frac{-(\Delta+S_2)-J_2-J_3+K_4+\omega_{\circ_1}-\omega_{\circ_2}-1}{2}}\,,
\end{align}
where $\mathbb{Q}_6$ and $\mathbb{Q}_{\tilde{6}}$ satisfy the same analytic constraints as $\mathbb{Q}_{2}$ and $\mathbb{Q}_{\tilde{2}}$. Let us, for completeness, record the Baxter equation:
\beq\la{baxterQ6appendix}
{\mathbb Q}_{\dot{\alpha}}^{[+2]}{
{\mathbb Q}_{1}^-
{\mathbb Q}_{3}^-
{\mathbb Q}_{\theta_2}^+
}
+
{\mathbb Q}_{\dot{\alpha}}{\mathbb T}
+
{\mathbb Q}_{\dot{\alpha}}^{[-2]}
{\mathbb Q}_{1}^+
{\mathbb Q}_{3}^+
{\mathbb Q}_{\theta_1}^-
=0\;\;,\;\;\dot{\alpha}=6,\tilde 6\;.
\eeq

\subsection{All the Massive Equations}
Having determined all the key Q-functions, we can easily derive the massive Bethe equations in the standard way. For example, evaluating
\eq{eq:QaiRelations} at the root of $\bP_1$ i.e. at $x=x_{1,k}$ we get
\begin{equation}
\left.\frac{Q^{+}_{1|1}}
    {Q^{-}_{1|1}}\right|_{x=x_{1,k}} =  1
\end{equation}
and using \eq{fig:ABBAgeneral} we get
\beq\la{QtoM}
\frac{{Q}_{1|1}^+}{{Q}_{1|1}^-}=
\left.\frac{
{\mathbb Q}_{2}^+
}{
{\mathbb Q}_{2}^-
}
\frac{\bB_{(-)}}{\bB_{(+)}}\frac
{\bar\kappa_2}
{
\kappa_1
}\right|_{x=x_{1,k}}=1\;,
\eeq
which is precisely the first equation in the Table~\ref{fig:ABBAgeneral}. Note that the second equation
in that table is \eq{baxterQ2appendix}. Similarly, the third equation can be obtained the same way as the first one, just this time we evaluate \eq{eq:QaiRelations} on the second sheet at the zero of $\tilde {\bf P}_1$ at 
$x=x_{3,k}$, producing
\beq
\left.\frac{
{\mathbb Q}_{2}^+
}{
{\mathbb Q}_{2}^-
}
\frac{\tilde\bB_{(-)}}{\tilde\bB_{(+)}}\frac
{\tilde{\bar\kappa}_2}
{
\tilde\kappa_1
}\right|_{x=x_{1,k}}=1\;.
\eeq
To bring it to the form appearing in Table~\ref{fig:ABBAgeneral}
we have to apply the ``cyclicity'' condition \eq{cyclicity}. 

Finally, for the middle-node equation we use the following exact QQ-relation
\beq
Q_{12 \mid 1} Q^{34 \mid 4}=Q_{12 \mid 12}^{+} Q_{12 \mid 13}^{-}-Q_{12 \mid 12}^{-} Q_{12 \mid 13}^{+}
\eeq
evaluated at $u=u_{4,k}\pm \tfrac i2$, which is zero of $Q_{12|12}$, and dividing by each other we get
\beq
\left.\frac{Q^+_{12 \mid 1} Q^{34 \mid 4+}}{
Q^-_{12 \mid 1} Q^{34 \mid 4-}
}\frac{Q^{[-2]}_{{12|12}}}{Q^{[+2]}_{{12|12}}}\right|_{u=u_{4,k}}=-1\;,
\eeq
then simply substituting \eq{eq:Q113ABBADef} and \eq{Q1213} we get
\begin{equation}
\frac{Q^+_{12 \mid 1} Q^{34 \mid 4+}}{
Q^-_{12 \mid 1} Q^{34 \mid 4-}}
    \frac{Q^{[-2]}_{12|13}}{Q^{[+2]}_{12|13}} = 
\(\frac{x_{4,k}^+}{x_{4,k}^-}\)^{L+2+N}
\frac{B^+_{1}R^+_{3}}{B^-_{1}R^-_{3}}
\frac{B^+_{7}R^+_{5}}{B^-_{7}R^-_{5}}
\(\frac{
\sigma_p^-\sigma^-_{\circ}\sigma^-_\bullet
}
{\sigma_p^+\sigma^+_{\circ}\sigma^+_\bullet}
\)^2
    \frac{
    \betheQ^{[-2]} \, }{
    \betheQ^{[+2]} \, }
    \frac{\kappa^{-}\bar{\kappa}^{-}}{{\kappa}^{+}\bar{\kappa}^{+}}\,
\end{equation}
which, after using 
\beq
\frac{\kappa^-\bar\kappa^-}{\kappa^+\bar\kappa^+}
=\(\frac{x_{4,k}^-}{x_{4,k}^+}\)^N\frac{Q_\theta^-}{Q_\theta^+}\,,\qquad Q_\theta\equiv \prod_{i=1}^N(u-\theta_i)\;,
\eeq
becomes the middle node equation in the form given in the Table~\ref{fig:ABBAgeneral} after introducing the following notations for the dressing phases
\beq\la{massiveDF}
\begin{split}
    \prod_{k=1}^{K_4}\sig_{\bullet\bullet}(x,x_{4,k})& \equiv \frac{\sigma_{\bullet}(x^+)}{\sigma_{\bullet}(x^-)},
    \qquad
    \prod_{k=1}^{N}\sig_{\bullet\circ}(x,z_k) \equiv \frac{\sigma_{\circ}(x^+)}{\sigma_{\circ}(x^-)},\\
    &\prod_{k=1}^{\frac{4-N}2}\sig_{\bullet p}(x,u_{0,k}) \equiv \frac{\sigma_{p}(x^+)}{\sigma_{p}(x^-)}
    \,.
    \end{split}
\eeq
The remaining equations are obtained in the same way, but with the Hodge dual Q-functions.

\subsection{The Massless Equation}
Whereas the above derivations in this appendix were generalizations of the original consideration in \cite{Gromov:2014caa}, the derivation of the equation, which would constrain the massless roots $z_i$, is totally novel. It served already as an inspiration for the AdS${}_3$ case in \cite{Ekhammar:2024kzp} where it also gave a prediction for the massless-massless case scattering, recently confirmed with independent methods in \cite{Frolov:2025ozz}.
The idea of the derivation was outlined in our previous paper \cite{Ekhammar:2024neh}, but here we present the complete derivation.

To derive the massless Bethe equations we start from the gluing equation \eqref{eq:TildeQComplex}. For our purposes, we only need two components of this equation
\begin{equation}\label{eq:GluingABBA}
    {\bQ}_\mu = \alpha_{\beta} \tilde{\bar{\bQ}}^{\dot\mu}\;\;,\;\;\mu=1,3\;\;,\;\;\dot\mu \equiv \mu+1=2,4\;.
\end{equation}
Above we have not derived the ABA form of the 
$\bQ_{\beta}$ so in order to impose
\eq{eq:GluingABBA} we will have to first find its expression.
The Q-function $\bQ_{\beta}$ can be obtained from the expression for 
$Q_{1|\beta}$ \eq{QtoMDef}
and $\bP_1$ \eq{eq:PABBADef} as from \eq{eq:QaiRelations} we have
\beq
\bQ_{\mu} = \frac{\left(Q^+_{1|\mu}-Q^-_{1|\mu}\right)}{\bP_1}\,,
\quad
\bQ^{\dot{\mu}} = \frac{\left(\left(Q^{4|\dot{\mu}}\right)^+-\left(Q^{4|\dot{\mu}}\right)^-\right)}{\bP^4}\,.
\eeq  
Combining all these equations together we get
\begin{align}
\bQ_{\mu} 
&\propto   \left(\prod_{n=1}^{\infty}\frac{\kappa^{[2n]}_1}{\bar{\kappa}^{[2n]}_{2}}\right) f^{[2]} \frac{\betheQ_{1|\mu}^+-\frac{\bfB_{(+)}}{\bfB_{(-)}}\frac{\kappa_1}{\bar{\kappa}_{2}}\betheQ_{1|\mu}^-}{(gx)^{-\frac{L+2}2}R_{1}B_{3}\sigma_p\sigma_{\bullet}\sigma_{\circ}}  \\
\bQ^{\dot\mu} &\propto  \left(\prod_{n=1}^{\infty}\frac{\kappa^{[2n]}_1}{\bar{\kappa}^{[2n]}_{2}}\right) f^{[2]} \frac{\left(\betheQ^{4|\dot{\beta}}\right)^+ -\frac{\bfB_{(+)}}{\bfB_{(-)}}\frac{\kappa_1}{\bar{\kappa}_{2}}\left(\betheQ^{4|\dot{\mu}}\right)^-}{(gx)^{-\frac{L+2}2}R_{7}B_{5}\sigma_p\sigma_{\bullet}\sigma_{\circ}} \;.
\end{align}
As the gluing involves complex conjugation and analytic continuation, let us write an expression for the complex conjugated $\bQ^{\dot\beta}$ evaluated at $1/x$
\begin{align}
\bar\bQ^{\dot\mu}(1/x) &\propto  \left(\prod_{n=1}^{\infty}\frac{\bar\kappa^{[-2n]}_1}{\kappa^{[-2n]}_{2}}\right) \bar f^{[-2]} 
    g^{L+2}
    \frac{\left(\bar{\betheQ}^{4|\dot{\mu}}\right)^--\frac{\bfR_{(-)}}{\bfR_{(+)}}\frac{\kappa_1}{\bar\kappa_{2}}\left(\bar{\betheQ}^{4|\dot{\mu}}\right)^+\[\prod\limits_{k=1}^{K_4}\frac{x^-_{4,k}}{x^+_{4,k}}\prod\limits_{i=1}^N\frac{1}{z_i}\]}{(gx)^{+\frac{L+2}2}B_{7}R_{5}\tilde\sigma_{\bullet}\tilde\sigma_p\tilde\sigma_{\circ}} \;,
\end{align}
which can be simplified by noticing that the factor in the square brackets is $1$ due to the cyclicity condition \eqref{cyclicity2}.

Now, by observation it is clear that $\bQ_\mu(x)$ and $\bar\bQ^{\dot \mu}(1/x)$ cannot be matched literary as written. 
However, we should remember that these expressions are only approximate and has the following limitations: strictly speaking the 
above expression for $\bQ_\mu$ is only accurate to wrapping order for $u\sim 1$. The approximation becomes essentially invalid when approaching the cut. In particular, the exact $\bQ$ has no poles anywhere on its Riemann surface, whereas the approximation above clearly has poles right on the cut.
Yet we have to match the two expressions on the cut. How can we do that?

The idea is to express $\bQ$'s as a Laurent series in $x$ in the region around the unit circle $\sum_{n=-\infty}^\infty {\bf c_n} x^n$. As the exact $\bQ$ should be regular on the annulus $1/R<|x|<R$ where $R=|x(2g+i/2)|\simeq \frac{1}{2g}$ which covers well the area on both sides of the cut. In other words we multiply the gluing condition \eq{eq:GluingABBA} by $x^{-n}$ and integrate around the cut
\begin{equation}\label{eq:GluingABBAint}
    {\bf c}_n\equiv \oint \frac{dx}{x} \frac{1}{x^n} {\bQ}_\mu(x) =\alpha_{\beta} \oint \frac{dx}{x} \frac{1}{x^n}  {\bar{\bQ}}^{\dot\mu}(1/x)\;,
\end{equation}
as the convergence radius is $\simeq \frac{1}{2g}$ we should have $|{\bf c}_n|\sim (2g)^{|n|}$
for $|n|$ large. The main trick is that in the above integrals \eq{eq:GluingABBAint} we can move the integration contour to the regime where the ABBA approximation for $\bQ$ is valid i.e. $u\sim 1$ or $|x|\sim \frac{1}{g}$. After replacing $\bQ$ by its ABBA approximation we can deform the contour to go inside the unit circle as much as possible i.e. until we hit the next singularity at $|x|\sim g$. However, while deforming the contour we have to pick the poles at $|x|\sim 1$. Let us estimate the relative scaling of these two contributions in $g$: due to the $x^{\frac{L+2}{2}+n}$ factor in $\bQ$ we see that the remaining contour integral would we additionally suppressed by $g^{\frac{L+2}{2}+n}$ and can be neglected if $n$ is not too large and negative. Similarly for the last term in \eq{eq:GluingABBAint} we start from the area where the argument of $\bar\bQ$ is such that $u\sim 1$ i.e. $x\sim g$ and expand the contour until $x\sim 1/g$.
Again on the way we encounter the poles, contributing with an opposite sign, and the remaining integral, which can be neglected if $n$ is not too large and positive.

In conclusion only the poles contribute for $n\sim 1$ and $L$ sufficiently large contribute. As there are only $N$ poles (with $N=4$ for HT, BA or $N=2$ for RA), coming from $\frac{\kappa_1}{\bar\kappa_2}{\mathbb Q}^-$, we conclude that for sufficiently large $L$ each individual residue in the r.h.s. should match (with the opposite sign) to the corresponding residue in the l.h.s. of \eq{eq:GluingABBAint} (up to a sign)
\beq\label{eq:ResidueEq}
\!\!\!\!{\rm res}\left(\prod_{n=1}^{\infty}\frac{\kappa^{[2n]}_1}{\bar{\kappa}^{[2n]}_{2}}\right)  \frac{f^{[2]}\frac{\bfB_{(+)}}{\bfB_{(-)}}\frac{\kappa_1}{\bar{\kappa}_{2}}\betheQ_{1|\mu}^-}{x^{-\frac{L+2}2}R_{1}B_{3}\sigma_p\sigma_{\bullet}\sigma_{\circ}} = 
-\alpha_\beta{\rm res}\left(\prod_{n=1}^{\infty}\frac{\bar\kappa^{[-2n]}_1}{\kappa^{[-2n]}_{2}}\right) 
    \frac{\bar f^{[-2]} 
    \frac{\bfR_{(-)}}{\bfR_{(+)}}\frac{\kappa_1}{\bar\kappa_{2}}\left(\bar{\betheQ}^{4|\dot{\mu}}\right)^+}{x^{+\frac{L+2}2}B_{7}R_{5}\tilde\sigma_p\tilde\sigma_{\bullet}\tilde\sigma_{\circ}}
\eeq
where we removed the terms which are regular on the unit circle and thus do not contribute to the residue.

Let us identify the poles more precisely. As follows from the Baxter equation the meromorphic function ${\mathbb Q}_{1|\beta}^-$ can only have poles at $u=\theta_i$ for $i=1,\dots,N/2$, which in the $x$ plane results into poles at $x=z_i$ and $x=1/z_i$ for $i=1,\dots,N/2$. However, the factor $\kappa_1/\bar \kappa_2$ cancels the first half of them and introduce additional poles at $1/z_i$ for $i=N/2+1,\dots,N$ so in summary the poles are simply at $1/z_i$ with the complete range $i=1,\dots,N$. The same arguments apply to the the r.h.s of 
\beq \label{eq:MasslessBethe}
\boxed{
z_i^{L+2}=
\left[-\frac{1}{\alpha_\beta}\frac{\betheQ_{1|\mu}^-}{(\bar{\betheQ}^{4|\dot{\mu}})^{+}}
\frac{R_7 B_5}{B_1R_3}\right]\, \frac{\betheQ^+_4}{\betheQ^-_4}\,
\left[\frac{\Gamma_1\bar{\Gamma}_2}{\bar{\Gamma}_1\Gamma_2}\prod^{\infty}_{n\neq 0}\frac{\kappa^{[2n]}}{\bar{\kappa}^{[2n]}}\right]^{1/2} 
\frac{\sigma_{p}}{\tilde{\sigma}_{p}} \,
\frac{\sigma_{\circ}}{\tilde{\sigma}_{\circ}} \, \frac{\sigma_{\bullet}}{\tilde{\sigma}_{\bullet}}\frac{f^{[2]}}{\bar{f}^{[-2]}}\bigg|_{x=z_i}
}\;.
\eeq
To obtain the above expression from \eq{eq:ResidueEq}
we used the identity
\beqa
\prod_{n=1}^{\infty}\frac{\kappa^{[2n]}_1}{\bar{\kappa}^{[2n]}_{2}}\frac{\kappa^{[-2n]}_{2}}{\bar\kappa^{[-2n]}_1}&\propto&
\left(\frac{\bar \Gamma_1}{\Gamma_1}
\frac{\Gamma_2}{\bar \Gamma_2}
\prod_{n\neq 0}^{\infty}\frac{\kappa^{[2n]}}{\bar{\kappa}^{[2n]}}\right)^{1/2}\;.
\eeqa
We presented this equation in Section~\eqref{sec:ABBA} in the main text with the following identifications
\begin{equation}\la{masslessDF}
    \sigma^2_{\circ\circ}(z) = \frac{\sigma^2_{\circ}(z)}{\sigma^2_{\circ}(\frac{1}{z})}\prod_{n=1}^{\infty} \frac{\kappa^{[2n]}}{\bar{\kappa}^{[2n]}}\,,
    \quad
    \sigma_{\circ\bullet}(x) = \frac{\sigma_{\bullet}(x)}{\sigma_{\bullet}(\frac{1}{x})} \frac{f^{[2]}}{\bar{f}^{[-2]}}
    \quad
    \sigma_{\circ p}(x) = \frac{\sigma_{p}(x)}{\sigma_{p}(\frac{1}{x})}\;. 
\end{equation}

We will show that these expressions agree with the integral representations presented in \eqref{eq:DressingPhasesIntegralDef} in Appendix~\ref{app:DressingPhases}.

\paragraph{Unimodularity.}
We need to check the self-consistency of the equations
\eqref{eq:MasslessBethe}. In particular, we assumed that $|z_i|=1$, which then implies that the r.h.s. of \eq{eq:MasslessBethe} should be 
uni-modular. Whereas some terms are uni-modular that is far from obvious that the whole expression in the r.h.s. is.

After removing all clearly unimodular terms, we notice that only the first square brackets is a potential problem. Its unimodularity is equivalent to the following identity
\beq\label{eq:unimod1}
\frac{\mQ_{1|\mu}^-\bar\mQ_{1|\mu}^+}{
(\mQ^{4|\dot\mu})^-(\bar{\mQ}^{4|\dot\mu})^+}\frac{\mQ_7\mQ_5}{\mQ_1\mQ_3}
= |\alpha_\beta|^2
\;\;,\;\;u=\theta_k\;,\;k=1,\dots,N\;,
\eeq
which indeed holds for the case $N=4$ as we show in Appendix~\ref{App:Qrelations}, see in particular \eqref{relationNeeded}, by using general properties of the Baxter equation \eq{baxterQ2appendix}. 
Using \eqref{eq:unimod1} allows us to write the expression in the first square bracket in \eq{eq:MasslessBethe} in an explicitly unimodular form (by the price of introducing a sign ambiguity)
\beq\la{BAEmassless}
\pm \[
\frac{\bar \alpha_\mu}{\alpha_\mu}\frac{\betheQ_{1|\mu}^-
{\betheQ}^{4|\dot\mu-}
B_{7}R_{5}B_{1}R_{3}}{
\bar\betheQ_{1|\mu}^+
\bar{\betheQ}^{4|\dot \mu+}R_{7}B_{5}R_{1}B_{3}}
\]^{1/2}\;.
\eeq
\paragraph{Equivalence of the massless equation for different $\beta$'s.}
Finally, we see that there are two copies of the equation for $\beta=1$ and $\beta=3$, which could be over-constraining. 
As the dependence on $\beta$ enters only through the terms in the square bracket, the ratio of the two gives 
\beq\label{eq:UniModConsistency}
\frac{\betheQ_{1|1}^-
\bar\betheQ_{1|3}^+
{\betheQ}^{4|2-}
\bar{\betheQ}^{4|4+}
}{
\bar\betheQ_{1|1}^+
\betheQ_{1|3}^-
\bar{\betheQ}^{4|2+}
{\betheQ}^{4|4-}
}\bigg|_{u=\theta_i}
=\frac{\alpha_1}{\bar \alpha_1}
\frac{\bar \alpha_3}{\alpha_3}\,,
\eeq
as we show in the Appendix~\ref{App:Qrelations}, this is indeed the case and the l.h.s. does not depend at which $\theta_i$ the expression is evaluate, implying consistency between the $\beta=1$ and $\beta=3$ variant of the equation, see \eqref{eq:ConsistencyBaxter}.

\paragraph{Invariance w.r.t. ordering of massless roots.}
Finally, in order to restore the symmetry w.r.t. the ordering of $\theta's$ we notice that we can us \eq{QtoM}
\beqa
\[
\frac{
\betheQ_{1|\beta}^-
{\betheQ}^{4|\dot\beta-}
}{
\bar\betheQ_{1|\beta}^+
\bar{\betheQ}^{1|\beta+}
}
\]^{1/2}
\prod_{n=1}^{\infty}\frac{\kappa^{[2n]}_1}{\bar{\kappa}^{[2n]}_{2}}\frac{\kappa^{[-2n]}_{2}}{\bar\kappa^{[-2n]}_1}&\propto&
\left[
\frac{
\betheQ_{1|\beta}^-
{\betheQ}^{4|\dot\beta-}
}{
\bar\betheQ_{1|\beta}^+
\bar{\betheQ}^{4|\dot\beta+}
}
\frac{\bar \Gamma_1}{\Gamma_1}
\frac{\Gamma_2}{\bar \Gamma_2}
\prod_{n\neq 0}^{\infty}\frac{\kappa^{[2n]}}{\bar{\kappa}^{[2n]}}
\right]^{1/2}\\
&\propto&
\left[
\frac{
{\mathbb M}_{1|\beta}^-
{{\mathbb M}}^{4|\dot\beta-}
}{
\bar{\mathbb M}_{1|\beta}^+
\bar{{\mathbb M}}^{4|\dot\beta+}
}
\frac{\Gamma}{\bar \Gamma}
\prod_{n\neq 0}^{\infty}\frac{\kappa^{[2n]}}{\bar{\kappa}^{[2n]}}
\right]^{1/2}\;,
\eeqa
where the proportionality coefficient does not depend on $z_i$ at which the equation is evaluated. We see that written in this form the dependence on the ordering of $\theta's$ disappear, as ${\mathbb M}$ is invariant under the permutation.

\section{The Dressing Phases}\label{app:DressingPhases}
In this appendix, we discuss the properties of the dressing phases $\sig_{\bullet\bullet},\sig_{\bullet\circ},\sigma_{\circ\bullet}$ and $\sigma_{\circ\circ}$,
as well as $\sigma_{p\circ}$ and $\sigma_{p\bullet}$, appearing in the 
ABBA equations in Figure~\ref{fig:ABBAgeneral}
in more detail. We also presented an explicit integral representations of the phases, and their derivation from the equations found in Appendix~\ref{app:ABBADerivation}. Finally, we also gives the weak coupling expansion of the phases in Subsection~\ref{app:DressingPhasesWeakCoupling}.

\subsection{Solving the Crossing Equations}\label{app:SolvingCrossing}
Let us recall the expressions obtained in Appendix~\ref{app:ABBADerivation}, see \eqref{massiveDF} and \eq{masslessDF}. We repeat them here for convenience
\begin{equation}\label{eq:DressingPhasesDefApp}
\begin{split}
    \prod_{k=1}^{K_4}\sig_{\bullet\bullet}(x,x_{4,k}) &= \frac{\sigma_{\bullet}(x^+)}{\sigma_{\bullet}(x^-)}\,,
    \qquad\;\;\,
    \prod_{k=1}^{N}\sig_{\bullet\circ}(x,z_k) = \frac{\sigma_{\circ}(x^+)}{\sigma_{\circ}(x^-)}\,,
    \\
    \prod_{k=1}^{K_4}\sig^2_{\circ\bullet}(x,x_{4,k}) &= \frac{\sigma_{\bullet}(x)}{\sigma_{\bullet}(\frac{1}{x})}\frac{f^{[2]}}{\bar{f}^{[-2]}}\,,
    \quad
     \prod_{k=1}^{N}\sig^2_{\circ\circ}(x,z_k) = \frac{\sigma_{\circ}(x)}{\sigma_{\circ}(\frac{1}{x})}\left[\prod_{n\neq 0} \frac{\kappa^{[2n]}}{\bar{\kappa}^{[2n]}}\right]^{1/2},\\
         \prod_{k=1}^{\frac{4-N}2}\sig_{\bullet p}(x,u_{0,k}) &=\frac{\sigma_{p}(x^+)}{\sigma_{p}(x^-)}\,,\qquad\;\;\,
             \prod_{k=1}^{\frac{4-N}2}\sig^2_{\circ p}(x,u_{0,k}) =\frac{\sigma_{p}(x)}{\sigma_{p}(1/x)},
\end{split}
\end{equation}
where the building blocks $\sigma_{\bullet},\sigma_{\circ}$and $\sigma_{p}$ are constrained to be short-cut functions with constant asymptotics satisfying the crossing equations, see \eqref{eq:CrossingABBADerivation},
\begin{equation}\label{eq:CrossingSigmaAppendix}
    \tilde{\sigma}_{p} \sigma_{p} \propto p\,,
    \quad
    \tilde{\sigma}_{\bullet} \sigma_{\bullet} \propto f^{[2]}\bar{f}^{[-2]}\,,
    \quad
    \tilde{\sigma}_{\circ} \sigma_{\circ} \propto \prod^{\infty}_{n=1} \left( \frac{\kappa}{x^{N/2}}\right)^{[2n]} \left(\frac{\bar{\kappa}}{x^{N/2}}\right)^{[-2n]}\,.
\end{equation}

To find explicit expressions for \eqref{eq:DressingPhasesDefApp} we first need to solve \eqref{eq:CrossingSigmaAppendix}. To do so we decompose $\sigma_{\bullet}$ and $\sigma_{\circ}$ over the massive and massless roots
\begin{equation}\label{eq:BuildingBlocksDressing}
    \sigma_{\bullet}(x) = \prod_{k=1}^{K_4} \varsigma_{\bullet}(x,x_{4,k})\,,
    \quad
    \sigma_{\circ}(x) = \prod_{i=1}^{N} \varsigma_{B}(x,z_i)\varsigma_{\circ}(x,z_i)\,.
\end{equation}
We have introduced a factor $\varsigma_{B}$ which we will tune so that $\varsigma_{\circ}$ can be interpreted as the massless limit of $\varsigma_{\bullet}$. To be precise we set
\begin{equation}\label{eq:SigmaBDef1}
    \varsigma_{B}(z,w) \propto \exp \left(\frac{1}{2}\sum_{n\neq 0}\log(1-\frac{1}{z \,w^{[2n]}}) \right)\,,
    \quad
    \prod_{i=1}^{N}\varsigma^2_{B}(x,z_i)\varsigma^2_{B}\(\frac{1}{x},z_i\) \propto  \prod_{n\neq 0} 
    \left(\frac{\kappa\bar{\kappa}}{x^{N}}\right)^{[2n]}\,,
\end{equation}
so that the building blocks defined in \eqref{eq:BuildingBlocksDressing} satisfy
\begin{equation}\label{eq:CrossingZeta}
\begin{split}
\tilde{\varsigma}_{\bullet}(x,y) \, \varsigma_{\bullet}(x,y) &\propto \prod_{n=1}^{\infty}\frac{1-\frac{1}{x^{[2n]}y^-}}{1-\frac{1}{x^{[2n]}y^+}}\frac{1-\frac{1}{x^{[-2n]}y^+}}{1-\frac{1}{x^{[-2n]}y^-}} \,,
    \\
    \tilde{\varsigma}^2_{\circ}(x,z) \varsigma^2_{\circ}(x,z) &\propto \prod_{n=1}^{\infty}\frac{1-\frac{z}{x^{[2n]}}}{1-\frac{1}{x^{[2n]}\,z}}\frac{1-\frac{1}{x^{[-2n]}z}}{1-\frac{z}{x^{[-2n]}}}\,, 
\end{split}
\end{equation}
where the proportionality is up to any function of $y^{\pm}$ or $z$. We can now indeed see that the second equation is obtained from the first by taking a massless limit $y^{\pm} \mapsto z^{\pm 1}$. The solution of \eqref{eq:CrossingZeta} is given by the standard dressing phase that appears in the spectral problem of local operators, see for example \cite{Volin:2009uv,Vieira:2010kb} for details. The solution is
\begin{equation}
   \log \varsigma_{\bullet}(x,y) = \chi(x,y^+)-\chi(x,y^-)\,,
    \quad
    \log \varsigma_{\circ}(x,z) = \frac{1}{2}\chi(x,z)-\frac{1}{2}\chi(x,\tfrac{1}{z})\,,
\end{equation}
where
\beq\label{eq:AppChiPhase}
\chi(x,y)=
-\oint \frac{dz}{2\pi \ii} 
    \frac{1}{x-z}
    \oint \frac{dw}{2\pi \ii} \frac{1}{y-w}\log\left( \frac{\Gamma\left(1+\ii u_{z}-\ii u_{w}\right)}{\Gamma\left(1-\ii u_{z}+\ii u_{w}\right)}\right)\,,
\eeq
where $u_{x} = g(x+\frac{1}{x})$. This identifies $\frac{\varsigma_{\bullet}(x^+,y)}{\varsigma_{\bullet}(x^-,y)}$ as the famous BES-dressing phase \cite{Beisert:2006ez}.

\paragraph{The ``boundary'' phase.}
We refer to the part of the phase coming from $\varsigma_{B}$ as a ``boundary'' phase since it looks like a scattering phase against some stationary integrable defect rather than another pseudo-particle.
From the definition of $\varsigma_{B}$ we find that it satisfies
\begin{equation}\label{eq:CrossingSigmaB}
    \tilde{\varsigma}^2_{B}(x,y)\, \varsigma^2_{B}(x,y) \propto \prod_{n\neq 0} (u_x-u_y+\ii n) \propto \frac{\sinh{\pi(u_x-u_y)}}{\pi\left(u_x-u_y\right)}\,,
\end{equation}
which is solved by
\begin{equation}\label{eq:SigmaBDef2}
    \varsigma_{B}(x,y) = \exp\left(\frac{1}{2}\oint \frac{dz}{2\pi \ii} \frac{1}{x-z}\log\frac{\sinh \pi (u_z-u_y)}{\pi(u_z-u_y)} \right)\,.
\end{equation}
As also remarked in the main text, $\varsigma^2_{B}(x,-x)$ is exactly the boundary piece that appeared when studying the cusp anomalous dimension of Wilson lines using the Thermodynamic Bethe Approach in both $\mathcal{N}=4$ \cite{Correa:2012hh,Drukker:2012de} and in ABJM \cite{Correa:2023lsm}. In analogy with \eqref{eq:AppChiPhase},
let us define
\begin{equation}
    \chi_B(x,y) = \oint \frac{dz}{2\pi \ii}\frac{1}{x-z} \log \frac{\sinh{\pi (u_z-u_y)}}{\pi (u_z-u_y)}\,,
    \quad
    \varsigma_{B}(x,z) = \exp\left(\frac{1}{2}\chi_B(x,z)\right)\,.
\end{equation}

\paragraph{The periodic phase.}

We can solve the crossing equation for $\varsigma_p$ in exactly the same way as for $\varsigma_B$, we find
\begin{equation}\la{sigmapdef}
    \varsigma_p(x) = \exp\left(\oint \frac{dz}{2\pi \ii}\frac{1}{x-z} \log p(u_z)\right)\;.
\end{equation}
as we discussed in Appendix~\ref{app:ABBADerivation} for HTs we have $N=4$ and $p\equiv 1$, rendering $\sigma_p$ trivial. However, for $N=2$ we have $p = \cosh{\pi(u-u_0)}$. In this case, we can relate the phase $\sigma_p(x)$ to the previously introduced $\varsigma_{B}(x,y)$. Perhaps the simplest way to do so is to see that
\begin{equation}
    \tilde{\varsigma}_p(x,x_0^\pm)\,\varsigma_p(x,x_0^\pm) \propto (u-u_0)\, \times \frac{\sinh{\pi(u-u_0)}}{\pi (u-u_0)}\,,
\end{equation}
where $x^\pm_0 = x(u_0\pm \frac{\ii}{2})$, this immediately fixes
\begin{equation}
    \varsigma_p(x,x_0^{\pm}) = \(1-\frac{1}{x\,x_0}\)\varsigma_B(x,x_0)\,. 
\end{equation}

\paragraph{Infinite products.}
It remains only to study the additional infinite products that appear in $\sig_{\circ\bullet}$ and $\sig_{\circ\circ}$, see \eqref{eq:DressingPhasesDefApp}. Using the product form of $\varsigma_B$, i.e \eqref{eq:SigmaBDef1}, it follows that
\begin{equation}\label{eq:fToSigmaB}
    \frac{f^{[2]}}{\bar{f}^{[-2]}} = \prod_{n\neq 0}\prod_{k=1}^{K_4} \frac{1-\frac{1}{x^{[2n]} x^-_{4,k}}}{1-\frac{1}{x^{[2n]}x^+_{4,k}}} = \prod_{k=1}^{K_4}\frac{\varsigma^2_{B}(x^-_{4,k},x)}{\varsigma^2_{B}(x^+_{4,k},x)}\;,
\end{equation}
and
\begin{equation}\label{eq:kappaToSigmaB}
    \prod_{n\neq 0} \frac{\kappa^{[2n]}}{\bar{\kappa}^{[2n]}} = \prod_{n\neq 0}\prod_{i=1}^{N}\frac{1-\frac{z_i}{x^{[2n]}}}{1-\frac{1}{z_ix^{[2n]}}} = \prod_{i=1}^{N} \frac{\varsigma^2_B(\frac{1}{z_i},x)}{\varsigma^2_B(z_i,x)}\;.
\end{equation}

\paragraph{Integral representation.}
Upon now combining \eqref{eq:BuildingBlocksDressing}, \eqref{eq:fToSigmaB} and \eqref{eq:kappaToSigmaB} we finally deduce the expressions for the dressing phases presented in \eqref{eq:DressingPhasesIntegralDef} which we repeat here for convenience
\begin{equation}
\begin{split}
   \sig_{\bullet\bullet}(x,y) &= \frac{\varsigma_{\bullet}(x^+,y)}{\varsigma_{\bullet}(x^-,y)}\,,
   \\
   \sig_{\bullet\circ}(x,z) &= \frac{\varsigma_{\circ}(x^+,z)}{\varsigma_{\circ}(x^-,z)}\frac{\varsigma_{B}(x^+,z)}{\varsigma_{B}(x^-,z)}\,,\\
    \sig^2_{\circ\bullet}(z,x) &= \frac{\varsigma_{\bullet}(z,x)}{\varsigma_{\bullet}(\frac{1}{z},x)}\frac{\varsigma^2_{B}(x^-,z)}{\varsigma^2_{B}(x^+,z)}\,,
   \\
   \sig^2_{\circ\circ}(z,w) &= \frac{\varsigma_{\circ}(z,w)}{\varsigma_{\circ}(\frac{1}{z},w)}\frac{\varsigma_{B}(z,w)}{\varsigma_{B}(\frac{1}{z},w)}\frac{\varsigma_B(\frac{1}{w},z)}{\varsigma_B(w,z)}\,,\\
   \sig_{\bullet p}(x,y) &= \frac{\varsigma_p(x^+,y)}{\varsigma_p(x^-,y)}\,, \\
   \sig^2_{\circ p}(z,y) &= \frac{\varsigma_p(z,y)}{\varsigma_p(\frac{1}{z},y)}\,.
\end{split}
\end{equation}

\paragraph{Unitarity.}
From its definition, it follows that $\chi(x,y) = -\chi(y,x)$. Thus, we have the following unitarity properties
\begin{align}
    \sig^2_{\bullet\bullet}(x,y) \sig^2_{\bullet\bullet}(y,x) = 1\,,
    \quad
    \sig^2_{\bullet\circ}(x,y) \sig^2_{\circ\bullet}(y,x) = 1\,,
    \quad
    \sig^2_{\circ\circ}(x,y)
    \sig^2_{\circ\circ}(y,x) = 1\,.
\end{align}

\subsection{Weak Coupling Expansion}\label{app:DressingPhasesWeakCoupling}
When solving the ABBA perturbatively, we need to be able to expand the dressing phases above perturbatively for $g\simeq 0$. To do so, one can use that $u=g(x+\frac{1}{x})$ and expand for small $g$ while keeping $x$ fixed and thereafter evaluate the contour integrals by picking the residues at $x=0$. We find
\begin{equation}
\begin{split}
    &\log \varsigma_{\circ}(x,z) = \left(\frac{z}{x^2}-\frac{1}{x^2 z}-\frac{z^2}{x}+\frac{1}{x z^2}\right)(\ii g)^3\zeta_3 \\
   &+\bigg(\frac{z}{x^4}-\frac{1}{x^4 z}-\frac{2 z^2}{x^3}+\frac{2}{x^3 z^2}+\frac{2
   z^3}{x^2}-\frac{2}{x^2 z^3}+\frac{10 z}{x^2}-\frac{10}{x^2 z}-\frac{z^4}{x}+\frac{1}{x
   z^4}-\frac{10 z^2}{x}+\frac{10}{x z^2} \bigg) (\ii g)^5 \zeta_5
\end{split}
\end{equation}
and 
\begin{equation}
\begin{split}
    \log &\varsigma_{B}(x,z) = \left(-\frac{1}{2 x^2}+\frac{z}{x}+\frac{1}{x z}\right)(\ii g)^2 \zeta_2 \\
    &+\bigg(-\frac{1}{4 x^4}+\frac{z}{x^3}+\frac{1}{x^3 z}-\frac{3 z^2}{2 x^2}-\frac{3}{2 x^2
   z^2}-\frac{4}{x^2}+\frac{z^3}{x}+\frac{1}{x z^3}+\frac{6 z}{x}+\frac{6}{x z}\bigg)(\ii g)^4 \zeta_4\,. 
\end{split}
\end{equation}
Upon combining these two expressions we find
\begin{equation}
\begin{split}
\log \sig^2_{\circ\circ}(x,z) &= \left(\frac{x^2}{2}-\frac{1}{2 x^2}-\frac{2 x}{z}+\frac{2 z}{x}-\frac{z^2}{2}+\frac{1}{2 z^2}\right)(\ii g)^2\zeta_2\\ 
   &+\left(-x^2 z+\frac{x^2}{z}+\frac{z}{x^2}-\frac{1}{x^2 z}+x z^2-\frac{x}{z^2}-\frac{z^2}{x}+\frac{1}{x z^2}\right)(\ii g)^3 \zeta_3 \\
   &+\bigg(\frac{x^4}{4}-\frac{1}{4 x^4}-\frac{2 x^3}{z}+\frac{2 z}{x^3}+\frac{3 x^2}{z^2}-\frac{3 z^2}{x^2}+4
   x^2-\frac{4}{x^2}-\frac{2 x}{z^3}\\
   &\quad\quad\quad\quad+\frac{2 z^3}{x}-\frac{12 x}{z}+\frac{12
   z}{x}-\frac{z^4}{4}+\frac{1}{4 z^4}-4 z^2+\frac{4}{z^2}\bigg)(\ii g)^4 \zeta_4 \\
   &+\dots\;.
\end{split}
\end{equation}
To find the piece $\sig_{\circ\bullet}(x,y)$ we use that $y$ is always associated to a massive momentum carrying root, so that $y=\frac{v+\sqrt{v+2g}\sqrt{v-2g}}{2g}$ and we assume $v\simeq \mathcal{O}(1)$. This gives
\begin{equation}
    \begin{split}
        \log \varsigma_\bullet(x,y) &= \frac{8 \zeta_3}{(1+4 v^2)x^2}g^4 + \frac{32 \left(12 v^2-1\right)\zeta_3}{\left(4 v^2+1\right)^3 x^2} g^6 -\frac{8 \left(10 x^2+1\right)\zeta_5}{\left(4 v^2+1\right) x^4} g^6 +\dots
    \end{split}
\end{equation}
and the full phase as
\begin{equation}
    \log \sig^2_{\circ\bullet}(x,y) =-\frac{8 i g^3 \left(x+1/x\right) \zeta_2}{4 v^2+1}+g^4 \left(-\frac{8 \left(x^4-1\right) \zeta_3}{\left(4 v^2+1\right) x^2}+\frac{32 i v
   \zeta_2}{\left(4 v^2+1\right)^2}\right) +\dots \;.
\end{equation}

The phases $\sig_{\bullet\bullet}(x,y)$ and $\sig_{\bullet\circ}(x,z)$ appear in the massive Bethe equations, in this case we should expand them for $x\simeq \frac{u}{g}+\dots,y\simeq \frac{v}{g}+\dots,z\simeq \mathcal{O}(1)$. The leading terms are
\begin{align}
    \log \sig_{\bullet \bullet}(x,y) &= 4\ii\zeta_3 \frac{(u-v)(u\, v-\frac{1}{4})}{(u^2+\frac{1}{4})^2(v^2+\frac{1}{4})^2}g^6 + \dots\,,
    \\
    \log \sig_{\bullet \circ}(x,z) &= 4 \ii \zeta_2 \frac{z+\frac{1}{z}}{4 u^2+1} g^3 + \dots 
\end{align}
with $u=g(x+\frac{1}{x}),v=g(y+\frac{1}{y})$.

Finally we have from \eqref{sigmapdef}
\begin{equation}
    \log \varsigma_{p}(x,y) = \zeta_2 g^2 \left(\frac{1}{ x^2}-2\frac{y+\frac{1}{y}}{x}\right)+\mathcal{O}(g^4)\,,
\end{equation}
so that 
\begin{align}
    \log \sig_{\bullet p}(x,y) &= g^3\frac{6 i \left(y+\frac{1}{y}\right)}{u^2+\frac{1}{4}}\zeta_2+\dots\;, \\
    \log \sig^2_{\circ p}(x,y) &=g^2\left(-3 x^2+6 y x+\frac{6 x}{y}-\frac{6 y}{x}-\frac{6}{y x}+\frac{3}{x^2}\right)\zeta_2 \;.
\end{align}
We conclude that all phases are suppressed at weak coupling.

\section{Solving the Q-System}\label{app:StartingUpNumerics}

In this appendix we provide additional details on how to reconstruct $\bP_a$ upon solving the ABBA, a procedure we discussed briefly in Section~\ref{subsec:StartingUp}. $\bP_{a}$ are significant since they are the input of the numerical algorithm of \cite{Gromov:2015wca}. To be precise, they take the form
\begin{equation}\label{eq:PAnsatzNumerics}
    \bP_{a} = \sum_{n=-{\tilde{M}}_{a}}^{\infty} \frac{\hat{c}_{a,n}}{x^{n}}\,.
\end{equation}
The primary objective is thus to deduce $\hat{c}_{a,n}$ from ABBA.

We use the $\bP \mu$ system \eqref{eq:PMu} which we repeat here for convenience
\begin{equation}\label{eq:PMuStartingNumerics}
    \tilde{\bP}_a = \mu_{ab}\bP^{b}\,,
    \quad
    \tilde{\bP}^{a} = \mu^{ab}\bP_{b} \,,
    \quad
    \quad
    \mu^{[2]}_{ab} = (\delta_{a}^{c}-\bP_{a}\bP^{c}) \mu_{cd}(\delta_{b}^{d}-\bP^{d}\bP_{b})\,,
    \quad
    \tilde{\mu}_{ab} = \mu_{ab}^{[2]}\,,
\end{equation}
which implies
\begin{equation}\label{eq:MuFiniteDifference}
    \mu^{[2]}_{ab}-\mu_{ab} = \bP_{a}\tilde{\bP}_b - \tilde{\bP}_{a}\bP_{b}\,.
\end{equation}

A powerful algorithm to solve these equations to any order for local operators was developed in \cite{Marboe:2018ugv}. However, the method of \cite{Marboe:2018ugv} relies on $Q_{a|i}$ being as a well-behaved rational function of $u$ times so-called $\eta$-functions introduced in \cite{Leurent:2013mr}. This is not true in our setting in general, and hence that method is not applicable. We will instead use a natural ansatz coming from our derivation of the ABBA that circumvents the need to introduce $Q_{a|i}$.

In Appendix~\ref{subapp:FindingMuOmega}, we found an analytic form of $\mu_{12}$ valid in the ABBA approximation. We can rerun the same argumentation to find $\mu_{ab}$
\begin{equation}\label{eq:MuAnsatzABBA}
    \mu^+_{ab} =\pf \, \betheQ_{ab}  f^+ \bar{f}^{-} \prod_{n=0}^{\infty} \left(\frac{\kappa}{x^{N/2}}\right)^{[2n+1]} \left( \frac{\bar{\kappa}}{x^{N/2}}\right)^{[-2n-1]}
\end{equation}
where we have introduced a constant $\pf$ which we importantly allow to scale with arbitrary powers of $g$ while all other terms in \eqref{eq:MuAnsatzABBA} must have a regular expansions, i.e $1+g+g^2\dots$. The new functions $\betheQ_{ab}$ are polynomials in $u$ and their degree can be inferred from the degree of $Q_{ab|13}$. For example, in the the simple sector of operators with conformal spin ($J_2=J_3=0, S_1 = -J_1+1+\omega$) we find 
\begin{equation}
    \{\betheQ_{12},\betheQ_{13},\betheQ_{14},\betheQ_{23},\betheQ_{24},\betheQ_{34}\} \sim \{u^{0},u^{J_1-1},u^{J_1},u^{J_1},u^{J_1+1},u^{2J+1}\}\,. 
\end{equation}
Just as for $\mu_{ab}$ we can repeat the arguments leading to the the expressions for $\bP_{1}$ and $\bP^{4}$ to find the remaining $\bP$-functions. This results in the following expressions
\begin{equation}\label{eq:PAnsatzABBA}
\begin{split}
    &\bP_{a} = \frac{\bp_{a}}{(g x)^{\Leff/2+1}}\, \sigma_{\bullet} \sigma_{\circ}\sigma_{p} \,,
    \quad
    \bP^{a} = \frac{\bp^{a}}{(g x)^{\Leff/2+1}}\, \sigma_{\bullet} \sigma_{\circ}\sigma_{p} \,,
\end{split}
\end{equation}
and $\bp_a,\bp^{a}$ are finite Laurent-series in the Zhukovsky variable $x$, we give their explicit form below in \eqref{eq:ScalingSmallPs} and \eqref{eq:ScalingSmallPHs}

We can now take the parametrisation in \eqref{eq:MuAnsatzABBA} and \eqref{eq:PAnsatzABBA} and use them in \eqref{eq:PMuStartingNumerics} to obtain
\begin{equation}\label{eq:TildePABBA}
    \tilde{\bp}_{a} = p\frac{\bfB_{(+)}}{\bfB_{(-)}}\frac{\pf\,\kappa}{x^{\Leff+2+N/2}}\left(\sigma_{\bullet} \sigma_{\circ}\sigma_{p}\right)^2\betheQ^{-}_{ab}\,\bp^{b}
\end{equation}
and
\begin{equation}\label{eq:FiniteDifferenceQABBA}
    \left(\frac{\bfB_{(-)}}{\bfB_{(+)}}\right)^{2}\frac{\bar{\kappa}}{\kappa}\betheQ^+_{ab} = \left(\delta^{c}_a-\bP_{a}\bP^{c} \right)\betheQ^-_{cd}\left(\delta^{d}_b-\bP^{d}\bP_{b} \right)\;.
\end{equation}
After taking an appropriate ansatz for $\bp_{a}$ and $\bp^{a}$ we can solve \eqref{eq:TildePABBA}, and \eqref{eq:FiniteDifferenceQABBA} order by order in $g$. Furthermore, in all examples we encountered these equations fixes all massive roots self-consistently with a perfect agreement with the results obtained from ABBA.  Thus in practice, these equations need only be supplemented with $z_i$ which enters through the explicit $\kappa$ dependence.

\paragraph{Weak Coupling Scaling of the Ansatz.}
The reason of the overall prefactor of $(x g)^{\Leff/2+1}$ in \eqref{eq:PAnsatzABBA} is to ensure that all coefficients in $\bp_{a}$ have a manageable $g$ scaling. We found for all examples considered in this paper, i.e $L=2,4$ for HT$_2$ that
\begin{align}\label{eq:ScalingSmallPs}
    \bp_{a} &= \sum_{m=0}^{\tilde{m}_{a}} (g x)^{m} d_{a,m} + \sum_{m=1}^{\tilde{n}_{a}} \left(\frac{g}{x} \right)^{m} c_{a,m}\,,
    \quad
    \begin{cases}
    \tilde{m}_{a} = -\tilde{M}_{a}+\Leff/2+1\,,
    \\
    \tilde{n}_a = -\tilde{M}_a+\Sigma-\Leff/2-N/2
    \end{cases} \\
    \bp^{a} &= \sum_{m=0}^{\tilde{m}^{\star}_{a}} (g x)^{m} d^{\star}_{a,m} + \sum_{m=1}^{\tilde{n}^{\star}_{a}} \left(\frac{g}{x} \right)^{m} c^{\star}_{a,m}\,,
    \quad
    \begin{cases}
    \tilde{m}_{a} = \tilde{M}_{a}+\Leff/2\,,
    \\
    \tilde{n}_a = \tilde{M}_a+\Sigma-\Leff/2-N/2-1
    \end{cases}\label{eq:ScalingSmallPHs}
\end{align}
where the explicit powers of $g$ have been added in order to ensure that the coefficients admit an expansion 
\begin{equation}
    d_{a,m} \sim \sum_{n=0}^{\infty} d_{a,m,n}\,g^{n}\,,
    \quad
    c_{a,m} \sim \sum_{n=0}^{\infty} c_{a,m,n}\,g^{n}\,,
\end{equation}
and the same is true for $d^{\star},c^{\star}$.

We can now compare powers of $g$ in \eqref{eq:TildePABBA} and see that in general we should have $\pf \sim \mathcal{O}(g^{-\Leff+2-N/2})$.

\subsection{Fitting Numerical Parameters} Finally, having obtained $\bp_{a}$ we now want to find $\hat{c}_{a,n}$ in \eqref{eq:PAnsatzNumerics}. To do so efficiently it is convenient to use both $\bP_{a}$ and $\tilde{\bP}_{a}$. Let us rewrite the numerical ansatz in the same form as \eqref{eq:ScalingSmallPs}
\begin{equation} \label{eq:PAnsatzNumericsV2}
    \bP_a = \frac{1}{(g\, x)^{\Leff/2+1}}\left(\sum_{n=0}^{\tilde{m}_{a}} \hat{d}_{a,n} (g\, x)^{n} + \sum_{n=1}^{\infty} \hat{c}_{a,n} \left(\frac{g}{x}\right)^{n}\right)\,,
\end{equation}
where $\hat{d}_{a,m},\hat{c}_{a,m} \sim \mathcal{O}(g^{0})$. Expanding for $x\simeq \frac{u}{g}+\dots$ and comparing with our ABBA ansatz we are able to fix the leading coefficients up to wrapping order but clearly this procedure will not fix most of the $\hat{c}_{a,n}$ since they are suppressed by explicit powers of $g$. To fix the entire tower of $\hat{c}_{a,n}$ we instead consider $\tilde{\bP}_{a}$ and expand first for small $g$ and then small $u$. For example, to leading order
\begin{equation}\label{eq:TildePC}
    \tilde{\bP}_{a} = \frac{u^{\Leff/2+1}}{g^{\Leff+2}}\left(\hat{d}_{a,n,0}+\sum_{n=1}^{\infty} \hat{c}_{a,n,0}\,u^{n}\right) + \mathcal{O}\(\frac{1}{g^{\Leff+1}}\)\,,
\end{equation}
and the higher orders are easily available from 
\eqref{eq:PAnsatzNumericsV2}. 

At the same time, from the crossing equations \eqref{eq:CrossingSigmaAppendix} and the explicit expression for $\bP$ in the ABBA regime we find
\begin{equation}\label{eq:TildePABBA}
    \tilde{\bP}_a = \left(\frac{x}{g}\right)^{\Leff/2+1}\frac{f^{[2]} \bar{f}^{[-2]}}{\sigma_{\bullet} \sigma_{\circ}\sigma_p}\,\left(\prod_{n=1}^{\infty} \left(\frac{\kappa}{x^2}\right)^{[2n]}\left(\frac{\bar{\kappa}}{x^2}\right)^{[-2n]} \right)\, \tilde{\bp}_{a}\;.
\end{equation}
It is straightforward to expand $\sigma_{\bullet},\sigma_{\circ}$ and $\sigma_{p}$ at weak coupling, see  Appendix~\ref{app:DressingPhases} for the explicit expressions. The infinite product over $\kappa$ and $\bar{\kappa}$ is divergent, but can be regularised as
\begin{equation}
\begin{split}
    \prod_{n=1}^{\infty} \left(\frac{\kappa}{x^{N/2}}\right)^{[2n]}\left(\frac{\bar{\kappa}}{x^{N/2}}\right)^{[-2n]} &\propto \exp\left(\int du \sum_{n=1}^{\infty} \partial_{u} \log \left(\frac{\kappa}{x^{N/2}}\right)^{[2n]} \left(\frac{\bar{\kappa}}{x^{N/2}}\right)^{[-2n]}  \right)\\
    &=\left(\frac{\sinh\(\pi u\)}{u}\right)^{N/2}+\mathcal{O}(g)\,,
\end{split}
\end{equation}
and similar for higher orders. This expression can be readily expanded for small $u$ and used to match \eqref{eq:TildePC} against \eqref{eq:TildePABBA} fixing an infinite number of coefficients.

\section{The Baxter Equation and its Solutions}\la{App:Qrelations}
In this appendix, we describe key relations for functions satisfying the Baxter equation 
\begin{equation}\label{eq:BaxterQQ}
    Q^{[2]}_{\beta}\,\betheQ^{-}_1 \betheQ^-_3 \betheQ^+_{\theta_2} + \mathbb{T}\,Q_{\beta} + Q^{[-2]}_{\beta} \,\betheQ^+_{1} \betheQ_{3}^+ \betheQ^-_{\theta_1} = 0\,,
    \quad
    \beta=2,\tilde{2}\,,
\end{equation}
and with pure asymptotics 
\begin{equation}\label{eq:AsymptoticsApp}
\begin{split}
Q_2 \simeq
u^{\frac{+(\Delta-S_2)-J_2+J_3+K_4-1+i\sum\limits_{k}^{N/2}(\theta_k-\theta_{N/2+k})}{2}}\;,
\\
Q_{\tilde{2}} \simeq
u^{\frac{-(\Delta-S_2)-J_2+J_3+K_4-1+i\sum\limits_{k}^{N/2}(\theta_k-\theta_{N/2+k})}{2}}
\;.
\end{split}
\end{equation}
For convenience we recall the notation used in \eqref{eq:BaxterQQ}:
\begin{equation}\label{eq:QsApp}
\begin{split}
\betheQ_1&=\prod_{k=1}^{K_1}(u-u_{1,k})\,,
\quad
\betheQ_3=\prod_{k=1}^{K_3}(u-u_{3,k})\,,
\\
\betheQ_{\theta_1}&=\prod_{k=1}^{N/2}(u-\theta_{k})\,,
\quad\;
\betheQ_{\theta_2}=\prod_{k=1}^{N/2}(u-\theta_{k+N/2})\,,
\end{split}
\end{equation}
and we will assume that $\mathbb{Q}_1,\mathbb{Q}_3,\mathbb{Q}_{\theta_1}$ and $\mathbb{Q}_{\theta_2}$ are real functions and $N$ is always even. We furthemore recall that there exist another Baxter equation for $Q_{6}$ and $Q_{\tilde{6}}$ with $\betheQ_{1}\betheQ_{3}$ replaced with  $\mathbb{Q}_5\mathbb{Q}_7$ and asymptotics
\beq\label{eq:AsymptoticsApp6}
\begin{split}
Q_6 &\simeq
u^{\frac{+(\Delta+S_2)-J_2-J_3+K_4-1+i\sum\limits_{k}^{N/2}(\theta_k-\theta_{N/2+k})}{2}}\;,\\
Q_{\tilde{6}} &\simeq
u^{\frac{-(\Delta+S_2)-J_2-J_3+K_4-1+i\sum\limits_{k}^{N/2}(\theta_k-\theta_{N/2+k})}{2}}
\;.
\end{split}
\eeq
see Section~\ref{sec:ABBA}. Since the analysis of the two Baxter equations are equivalent up to notation, we will focus on \eqref{eq:BaxterQQ}.

An important consequence of \eqref{eq:BaxterQQ} is the relation
\begin{equation}\label{eq:GeneralQQApp}
    \frac{\Wr^+(Q_2,Q_{\tilde{2}})}{\Wr^-(Q_2,Q_{\tilde{2}})} = \frac{\betheQ^+_1\betheQ^+_3}{\betheQ^-_1\betheQ^-_3} \frac{\betheQ^-_{\theta_1}}{\betheQ^+_{\theta_2}}\,,
\end{equation}
with $\Wr^{\pm}(f,g) = f^{[1\pm 1]}g^{[-1\pm 1]}-f^{[-1\pm 1]}g^{[1\pm 1]}$.

Simply defining $Q_{a}$ through \eqref{eq:BaxterQQ} and \eqref{eq:AsymptoticsApp} does not define the functions uniquely. To do so, we must also prescribe additional analytic properties. The most commonly used analytic constraint in this paper is to require Q-functions to be upper-half-plane analytic. We follow the notation in the main text and write $\betheQ_{2} = Q_2,\betheQ_{\tilde{2}}= Q_{\tilde{2}}$ with $\betheQ_{\beta}\,,\,\beta=2,\tilde{2}$ upper half-plane-analytic functions.

\subsection{Wronskian Identities}
Let us denote by $\theta^{(1)}$ any $\theta_{i\leq N/2}$ and by $\theta^{(2)}$ any $\theta_{i > N/2}$. $\betheQ_{\beta}$ cannot, by definition, have poles in the upper-half-plane, but from Baxter it follows that they can have poles at $\theta^{(1)}-\ii n - \frac{\ii}{2},n=0,1\dots$. Using the analytic properties of $\betheQ_{\beta}$ we can solve \eqref{eq:GeneralQQApp} to obtain the following Wronskian identity:
\beq\la{wronskianQ}
{\mathbb Q}_{2}^+
{\mathbb Q}_{\tilde{2}}^-
-
{\mathbb Q}_{\tilde{2}}^+
{\mathbb Q}_{2}^-
=w_{2\tilde{2}}\,
\frac{\Gamma_1}
{\Gamma_2^{++}}{\mathbb Q}_{1}{\mathbb Q}_{3}\,,
\eeq
where $w_{13}$ is a constant needed to ensure compatability with \eqref{eq:AsymptoticsApp} and \eqref{eq:QsApp}. We recall that
\begin{equation}
    \Gamma_1 = \prod_{k=1}^{N/2}\Gamma(-\ii u+\ii \theta_k)\,,
    \quad
    \Gamma_{2} = \prod_{k=1}^{N/2}\Gamma(-\ii u+\ii \theta_{k+N/2})\,.
\end{equation}
At large $u$ we have  $\Gamma_1/\Gamma_2^{++}\simeq (-i u)^{-N/2+i\sum_k(\theta_k-\theta_{K+k})}$,
so one can determine the constant $w_{13}$ explicitly to be $(-i)^{-\ii \sum_{k=1}^{N/2}\left(\theta_k - \theta_{N/2+k}\right)+N/2+1} (S_2-\Delta )$.

As $\Gamma_2$ has poles at $i\theta^{(2)}-i n$ with $n=0,1,\dots$ we have from \eqref{wronskianQ}
\beq
\frac{{\mathbb Q}_{2}(\theta^{(2)}-\tfrac{i}{2}-i n)}{
{\mathbb Q}_{\tilde{2}}(\theta^{(2)}-\tfrac{i}{2}-i n)
}
=
\frac{{\mathbb Q}_{2}(\theta^{(2)}-\tfrac{i}{2}-i (n+1))}
{{\mathbb Q}_{\tilde{2}}(\theta^{(2)}-\tfrac{i}{2}-i (n+1))}\;\;,\;\;n=0,1,\dots
\eeq

The complex conjugate $\bar{\mQ}_{\beta}$  does not satisfy the same Baxter equation as $\mQ_{\beta}$, the correct equations instead have the two groups of $\theta$ interchanged. This prompts us to introduce 
\beq
\hat\mQ_{\beta}=(-i)^{-i \sum_{k=1}^{N/2}\left(\theta_{k}-\theta _{N/2+k}\right)}
\bar\mQ_{\beta}\frac{\Gamma_1^+}{\Gamma_2^+}\;,
\eeq
which does satisfies exactly the same Baxter equation as $\betheQ_{\beta}$, i.e \eqref{eq:BaxterQQ}. It furthermore satisfies the Wronskian identity
\begin{equation}
\hat{\betheQ}^+_{2}\hat{\betheQ}_{\tilde{2}}^- - \hat{\betheQ}^+_{\tilde{2}}\hat{\betheQ}_{2}^- = -\bar{w}_{13}\left(\frac{\bar{\Gamma}_1\Gamma^{[2]}_1}{\bar{\Gamma}_2^{[-2]}\Gamma_{2}}\right) \frac{\Gamma_1}{\Gamma_2^{[2]}}\betheQ_1\betheQ_3\,,
\end{equation}
where the expression within parentheses is a periodic function as expected.

Using both $\betheQ$ and $\hat{\betheQ}$ we can build 
\beq\la{wronskianQbar}
{\mathbb Q}_{\alpha}^+
\hat{\mathbb Q}_{\beta}^-
-
\hat{\mathbb Q}_{\beta}^+
{\mathbb Q}_{\alpha}^-
= P_{\alpha,\beta}
\frac{\Gamma_1}
{\Gamma_2^{++}}{\mathbb Q}_{1}{\mathbb Q}_{3}\;,
\eeq
where $P_{\alpha,\beta}$ is a periodic function.
Written in terms of the complex conjugate function we get
\beq\la{wronskianQbar2}
{\mathbb Q}_{\alpha}^+
\bar{\mathbb Q}_{\beta}^-
\mQ_{\theta_2}
-
\bar{\mathbb Q}_{\beta}^+
{\mathbb Q}_{\alpha}^-
\mQ_{\theta_1}
=(-1)^{N/2}  P_{\alpha,\beta}
{\mathbb Q}_{1}{\mathbb Q}_{3}\;,
\eeq
we see that $P_{\alpha,\beta}$ only has simple poles at $u=\theta^{(1)}+i {\mathbb Z}$. We also notice that for under the complex conjugation $\bar P_{\alpha,\beta}=P_{\beta,\alpha}$, so we can write
\beq\la{Pansz}
P_{\alpha,\beta}=\frac{
\sum_{n=0}^{N/2} w^{(n)}_{\alpha,\beta}e^{2\pi n u}
}{\prod_{k=1}^{N/2}(1-e^{2\pi (u-\theta_k)})}\,,
\eeq
where $w^{(n)}_{\alpha,\beta}$ are some constants. These constant can be constrained further by comparing large $u$ asymptotics. 

\paragraph{The Special Case of $N=4$ and Unitarity of the Massless Bethe Equation} Let us now specialize to the case $N=4$ and $\alpha=\beta$. In this case both $\mQ_{\beta}$ and $\bar\mQ_{\beta}$ have the same large $u$ asymptotic and are equal to each other with exponential precision at large $u$, meaning that $P_{\alpha,\alpha}$ should decay exponentially for both $u\to \pm\infty$, which sets $\omega_{\alpha,\alpha}^{(0)}=\omega_{\alpha,\alpha}^{(2)}=0$ so for $N=4$ the function $P_{\alpha,\alpha}$ is fixed up to an overall real constant as
\beq
P_{\alpha,\alpha}=\frac{w_{\alpha}}{\prod_{n=1}^{N/2}
\sinh(\pi (u-\theta_n))}\;.
\eeq

Note that this means that for Baxter equation associated to node $6$ the Wronskians are thus also has the same periodic factor and we can write
\beq
\frac{
{\mathbb Q}_{\alpha}^+
\bar{\mathbb Q}_{\alpha}^-
\mQ_{\theta_2}
-
\bar{\mathbb Q}_{\alpha}^+
{\mathbb Q}_{\alpha}^-
\mQ_{\theta_1}
}{
\mathbb{Q}_{\dot{\beta}}^+
\bar{\mathbb{ Q}}_{\dot{\beta}}^-
\mQ_{\theta_2}
-
\bar{\mathbb{ Q}}_{\dot{\beta}}^+
\mathbb{Q}_{\dot{\beta}}
\mQ_{\theta_1}
} = \frac{w_{\alpha}\mQ_1\mQ_3}{ w_{\dot{\beta}}\mQ_7\mQ_5}\,,
\quad
\alpha=2,\tilde{2}\;,
\;\;
\dot{\beta} = 6,\tilde{6}\;,
\eeq
so evaluating the above relation at $u=\theta^{(2)}$ the first term is suppressed as it is multiplied by $Q_{\theta_2}$.
At the same time for $u\to\theta^{(1)}$ the first term is regular, whereas the second term has a first order pole, so again only the second terms in numerator and denominator survive, giving
\beq\la{relationNeeded}
\frac{
\bar{\mathbb Q}_{\alpha}^+
{\mathbb Q}_{\alpha}^-
}{
\bar{\mQ}_{\dot{\beta}}^+
\mathbb Q_{\dot{\beta}}^-
} = \frac{w_{\alpha}\mQ_1\mQ_3}{ w_{\dot{\beta}}\mQ_7\mQ_5}\bigg|_{u=\theta_k}\;\;,\;\;k=1,\dots,N\;.
\eeq

At this time, we can return to the unitarity of the massless equation. From the fact that $w_{\alpha}$ and $\dot{w}_{\dot{\beta}}$ are real it follows that the massless equation is unimodular, see \eqref{eq:unimod1}.

Similarly, we can consider the ratio of the Wronskians for 
$\alpha=2$ and $\alpha=\tilde{2}$ to get
\beq
\frac{
{\mathbb Q}_{2}^+
\bar{\mathbb Q}_{2}^-
\mQ_{\theta_2}
-
\bar{\mathbb Q}_{2}^+
{\mathbb Q}_{2}^-
\mQ_{\theta_1}
}{
{\mathbb Q}_{\tilde{2}}^+
\bar{\mathbb Q}_{\tilde{2}}^-
\mQ_{\theta_2}
-
\bar{\mathbb Q}_{\tilde{2}}^+
{\mathbb Q}_{\tilde{2}}^-
\mQ_{\theta_1}
} = \frac{w_{2}}{w_{\tilde{2}}}\,.
\eeq
Applying the same argument as above we arrive to
\beq\la{relation2}
\frac{
\bar{\mathbb Q}_{2}^+
{\mathbb Q}_{2}^-
}{
\bar{\mathbb Q}_{\tilde{2}}^+
{\mathbb Q}_{\tilde{2}}^-
}\bigg|_{u=\theta_k} = \frac{w_{2}}{w_{\tilde{2}}}\;\;\;,\;\;k=1,\dots,N\;.
\eeq

\subsection{The Conjugation Matrix $\Omega$ for $N=4$}\label{app:ConjugationOmega}

Since $\hat\mQ_{\beta}$ solves the same Baxter equation as $\mQ_{\beta}$ we can re-expand it as
\beq
\hat\mQ_{\beta}=\Omega_{\beta}{}^{\alpha}\;\mQ_{\alpha}\,,
\eeq
where $\Omega$ is a periodic matrix.  
By plugging this definition into \eqref{wronskianQbar} and using \eqref{wronskianQ} aswell as periodicity of $\Omega_{\beta}{}^{\alpha}$ we find
\begin{equation}\la{PtoOme}
    P_{\alpha\beta} = w_{2\tilde{2}}\,(\Omega^+)_{\alpha}{}^{\gamma}\epsilon_{\gamma\beta}\;.
\end{equation}

We now fix $\Omega$ as much as possible. First we use the following ansatz form $\Omega$, similar to \eq{Pansz}
\beq\la{Wansatz}
\Omega_{\alpha}{}^{\beta}=\frac{
\sum\limits_{n=0}^2 W^{(n)}_{\alpha,\beta}e^{2\pi n u}
}{(e^{2\pi u}+e^{2\pi\theta_1})(e^{2\pi u}+e^{2\pi\theta_2})}\;,
\eeq
where we have used that $N=4$ in this subsection. To constrain $\Omega$ we now proceed to find identities it must satisfy.

First let us write the effect of complex conjugation of $\mQ_{\beta}$ in terms of the initial $\mQ_{\beta}$ and $\Omega$
\beqa\la{app:conj}
\bar\mQ_{\beta}&=&
(-i)^{-i \left(\theta _1+\theta _2-\theta _3-\theta _4\right)}
\frac{\Gamma_2^+}{\Gamma_1^+}\Omega_{\beta}{}^{\alpha}\;\mQ_{\alpha}\;,
\eeqa
and applying this relation twice, we should get back to the initial $\mQ_{\beta}$ 
\beqa
(+i)^{-i \left(\theta _1+\theta _2-\theta _3-\theta _4\right)}
\frac{\bar\Gamma_1^-}{\bar\Gamma_2^-}\mQ_{\beta}&=&
\bar\Omega_{\beta}{}^{\alpha}\;\bar\mQ_{\alpha}=
(-i)^{-i \left(\theta _1+\theta _2-\theta _3-\theta _4\right)}
\frac{\Gamma_2^+}{\Gamma_1^+}\bar\Omega_{\beta}{}^{\alpha}\;\Omega_{\alpha}{}^{\gamma}\;\mQ_{\gamma}
\;,
\eeqa
which implies the following non-trivial identity
\beq\la{OmOmbar}
\bar\Omega_{\beta}{}^{\alpha}\;\Omega_{\alpha}{}^{\gamma}
=\delta_{\beta}^\gamma
e^{-\pi  \left(\theta _1+\theta _2-\theta _3-\theta _4\right)}
 \prod_{k=1}^{N/2}\frac{\cosh(\pi(u-\theta_{k+N/2}))}{\cosh(\pi(u-\theta_{k}))}\;.
\eeq
Furthermore, an additional constraints on $\Omega$ follow from large $u$ asymptotics of $\mQ_{\alpha}$ and $\hat\mQ_{\alpha}$. Using that $\frac{\Gamma_1}
{\Gamma_2}\simeq (i u)^{-i\sum_k(\theta_k-\theta_{K+k})}$ and being careful when deducing the asymptotics of $\hat \betheQ$ at $-\infty$ we find
\beqa
{\mathbb Q}_{1|1},\;{\mathbb Q}_{1|3} &\simeq_{u\rightarrow -\infty}&
e^{i\pi\alpha^\pm_{1,2}}|u|^{\alpha^\pm_{1,2}}\;,\\
\hat{\mathbb Q}_{1|1},\;\hat{\mathbb Q}_{1|3} &\simeq_{u\rightarrow -\infty}&
e^{2 \pi  \left(-\theta _1-\theta _2+\theta _3+\theta _4\right)}
e^{-i\pi\alpha^\pm_{1,2}}|u|^{\alpha^\pm_{1,2}}\;,
\eeqa
with
\begin{equation}
    \alpha^{\pm}_{12} = \frac{1}{2}\bigg(\pm(\Delta-S_2)-J_2+J_3+K_4-1+\ii \left(\theta_1+\theta_2-\theta_3-\theta_4\right)\bigg)\,.
\end{equation}
Which implies that we have
\begin{equation}\label{eq:AsymptoticsOmega}
    \Omega \simeq_{u \rightarrow \infty} \begin{pmatrix}
        1 & 0 \\
        0 & 1
    \end{pmatrix}\,,
    \quad
    \Omega \simeq_{u\rightarrow -\infty} \begin{pmatrix}
        e^{2\pi(-\theta_1-\theta_2+\theta_3+\theta_4)-2\pi \ii \alpha^+_{1,2}} & 0 \\
        0 & e^{2\pi(-\theta_1-\theta_2+\theta_3+\theta_4)-2\pi \ii \alpha^-_{1,2}} 
    \end{pmatrix}\,,
\end{equation}
in a basis $\betheQ_{\beta} = \{\betheQ_2,\betheQ_{\tilde{2}}\}$. We can repeat the same analysis for $Q_{6},Q_{\tilde{6}}$ and the Baxter equation they satisfy, let us write $\dot{\Omega}$ and $\dot{P}$ for the relevant matrices here for clarity. When deducing properties of this matrices we we will get the same result with $S_2,\;J_3\to -S_2,\;-J_3$. The crucial observation is that since $S_2$ and $J_3$ are integer numbers the diagonal components are related as $\Omega_{2}{}^{2}=\dot{\Omega}_{6}{}^{6}$ and $\Omega_{\tilde{2}}{}^{\tilde{2}}=\dot{\Omega}_{\tilde{6}}{}^{\tilde{6}}$! This in particular implies that, due to \eq{PtoOme}
\beq
\frac{P_{2,\tilde{2}}}{ \dot{P}_{6,\tilde{6}}}=\frac{w_{2\tilde{2}}}{\dot w_{6\tilde{6}}}=\frac{
\mQ_{2}^+\bar{\mQ}_{\tilde{2}}^-\mQ_{\theta_2}
-
\bar\mQ_{\tilde{2}}^+\mQ_{2}^-\mQ_{\theta_1}
}{{\mQ}_{6}^+\bar{\mQ}_{\tilde{6}}^-\mQ_{\theta_2}
-
\bar{\mQ}_{\tilde{6}}^+{\mQ}_{6}^-\mQ_{\theta_1}}
\frac{\mQ_7\mQ_5}{\mQ_1\mQ_3}\;.
\eeq
Giving
\begin{equation}
\frac{\bar{\betheQ}^+_{\tilde{2}}\betheQ^-_{2}}{\bar{\betheQ}^+_{\tilde{6}}\betheQ^-_6}\frac{\betheQ_7\betheQ_5}{\betheQ_1\betheQ_3}\bigg|_{u=\theta_k} = \frac{w_{2\tilde{2}}}{\dot{w}_{6\tilde{6}}}\,,
\quad
\frac{\bar{\betheQ}^+_{2}\betheQ^-_{\tilde{2}}}{\bar{\betheQ}^+_{6}\betheQ^-_{\tilde{6}}}\frac{\betheQ_7\betheQ_5}{\betheQ_1\betheQ_3}\bigg|_{u=\theta_k} = \frac{w_{2\tilde{2}}}{\dot{w}_{6\tilde{6}}}\,,
\end{equation}
where the second equality comes from considering also $P_{\tilde{2},2}/P_{\tilde{6},6}$. Dividing these two equations with each other we obtain
\begin{equation}\label{eq:ConsistencyBaxter}
    \frac{\betheQ_2^-}{\bar{\betheQ}_2^+}\frac{\betheQ_{\tilde{6}}^-}{\bar{\betheQ}_{\tilde{6}}^+}\frac{\bar{\betheQ}_{\tilde{2}}^+}{\betheQ_{\tilde{2}}^-}\frac{\bar{\betheQ}_{6}^+}{\betheQ_{6}^-}\bigg|_{u=\theta_k} = 1\,,
\end{equation}
which then ensures the consistency of the massless Bethe equations \eq{BAEmassless}, see \eqref{eq:UniModConsistency}.

\subsubsection{Explicit Expression for $\Omega$ }\label{app:ExplicitConjugationOmega}
Let us finally consider the explicit form of $\Omega$ obtained upon imposing \eqref{OmOmbar} and \eqref{eq:AsymptoticsOmega}. We use notation
\begin{equation}
    \Omega = \begin{pmatrix}
        \Omega_2{}^{2} & \Omega_{2}{}^{\tilde{2}} \\
        \Omega_{\tilde{2}}{}^{2} & \Omega_{\tilde{2}}{}^{\tilde{2}}
    \end{pmatrix}\,.
\end{equation}
To slightly simplify notation we introduce $d$, (for denominator),
\beq
d=\left(-1+e^{2 i \left.\text{($\Delta \pm $}S_2\right) \pi }\right) \left(e^{2 \pi  u}+e^{2 \pi  \theta _1}\right) \left(e^{2 \pi  u}+e^{2 \pi  \theta _2}\right)\,,
\eeq
then we find
\beqa
&&\Omega_{2}{}^{2}\; d=
-e^{4 \pi  u}+e^{2 i \pi  \left.\text{($\Delta \pm $}S_2\right)+4 \pi  u}-e^{2 \pi  \left(u+\theta _1\right)}-e^{2 \pi  \left(u+\theta _2\right)}-e^{2 \pi  \left(u+\theta _3\right)}\\&&
\nonumber -e^{\pi  \left(i \left.\text{($\Delta \pm $}S_2\right)+2 u+\theta _1+\theta _2+\theta _3-\theta _4\right)}
-e^{2 \pi  \left(u+\theta _4\right)}-e^{\pi  \left(i \left.\text{($\Delta \pm $}S_2\right)+2 u+\theta _1+\theta _2-\theta _3+\theta _4\right)}\\&&
\nonumber -e^{\pi  \left(i \left.\text{($\Delta \pm $}S_2\right)+2 u+\theta _1-\theta _2+\theta _3+\theta _4\right)}-e^{\pi  \left(i \left.\text{($\Delta \pm $}S_2\right)+2 u-\theta _1+\theta _2+\theta _3+\theta _4\right)}\\&&
\nonumber +e^{\pi  \left(-i \left.\text{($\Delta \pm $}S_2\right)+\theta _1+\theta _2+\theta _3+\theta _4\right)}-e^{\pi  \left(i \left.\text{($\Delta \pm $}S_2\right)+\theta _1+\theta _2+\theta _3+\theta _4\right)}
\eeqa
\beqa
&&\frac{\Omega_{2}{}^{\tilde{2}}\; d}{i w e^{+2\pi u}(e^{2 i \left.\text{($\Delta \pm $}S_2\right) \pi }-1)}=1
\eeqa
\beqa
&&\Omega_{\tilde{2}}{}^{2}\; d\;i w e^{-2\pi u}(e^{2 i \left.\text{($\Delta \pm $}S_2\right) \pi }-1)=-e^{2 i \pi  \left.\text{($\Delta \pm $}S_2\right)+4 \pi  \theta _1}-e^{2 \pi  \left(\theta _1+\theta _2\right)}-2 e^{2 \pi  \left(i \left.\text{($\Delta \pm $}S_2\right)+\theta _1+\theta _2\right)}\\&&
\nonumber-e^{2 \pi  \left(2 i \left.\text{($\Delta \pm $}S_2\right)+\theta _1+\theta _2\right)}-e^{2 i \pi  \left.\text{($\Delta \pm $}S_2\right)+4 \pi  \theta _2}-e^{2 \pi  \left(\theta _1+\theta _3\right)}-2 e^{2 \pi  \left(i \left.\text{($\Delta \pm $}S_2\right)+\theta _1+\theta _3\right)}-e^{2 \pi  \left(2 i \left.\text{($\Delta \pm $}S_2\right)+\theta _1+\theta _3\right)}\\&&
\nonumber-e^{2 \pi  \left(\theta _2+\theta _3\right)}-2 e^{2 \pi  \left(i \left.\text{($\Delta \pm $}S_2\right)+\theta _2+\theta _3\right)}-e^{2 \pi  \left(2 i \left.\text{($\Delta \pm $}S_2\right)+\theta _2+\theta _3\right)}-e^{2 i \pi  \left.\text{($\Delta \pm $}S_2\right)+4 \pi  \theta _3}\\&&
\nonumber-e^{2 \pi  \left(i \left.\text{($\Delta \pm $}S_2\right)+\theta _1+\theta _2+\theta _3-\theta _4\right)}-e^{\pi  \left(i \left.\text{($\Delta \pm $}S_2\right)+3 \theta _1+\theta _2+\theta _3-\theta _4\right)}-e^{\pi  \left(3 i \left.\text{($\Delta \pm $}S_2\right)+3 \theta _1+\theta _2+\theta _3-\theta _4\right)}\\&&
\nonumber-e^{\pi  \left(i \left.\text{($\Delta \pm $}S_2\right)+\theta _1+3 \theta _2+\theta _3-\theta _4\right)}-e^{\pi  \left(3 i \left.\text{($\Delta \pm $}S_2\right)+\theta _1+3 \theta _2+\theta _3-\theta _4\right)}-e^{\pi  \left(i \left.\text{($\Delta \pm $}S_2\right)+\theta _1+\theta _2+3 \theta _3-\theta _4\right)}\\&&
\nonumber-e^{\pi  \left(3 i \left.\text{($\Delta \pm $}S_2\right)+\theta _1+\theta _2+3 \theta _3-\theta _4\right)}-e^{2 \pi  \left(\theta _1+\theta _4\right)}-2 e^{2 \pi  \left(i \left.\text{($\Delta \pm $}S_2\right)+\theta _1+\theta _4\right)}-e^{2 \pi  \left(2 i \left.\text{($\Delta \pm $}S_2\right)+\theta _1+\theta _4\right)}\\&&
\nonumber-e^{2 \pi  \left(\theta _2+\theta _4\right)}-2 e^{2 \pi  \left(i \left.\text{($\Delta \pm $}S_2\right)+\theta _2+\theta _4\right)}-e^{2 \pi  \left(2 i \left.\text{($\Delta \pm $}S_2\right)+\theta _2+\theta _4\right)}-e^{2 \pi  \left(i \left.\text{($\Delta \pm $}S_2\right)+\theta _1+\theta _2-\theta _3+\theta _4\right)}\\&&
\nonumber-e^{\pi  \left(i \left.\text{($\Delta \pm $}S_2\right)+3 \theta _1+\theta _2-\theta _3+\theta _4\right)}-e^{\pi  \left(3 i \left.\text{($\Delta \pm $}S_2\right)+3 \theta _1+\theta _2-\theta _3+\theta _4\right)}-e^{\pi  \left(i \left.\text{($\Delta \pm $}S_2\right)+\theta _1+3 \theta _2-\theta _3+\theta _4\right)}\\&&
\nonumber-e^{\pi  \left(3 i \left.\text{($\Delta \pm $}S_2\right)+\theta _1+3 \theta _2-\theta _3+\theta _4\right)}-e^{2 \pi  \left(\theta _3+\theta _4\right)}-2 e^{2 \pi  \left(i \left.\text{($\Delta \pm $}S_2\right)+\theta _3+\theta _4\right)}-e^{2 \pi  \left(2 i \left.\text{($\Delta \pm $}S_2\right)+\theta _3+\theta _4\right)}\\&&
\nonumber-e^{2 \pi  \left(i \left.\text{($\Delta \pm $}S_2\right)+\theta _1-\theta _2+\theta _3+\theta _4\right)}-e^{\pi  \left(i \left.\text{($\Delta \pm $}S_2\right)+3 \theta _1-\theta _2+\theta _3+\theta _4\right)}-e^{\pi  \left(3 i \left.\text{($\Delta \pm $}S_2\right)+3 \theta _1-\theta _2+\theta _3+\theta _4\right)}\\&&
\nonumber-e^{2 \pi  \left(i \left.\text{($\Delta \pm $}S_2\right)-\theta _1+\theta _2+\theta _3+\theta _4\right)}-e^{\pi  \left(-i \left.\text{($\Delta \pm $}S_2\right)+\theta _1+\theta _2+\theta _3+\theta _4\right)}-3 e^{\pi  \left(i \left.\text{($\Delta \pm $}S_2\right)+\theta _1+\theta _2+\theta _3+\theta _4\right)}\\&&
\nonumber-3 e^{\pi  \left(3 i \left.\text{($\Delta \pm $}S_2\right)+\theta _1+\theta _2+\theta _3+\theta _4\right)}-e^{\pi  \left(5 i \left.\text{($\Delta \pm $}S_2\right)+\theta _1+\theta _2+\theta _3+\theta _4\right)}-e^{\pi  \left(i \left.\text{($\Delta \pm $}S_2\right)-\theta _1+3 \theta _2+\theta _3+\theta _4\right)}\\&&
\nonumber-e^{\pi  \left(3 i \left.\text{($\Delta \pm $}S_2\right)-\theta _1+3 \theta _2+\theta _3+\theta _4\right)}-e^{\pi  \left(i \left.\text{($\Delta \pm $}S_2\right)+\theta _1-\theta _2+3 \theta _3+\theta _4\right)}-e^{\pi  \left(3 i \left.\text{($\Delta \pm $}S_2\right)+\theta _1-\theta _2+3 \theta _3+\theta _4\right)}\\&&
\nonumber-e^{\pi  \left(i \left.\text{($\Delta \pm $}S_2\right)-\theta _1+\theta _2+3 \theta _3+\theta _4\right)}-e^{\pi  \left(3 i \left.\text{($\Delta \pm $}S_2\right)-\theta _1+\theta _2+3 \theta _3+\theta _4\right)}-e^{\pi  \left(i \left.\text{($\Delta \pm $}S_2\right)+\theta _1+\theta _2-\theta _3+3 \theta _4\right)}\\&&
\nonumber-e^{\pi  \left(3 i \left.\text{($\Delta \pm $}S_2\right)+\theta _1+\theta _2-\theta _3+3 \theta _4\right)}-e^{\pi  \left(i \left.\text{($\Delta \pm $}S_2\right)+\theta _1-\theta _2+\theta _3+3 \theta _4\right)}-e^{\pi  \left(3 i \left.\text{($\Delta \pm $}S_2\right)+\theta _1-\theta _2+\theta _3+3 \theta _4\right)}\\&&
\nonumber-e^{\pi  \left(i \left.\text{($\Delta \pm $}S_2\right)-\theta _1+\theta _2+\theta _3+3 \theta _4\right)}-e^{\pi  \left(3 i \left.\text{($\Delta \pm $}S_2\right)-\theta _1+\theta _2+\theta _3+3 \theta _4\right)}-e^{2 i \pi  \left.\text{($\Delta \pm $}S_2\right)+4 \pi  \theta _4}
\eeqa
\beqa
\nonumber&&\Omega_{\tilde{2}}{}^{\tilde{2}}d=
e^{2 i \pi  \left.\text{($\Delta \pm $}S_2\right)+4 \pi  u}+e^{2 \pi  \left(i \left.\text{($\Delta \pm $}S_2\right)+u+\theta _1\right)}+e^{2 \pi  \left(i \left.\text{($\Delta \pm $}S_2\right)+u+\theta _2\right)}+e^{2 \pi  \left(i \left.\text{($\Delta \pm $}S_2\right)+u+\theta _3\right)}\\
&&+e^{\pi  \left(i \left.\text{($\Delta \pm $}S_2\right)+2 u+\theta _1+\theta _2+\theta _3-\theta _4\right)}+e^{2 \pi  \left(i \left.\text{($\Delta \pm $}S_2\right)+u+\theta _4\right)}+e^{\pi  \left(i \left.\text{($\Delta \pm $}S_2\right)+2 u+\theta _1+\theta _2-\theta _3+\theta _4\right)}\\
\nonumber&&+e^{\pi  \left(i \left.\text{($\Delta \pm $}S_2\right)+2 u+\theta _1-\theta _2+\theta _3+\theta _4\right)}+e^{\pi  \left(i \left.\text{($\Delta \pm $}S_2\right)+2 u-\theta _1+\theta _2+\theta _3+\theta _4\right)}+e^{\pi  \left(i \left.\text{($\Delta \pm $}S_2\right)+\theta _1+\theta _2+\theta _3+\theta _4\right)}\\
\nonumber&&-e^{\pi  \left(3 i \left.\text{($\Delta \pm $}S_2\right)+\theta _1+\theta _2+\theta _3+\theta _4\right)}-e^{4 \pi  u}
\eeqa

\bibliographystyle{JHEP}
\bibliography{ref}

\end{document}